\documentclass[pra,aps,twocolumn,10pt,floatfix,superscriptaddress]{revtex4-2}
\usepackage{amsmath}
\usepackage{amssymb}
\usepackage{graphicx}
\usepackage{color}
\usepackage{soul}
\usepackage[utf8]{inputenc}
\usepackage[T1]{fontenc}
\usepackage{hyperref}
\usepackage{array}
\usepackage{ulem}
\usepackage{comment}
\usepackage{multirow}
\usepackage[caption=false]{subfig}
\usepackage{hyperref}
\hypersetup{
	colorlinks=true,
	linkcolor=blue,
	filecolor=magenta,     
	urlcolor=cyan,
	pdftitle={Overleaf Example},
	pdfpagemode=FullScreen,
}

\begin{document}

\title{Evidence  of Kolmogorov like scalings and multifractality  in the rainfall events }

\author{Joya GhoshDastider}
\affiliation{Department of Physics, Indian Institute of Technology, Guwahati 781039, Assam, India} 

\author{Dilip Pal}
\affiliation{Department of Physics, Indian Institute of Technology, Guwahati 781039, Assam, India} 

\author{Pankaj Kumar Mishra}
\affiliation{Department of Physics, Indian Institute of Technology, Guwahati 781039, Assam, India}

\date{\today}

\begin{abstract} 


In this paper we present a detailed statistical analysis related to the characterization of the spatial and temporal fluctuations present in the rainfall  patterns of North-East region ($26.05^{\circ}N-26.95^{\circ}N$, $88.05^{\circ}E-94.95^{\circ}E$) of India using half hourly rainfall data over the last 20 years for the range 2001-2020. We analyze the  nature of the distribution by computing the mean, second moment of the fluctuation, skewness and kurtosis of the temporal rainfall data that indicate the presence of heavy tail in the right skewed distribution a typical feature of the presence of rare events.  We find that the temporal distribution of the rainfall data follow the multiplicative Log-Normal probability distribution. Further we compute the spatial and temporal correlation of the rainfall in this region indicate that the rainfall events are correlated in the spatial direction of about 70 Km. The Power spectral density of temporal rainfall shows power law behaviour with  frequency with an exponent $\sim -1.5$ close to the Kolmogorov exponent ($-1.67$) exhibited for the turbulent passive scalar driven by the mean flow. Our wavelet analysis reveals the evidence of multiple frequencies in the rainfall pattern which can attributed to different short and long range factors responsible for the rainfall. We have also used the Hilbert Huang transformation to identify the frequencies corresponding fluctuating part of the rainfall time series. Using multifractal detrended fluctuation analysis, finally we establish the multifractal nature of the rainfall pattern with Hurst exponent close to $0.65$ .

\end{abstract}

\flushbottom

\maketitle

\section{Introduction}\label{sec1}





In recent years the research on complex system has seen the unprecedented growth in number due to its manifestation in different spectrum of natural phenomena~\cite{PhysRevX.11.011011,PhysRevB.107.094307,PhysRevApplied.17.064021,PhysRevLett.128.024502,PhysRevE.66.036120,PhysRevResearch.2.013264,moustakis2020atmospheric, PhysRevLett.86.4286, parisi1999complex, papo2023does,PhysRevLett.125.048105, PhysRevResearch.2.043352}. Among them Rainfall is a very complex process owing to its dependency on several atmospheric components, like, temperature, pressure, humidity, wind flow direction along with the topography of earth surface, geographical position, etc.~\cite{neelin2022precipitation}. Similar to other complex phenomena, rainfall is also highly nonlinear encompassing the several length and time scale which is difficult to quantify through the dynamical models~\cite{holovatch2017complex}.

In recent years there are several models that have been used to describe the rainfall processes. In this direction, researchers have mainly adopted two class of deterministic models that include Global Climate~\cite{mandal2019reservoir} and Statistical Dynamical Model~\cite{kurihara1970statistical}. However, stochastic nature of the rainfall is least explored~\cite{palmer2019stochastic}.  Hasselmann~\cite{hasselmann1976stochastic,frankignoul1977stochastic,lemke1977stochastic} introduced the stochastic model to describe the phenomenology of the sea surface temperature. Although these seminal models yield the promising nature of the events, they lack the essence of exact physical details and mechanisms involved in the rainfall process that imposes a limitation on the accuracy and performance of these models~\cite{raju2020review, vallis1982statistical}. Similar to these models, there are several works that have been proposed to unravel the underlying mechanism  of the rainfall. Some of the analytical models consider that the convective precipitation gets triggered by nonlinear continuous phase transition when the accumulated water vapour results in intense precipitation as it increases beyond a certain threshold value~\cite{peters2006critical}. The matching of output of these models with the actual rainfall process very much depend upon many free parameters. This particular feature again posses some limitation in the robustness of these models. Therefore, it is quite daunting task to come up with a model with minimal free parameter that can accurately describe the real time behaviour of the rainfall. In past few years the time-series analysis of the rainfall data has become an essential ingredient to characterize the rainfall data and come up with a model that can capture the real time process in the adequate manner. In this paper we have analyzed the statistical behaviour of the rainfall in the north east region of India which has typical topography.



Rainfall is a multiscale phenomena. This whole phenomenon is far from equilibrium and can be characterized using the two time scales: first the slow time scale in which the heat is transferred from the solar radiation to the atmosphere and  second one is the fast convective flow resulting the increase of moisture in the atmosphere that rises against the gravity and results as the rain~\cite{neelin2022precipitation}. Having these two time scales embedded in the rainfall event, in recent years it has been modeled in the similar line as for self organization criticality (SOC)~\cite{PhysRevLett.59.381}. The slowly driven convective flow resulting in an avalanche-like rainfall event encompasses several features which can be identified as the SOC like behaviour~\cite{PhysRevLett.88.018701}. Using the SOC like model for rainfall event it has been shown that the event size distribution follows the  power law behaviour at different stations from diverse locations indicating the presence of universality in the rainfall event~\cite{peters2010universality}. There are several works that consider the rainfall as a stochastic process.  Depending on the continuous phase transition phenomena of rainfall process~\cite{peters2006critical}, two prototype models relating water column vapour and precipitation  have been proposed that consider the two state model for convective onset and three state model with stratiform precipitation. The critical exponents obtained using SOC model and stochastic forcing have been further utilized to connect the rainfall process as a first-passage-time problem~\cite{stechmann2014first}.  

Characterizing the nature of the rainfall distribution is quite non-trivial task as it depends upon a lot of factors. There are many studies that have been undertaken to unfold the rainfall patterns in different part of world using the statistical models, however a clear pattern for the rainfall is still lacking. Some studies have used the extreme event statistics to characterize the rainfall~\cite{yalcin2016extreme} and proposed that the rainfall events can be described better with the Poisson model. On the other hand the presence of skewed distribution of the rainfall in other parts of the word has been reported using principal component analysis and three coherent regions of rainfall has been identified~\cite{ibebuchi2023rainfall}. Using twenty years rainfall data (1999-2018) for Jakarta it was shown that the Log-Normal distribution with three free parameters and Log-Pearson type III distribution appear to fit the rain distribution better than other closely associated distribution such as, Gumbel, Pearson 3, Gamma, Log-Gamma, Normal, Log-Normal ( with 2 parameters), etc.~\cite{kurniawan2019distribution}. 
However, studies performed for the rainfall of the Australian continent do not show mixture of both Gamma and Log-normal distribution for the rare rainfall events.  On the other hand, Cho \textit{et al.} demonstrated that the Log-Normal distribution fits better in the dry region while Gamma distribution fits better in the wet region of the rainfall for the tropical rain~\cite{cho2004comparison}.
 
One of the pertinent question that arise in the analysis of the rainfall is what are the control parameters which govern the distribution and statistics of the rainfall in a particular region. India mainly receives convective rainfall during the summer-monsoon season that typically spans between June-September of the year. There are different hypothesis available to describe the prominent cause of rainfall in Indian peninsula. Some of the recent works indicate it as a manifestation of the seasonal migration of the intertropical convergence zone~\cite{rajeevan2010active}. Indian summer-monsoon rainfall inherits two kind of characteristics. First is the intermittent occurrence of the rainfall happens in the quasi-periodic manner in which the whole event can be described as a combination of the active and break spells termed as intraseasonal variation of the rainfall at the supersynoptic scale. Second one is the northward drifting of the effective rainfall region in a period of two to six weeks of time~\cite{gadgil2003indian}. In this paper explore the rainfall in the topography region mainly North East part of India which is is mainly wet in nature.

Most of the studies~\cite{mitra2018spatio, karmakar2017increased} reveal that due to its unique geographical position North-East part of India behaves differently than the other main land part of India. One of the most prominent features of rainfall is the presence of extreme events during the prolonged summer-monsoon season in this region, which lacks in the rest parts of India~\cite{goswami2010multiscale}. Through different statistical analysis researchers have obtained that North-East rainfall pattern in India in general appears to be negatively correlated from the rainfall intensity in the rest part of India~\cite{rajeevan2010active}. Due to the presence of these anomalies it requires some special care while proposing a comprehensive rainfall model in this region. Due to the presence of many rare events, models based on the studies are not so robust and depend very much on the spatial location as well as the year of the rainfall event. Some statistical studies employing Sen's estimator and Mann-Kendall test for the North-East region of India for the time span of one hundred and thirty seven years (1871-2008) reported discrete trends for seasons and hydro-meteorological subdivisions but could not find any general trend as a whole for the entire region~\cite{jain2013analysis}. In another study, five different probability distribution functions (Normal, Log-normal, Gumbel, Log-logistic and Exponential) were examined and none of them found to be suitable. All these attributes need more quantitative investigation in order to understand the driving factors of these kind of anomalous behaviours in the North-East region of India. 


So far most of the model related to the rainfall is based on the fact that the event is linear in nature where the fluctuating part of the event is smaller than the mean part. However, the whole process that is responsible for the rainfall is highly non-linear and hence a proper analysis for the noise embedded in the rainfall is sought in order to come up with a robust model. In this paper we bridge the gap by providing a detailed statistical as well spectral analysis of the rainfall data of North-East region of India and propose some universal nature of the distribution of the rainfall as well the characteristics of the noise. We consider the real time rainfall data and through the statistical analysis like, mean, second moment of the fluctuation, skewness and kurtosis of the temporal rainfall data find presence of skewed distribution of the rare events which shows the Log Normal probability distribution. Our spectral analysis of the  temporal rainfall shows Kolmogorov like scaling for the power spectral density (with an exponent $-1.5$) for the rainfall event during one day of time which is quite universal for all the year and spatial points in this region. We also analyze the local frequency present in the rainfall event using the wavelet analysis that reveals different characteristic time scale which we have related with many natural events, like, interseasonal variation of the monsoon season, time scale for the cyclone, depression, etc.  Further we establish the presence of multifractality nature of rain event in this north-east region.

The paper is structured as follows. In Sec.~\ref{sec2} we present our analysis related to the statistical distribution, spectral calculation, wavelet, Hilbert Huang transformation and multifractal analysis. Finally we conclude our paper in Sec.~\ref{sec8}. We provide the relevant details of distribution fitting, methodology for wavelet and Hilbert Huang transformation and multifractal analysis respectively in the Appendix~\ref{app:a}, Appendix~\ref{app:b}, Appendix~\ref{app:c}, Appendix~\ref{app:d}.

\section{Results}\label{sec2}
In this paper one of the main aim is to characterize the spatio  and temporal fluctuation present in the rainfall event in the North-East region of India. To unveil the prominent characteristics and to understand the underlying mechanism of rainfall process, we consider the rainfall time series of the North-East region ($26.05^{\circ}N-26.95^{\circ}N$, $88.05^{\circ}E-94.95^{\circ}E$) of India with half hourly rainfall data from May to October month of last 20 years (2001-2020). The North-East region of India contains two highest raining places of the world, Mawsynram (11871 mm/year) and Cherrapunji (11777 mm/year) which are extended Himalayan part of India that witnesses a long monsoon season along with a few intense patches of shower that leads to flood in Brahmaputra river every year. Unique geographical position and different weather variables contribute to the distinctive patterns along with extreme events in rainfall of this region. The onset of monsoon in the North-East region in general happens from May~\cite{sharma2023variability} and it extends until beginning of October. As we are interested in analyzing the rainfall during the Monsoon season of North East part of India, in this paper we consider the rainfall data from May to October month of the year that spans for twenty years (2001-2020). 


For all our analysis we have considered the half hourly rainfall data recorded (in  millimeter unit) which have been obtained from the NASA project known as ``Integrated Multi-satellitE Retrievals for Global Precipitation Measurement'' (GPM IMERG)~\cite{datasets}, NASA, USA. The data have $0.1^{\circ} \times 0.1^{\circ}$ spatial resolution on Cartesian grid (approximately 10 km $\times$ 10  km) procured using passive microwave sensors situated at various precipitation-relevant satellite. To differentiate between the dry and wet days, we have used a threshold cutoff on the half-hourly rainfall. If the threshold value of the rainfall intensity is lower than 0.001 mm we consider it dry half hour otherwise it is a wet half hour.


\begin{figure}  [!htp]     
\centering
\includegraphics[width=0.48\textwidth]{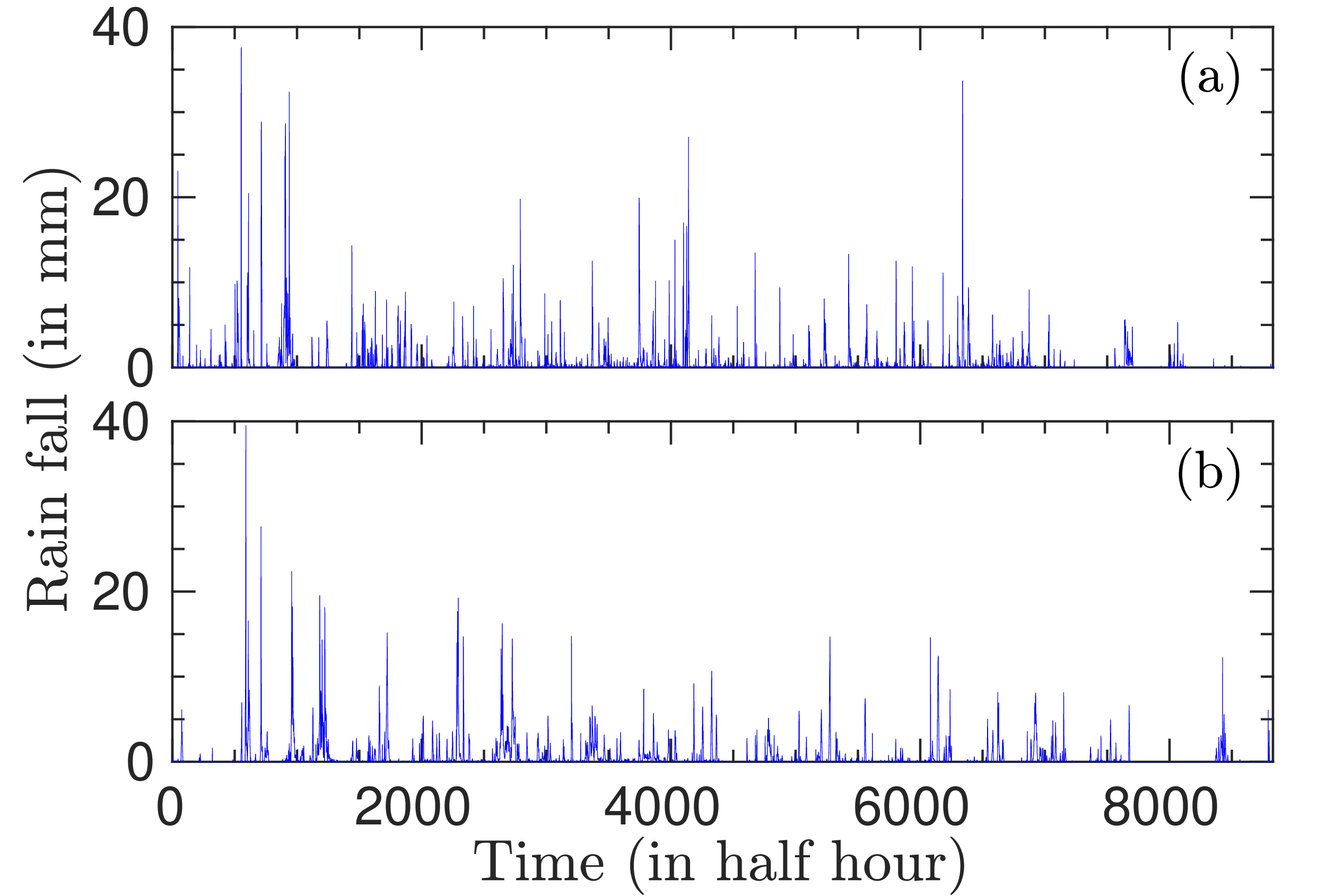}
\caption{Temporal evolution of rainfall of (a) station $26.05^{\circ}$N, $88.05^{\circ}$E, year 2010 and (b) station $26.55^{\circ}$N, $91.65^{\circ}$E, year 2020 respectively. The time-series of rainfall event  for both the stations exhibits the stochastic variation accompanied by intermittent jump of the rain intensity suggesting the presence of nonlinear nature of the rainfall events.} \label{fig1}
\end{figure}

In Fig.~\ref{fig1} we show the temporal evolution of rainfall data for two different stations (a) $26.05^{\circ}$N, $88.05^{\circ}$E  and  (b) $26.55^{\circ}$N, $91.65^{\circ}$E for two different years 2010 and 2020 respectively. The time series exhibits that the rainfall events are random with time and shows intermittent jumps that indicate towards the presence of complicated nonlinear fluctuations in the rainfall events. In this work we are intended to analyze the detailed nature of these nonlinear fluctuations which happens to be multifractal in nature.  




\subsection{Statistical analysis of rainfall time-series } \label{sec3}
\begin{figure*} [!htp]      
\centering
\includegraphics[width=1\textwidth]{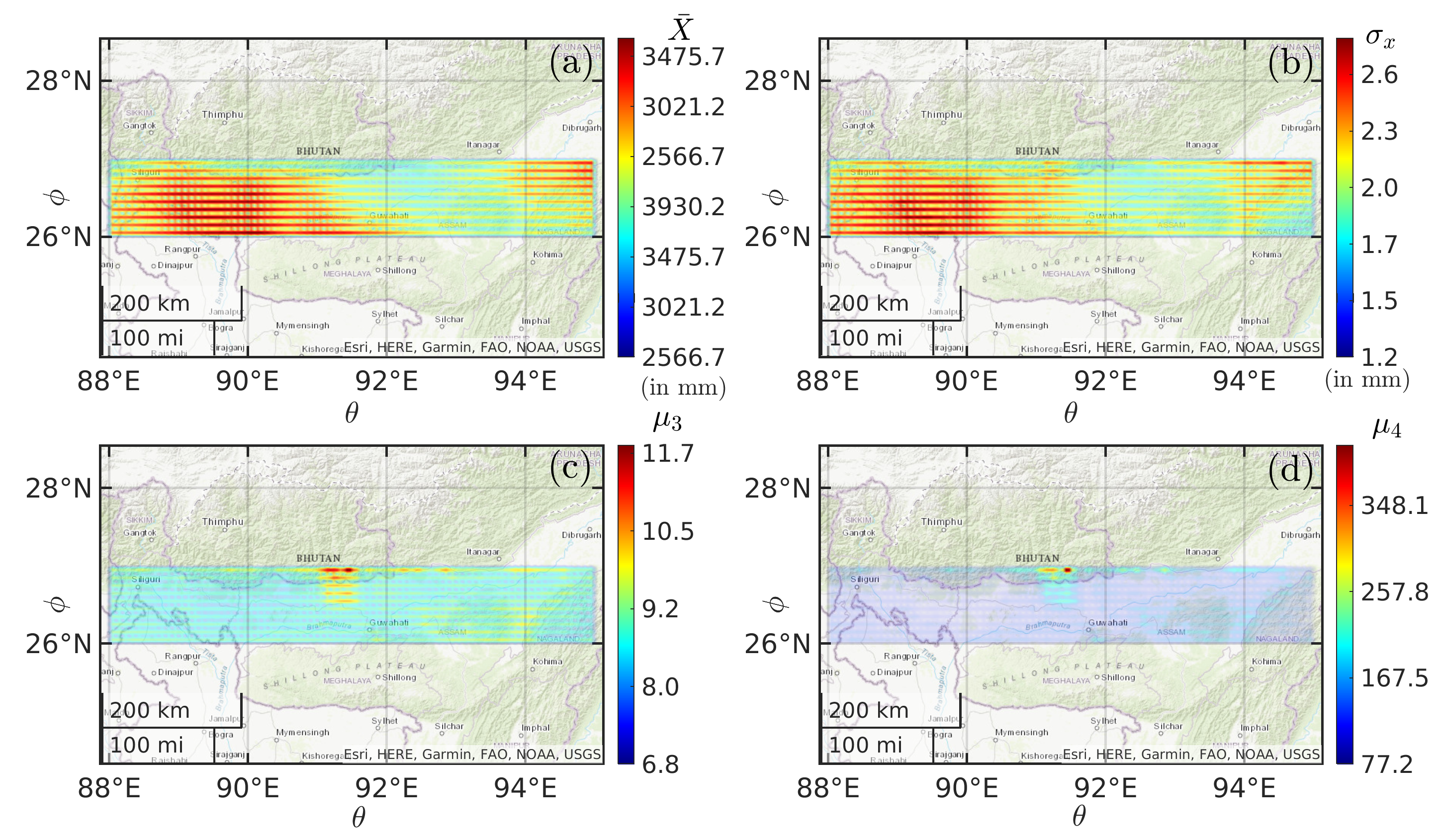}
\caption{Different statistical variables averaged over twenty years (2001-2020) for the selected area under investigation. (a) mean, (b) standard deviation, (c) skewness and (d) kurtosis  on the geographic map in the $\theta - \phi$ plane. Color bar represents the magnitude of different statistical quantities. These values indicate that the distribution of rainfall intensity is right skewed and possess a heavy tail.}  \label{fig2}
\end{figure*}

In this section, we perform detailed statistical analysis of the rainfall time series. First we calculate different moments of the rainfall time series to understand its statistical behaviour. After that we determine the suitable probability distribution function (PDF) of the rainfall for this region. The proposed PDF is Log Normal in nature that has been also confirmed by measuring the goodness of fit tests. We have further presented the spatial and temporal correlation of the rainfall data that show the presence of weak correlation in time while relatively string correlation in space.  

We begin our analysis by calculation the different statistical moment of  the real time rainfall data over the region. For this we have computed the  mean ($\Bar{X}$), standard deviation ($\sigma_{x}$), skewness ($\mu_{3}$) and kurtosis ($\mu_{4}$) of the rainfall and extracted a detailed feature of rainfall distribution. In Table~\ref{table1} we provide the formula for the different moments of the rainfall event used for our studies.
\begin{table}[!htp] 
\centering
\caption{Definition of different moments used to characterize the statistics of the rainfall data~\cite{samui2019handbook}.} \label{table1}
\begin{tabular}{m {4 cm} m {3 cm}}
\hline \\
Statistical variable & Formula \\ \\
\hline\hline \\
Mean ($\Bar{X}$) & $\frac{1}{N}\sum_{i=1}^N X_{i}$\\ \\
Standard deviation ($\sigma_{x}$) & $\sqrt{\frac{\sum_{i=1}^N X_{i} - \Bar{X}}{N}}$\\ \\
Skewness ($\mu_{3}$) & $\frac {\sum_{i=1}^N (X_{i}-\Bar{X})^{3}}{(N-1) \sigma^{3}_{X}}$ \\ \\
Kurtosis ($\mu_{4}$) & $\frac {\sum_{i=1}^N (X_{i}-\Bar{X})^{4}}{(N-1) \sigma^{4}_{X}}$\\ \\
\hline
\end{tabular}
\end{table}
Here, $X$ is the random variable corresponding to the intensity of rainfall and $N$ is the total number of data points in rainfall time series. In Fig.~\ref{fig2} we depict all the computed moments (in pseudo color) of the rainfall data at different stations averaged over 20 years on  actual geographical map choosing longitude ($\theta$) as x-axis and latitude ($\phi$) as y-axis.  We notice that the maximum mean rainfall over these 20 years is 3475.7 mm and the minimum of it is 2566.7 mm over the six months [see Fig.~\ref{fig2}(a)]. The variance measures the dispersion from the mean of the dataset. The mean standard deviation (square root of variance) has a maximum value 2.6 mm and minimum value 1.2 mm [see Fig.~\ref{fig2}(b)]. Skewness is the third moment of rainfall data and it shows the asymmetry in the data about its mean. Here for our datasets the Skewness values are scattered between 11.7 (maximum) to 6.8 (minimum) [see Fig.~\ref{fig2}(c)]. The fourth moment of data is known as kurtosis which gives the relative idea whether the data is concentrated at center or it is spread over its tails. The Kurtosis values of the whole datasets are spread from 348.1 (maximum) to 77.2 (minimum) [see Fig.~\ref{fig2}(d)]. 

The Skewness and kurtosis values for a normal distribution is 0 and 3 respectively. In Fig.~\ref{fig2}(c), we find that Skewness value of all the stations averaged over all the years turns out to be positive which implies that the rainfall distribution is skewed at the right side. All the Kurtosis values for all the stations are greater than 3 implying the fall of the distribution is lower than that those for the normal distribution indicating the presence of heavy tail in the distribution of the rainfall data [see Fig.~\ref{fig2}(d)]. These results reveal that rainfall data does not follow Gaussian statistics. Fat tailed distribution is one of the main characteristics of complex systems~\cite{holovatch2017complex}. So it is quite necessary to obtain a detailed nature of probability distribution function of the rainfall event in order to characterize it in a better way.

\subsubsection{Probability Distribution Function of rainfall in North-East Region}  \label{sec4}
\begin{figure*} [!htp]      
\centering
\includegraphics[width=1\textwidth]{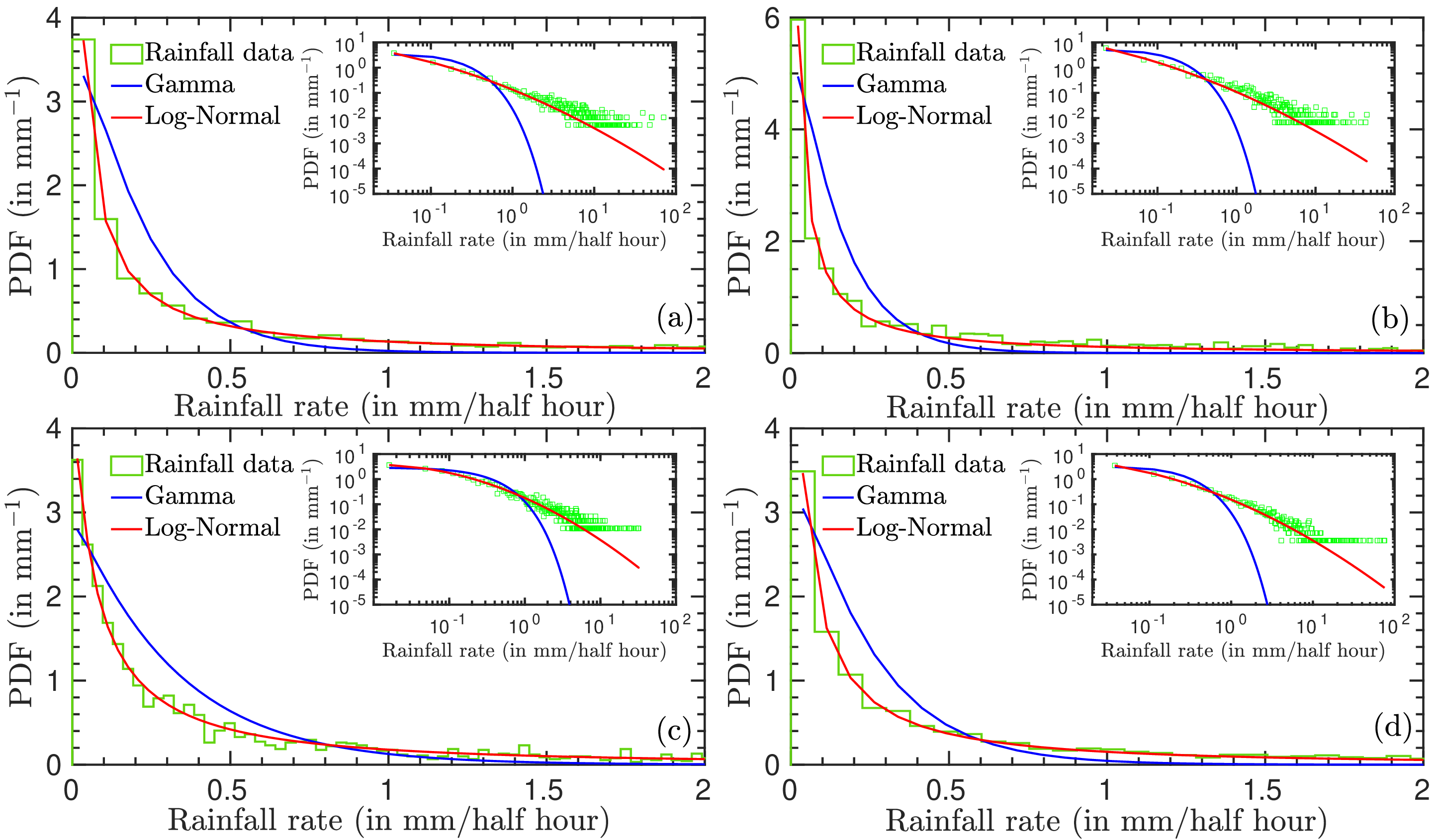}
\caption{PDF of the wet half hour rainfall data. In the left panels (a)for  station $26.05^{\circ}$N, $88.05^{\circ}$E, year 2005 and (c) for station $26.05^{\circ}$N, $88.05^{\circ}$E, year 2015 while in the right panel (b) for station $26.55^{\circ}$N, $91.65^{\circ}$E, year 2005 and (d) for station $26.55^{\circ}$N, $91.65^{\circ}$E, year 2015. For all the stations over all the year, Log Normal PDF (red curve) is found to be better fitted than the Gamma PDF (blue curve). The inset of each figure contains the rainfall intensity data (green square) fitted with these two distribution functions in log-log scale suggesting the better fitted Log-normal distribution.} \label{fig3}
\end{figure*}
Among different probability distribution functions (PDF), both Gamma and Log Normal distribution function possess right skewed and heavy tail features along with validity for positive arguments only (as rainfall magnitude can take zero or positive values only). There are several studies that point out both log normal or Gamma distribution for the rainfall data is suitable depending on different locations and other control parameters~\cite{deka2008use,cho2004comparison}. The origin of Log-Normal distribution is multiplicative process. Both Log- Normal and Gamma PDF belongs to Gumbel extreme distribution class. Based on the results of statistical analysis, we find Log Normal and Gamma PDF as the two possible contender for the rainfall distribution in the North-east region of India. In what follows provide a detailed analysis related to the rain distribution of the rainfall data. 

Log Normal PDF is defined as~\cite{forbes2011statistical}
\begin{multline}
    f_{LN}(x;\theta,\sigma_{x}) = \frac{1}{x \sigma_{x} \sqrt{2\pi}} \exp(-\frac{(ln(x/\theta))^2}{2 \sigma^{2}_{x}})
    \\ \forall x>0,~~\theta,\sigma_{x} > 0 \label{eq5}
\end{multline} 
where $x$ is the random variable, $\theta$ is the scale parameter and $\sigma_{x}$ is the shape parameter. Gamma PDF can be expressed as~\cite{forbes2011statistical} 
\begin{multline}
     f_{GA}(x;\alpha,\beta) = \frac{\beta^{\alpha}}{\Gamma(\alpha)} x^{\alpha -1} \exp(-\beta x)  \\ \forall x>0,~~\alpha,\beta > 0 \label{eq6}
\end{multline}
where $x$ is the random variable and $\alpha$ and $\beta$ are the shape and rate parameter respectively. We have utilized the least square fitting method to ascertain the accuracy of our method used to the determine the distribution. In Fig.~\ref{fig3} we show the histogram of wet rainfall data fitted  with both Gamma (blue curve) and Log normal (red curve) PDF for  two  different stations with coordinate $26.05^{\circ}N$, $88.05^{\circ}E$ (left panel) and  $26.55^{\circ}$N, $91.65^{\circ}$E (right panel). Figure~\ref{fig3}(a)-(b) shows the distribution for the year 2005 and Fig.~\ref{fig3}(c)-(d) represents the same for the year 2015. In the inset of each figure, the same graphs are plotted in log-log scale just to demonstrate the fitting for the higher values of rain intensity which contains the extreme events. We find the evidence of Log Normal PDF for the rainfall data that fits better than the Gamma distribution. In Table~\ref{table2} we list out the parameters of the distribution fitted with Log-normal and Gamma PDF as shown in Fig.~\ref{fig3}.

\begin{table} 
	\centering	
 	\caption{The fitting parameters for both Log Normal and Gamma PDF shown in fig~\ref{fig3}.}   \label{table2}
	\resizebox{0.48\textwidth}{!}{\begin{tabular}{c c c c c c}
			\hline \hline
			\multirow{2}{*}{Station} & \multirow{2}{*}{Year} &\multicolumn{2}{ c }{Log-Normal}&\multicolumn{2}{ c }{Gamma}\\
			& &  $\theta$ & $\sigma_{x} $  & $\alpha$ & $\beta$ \\ 
			\hline \hline

        $26.05^{\circ}$ N, $88.05^{\circ}$ E & 2005 & -1.6 & 2.3 & 1.2 & 0.2 \\ 
        $26.05^{\circ}$ N, $88.05^{\circ}$ E & 2015 & -1.2 & 1.8 & 1.0 & 0.3 \\ 
        $26.55^{\circ}$ N, $91.65^{\circ}$ E & 2005 & -2.2 & 2.5 & 1.1 & 0.1 \\ 
        $26.55^{\circ}$ N, $91.65^{\circ}$ E & 2015 & -1.4 & 2.0 & 1.1 & 0.2 \\ 
         \hline \hline
\end{tabular}}

\end{table}

To find out the best possible PDF between these two, we have computed Root Mean Square Error ($\mathcal{E}_{RMSE}$) and Mean Absolute Error ($\mathcal{E}_{RMSE}$) values for both of them. $\mathcal{E}_{RMSE}$ and $\mathcal{E}_{MAE}$ are calculated using the following formulae respectively~\cite{sagar2023encyclopedia}: 
\begin{equation}
   \mathcal{E}_{RMSE}= \sqrt{\frac{1}{n} \sum_{i=1}^n |A_{i}-F_{i}|^{2} } \label{eq7}
\end{equation}
\begin{equation}
   \mathcal{E}_{MAE}= \frac {\sum_{i=1}^n |A_{i}-F_{i}|} {n} \label{eq8}
\end{equation}
where $n$ is the number of data points, $A_{i}$ is actual data array and $F_{i}$ is the predicted data array. By calculating both the error factors, we find that the Log Normal PDF fits better than the Gamma PDF. The $\mathcal{E}_{RMSE}$ and $\mathcal{E}_{MAE}$ values for the plots shown in Fig.~\ref{fig3} can be found in Table~\ref{table4} in the appendix. The mean values of the difference of RMSE value for Gamma and Log-Normal fit and the same for MAE values over twenty years are displayed for all the seven hundred stations on $\theta -\phi$ plane in the fig.~\ref{fig13}(a) and (b) respectively in the appendix~\ref{app:a}. This suggests Log Normal PDF as the most suited distribution function of our wet rainfall data for the region under consideration. Our procured result is inline with other studies \cite{sreedhar2019fitting, kedem1987lognormality, cho2004comparison, foster2006precipitable} which have also shown the log normal behaviour of rain distribution in different regions of the world. In discriminating between two competing distribution functions, ratio of maximized likelihoods (RML) serves as a reliable quantity~\cite{atkinson1970method}. The likelihood function for the dataset following Log normal PDF can be expressed as~\cite{rohde2014introductory} 
\begin{equation}
    L_{LN}(\theta,\sigma_{x}) = \prod_{i=1}^n f_{LN}(x_{i};\theta,\sigma_{x}) \label{eq9}
\end{equation} 
and the same assuming the dataset follows Gamma PDF can be written as (LN and GA subscript stands for log-normal and gamma PDF respectively)~\cite{rohde2014introductory} 
\begin{equation}
    L_{GA}(\alpha,\beta) = \prod_{i=1}^n f_{GA}(x_{i};\alpha,\beta) \label{eq10}
\end{equation} 
Using these, the RML can be defined as~\cite{casiraghi2021likelihood} 
\begin{equation}
    RML=\frac{L_{LN}(\hat\theta,\hat\sigma_{x})}{ L_{GA}(\hat\alpha,\hat\beta)} \label{eq11}
\end{equation}
where ($\hat\theta,\hat\sigma_{x}$) and ($\hat\alpha,\hat\beta)$) are the maximum likelihood estimators of ($\theta,\sigma_{x}$) and ($\alpha,\beta)$) respectively based on the observed dataset ${X_{1},....,X_{n}}$. The natural logarithm of RML (denoted by $T$) can be expressed as 
\begin{multline}
T=n \left[\ln(\frac{\Gamma(\hat\alpha)}{\hat\sigma_{x}}) - \hat\alpha \ln(\hat\beta \tilde X) + \hat\beta \bar X  \right. \\\left.-\frac{1}{2 \hat\sigma^{2}_{x} n} \sum_{i=1}^n(\ln(\frac{X_{i}}{\hat\theta}))^{2} - \frac{1}{2}\ln(2\pi) \right] \label{eq12}
\end{multline} 
where $\tilde X$ and $\bar X$ are the geometric and arithmetic mean of the rainfall data sets ${X_{1},....,X_{n}}$ defined as $\tilde X= (\prod_{i=1}^n X_{i})^{(\frac{1}{n})}$ and $\bar X= \frac{1}{n} \sum_{i=1}^n X_{i} $ respectively. For determine goodness of the fitting we also compute the  maximum likelihood estimators as
$\hat\theta=\tilde X$, $\hat\sigma_{x}= \sqrt{\frac{1}{n} \sum_{i=1}^n ln(X_{i}/ \tilde X)^2 }$ and $\hat\beta= \hat{\alpha} / \bar X $. Here magnitude of $T$ decides about the robustness of more suited distribution. For instance as  $T>0$ the Log Normal PDF is best suitable nature of the distribution, however for $T<0$, the Gamma distribution fits better with the rainfall data. Using our analysis performed on all the stations at different years we find that $T$  is positive implying Log Normal PDF nature of the rainfall in this region consistent with earlier studies. The more detailed nature of the averaged $T$ over twenty years for all the stations has been provided on the $\theta - \phi$ plane in the Appendix~\ref{app:a}.

Low mean value, comparatively large variance and the positive values of rainfall intensity yields an skewed distribution of non zero intensity values of rainfall and in our investigation, we find it to follow log normal PDF for the particular region under consideration. Several other complex systems also obey Log-Normal distribution such as the dissipation of kinetic energy by horizontal friction in high resolution global ocean models~\cite{pearson2018log}, the moments of resistance of weakly disordered systems~\cite{slevin1990log}, the concentration of the elements in the earth's crust and their radioactivity, the size distribution of aerosols in the air, the local strain of plastic deformation in material~\cite{tang2020lognormal} etc. A discussion of the physical origin of the log normal PDF is presented in~\cite{hosoda2011origin}. This particular distribution results from a balance between the growth i.e. continuous additive increase (in our case, the increase of rain intensity within an active spell) and stochastic jumps i.e. discrete multiplicative decrease (in our case, the sudden change in rain intensity). In~\cite{manola2018future}, future precipitation pattern was simulated base on a historic rainfall event of Netherlands and the simulated rainfall intensity has been found to follow log normal distribution. PDF describes how rain intensity is distributed at any particular station at a specific year. But rainfall possess a broad spatio-temporal connection and to understand that relation, we perform spatial and temporal correlation analysis. 

\begin{figure*}  [!htp]      
\centering
\includegraphics[width=1\textwidth]{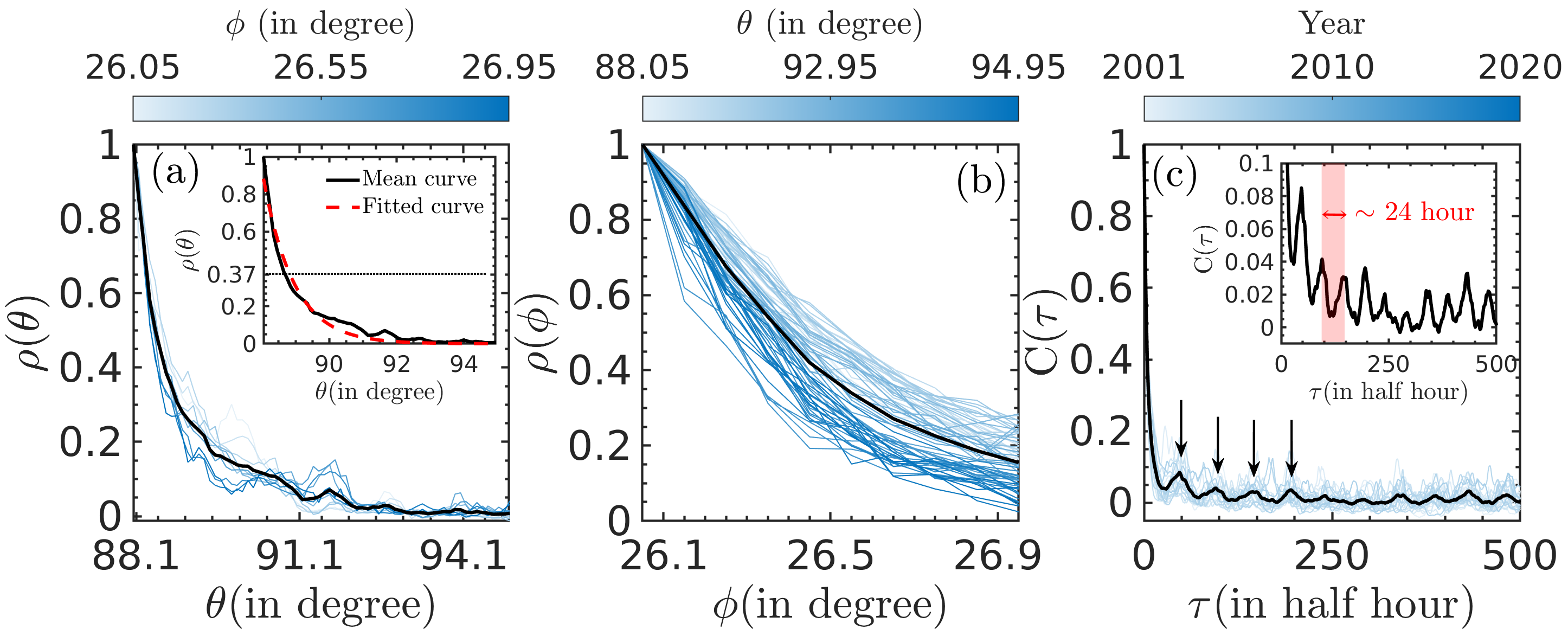}
\caption{Spatial and temporal correlation plots. (a) Spatial correlation coefficients ($\rho(\theta)$) for the stations having same latitude values but varying longitudes for the year 2005. The thin graphs (from light blue to dark blue) denote the correlation coefficients for a fixed latitude and the thick black graph shows the average behaviour of all these graphs. Inset shows the exponential fitting (dotted red) to the mean graph indicating the correlation length. (b) Spatial correlation coefficients for the stations having same longitude values but varying latitudes for the year 2005. The thin graphs (from light blue to dark blue) shows the correlation coefficients variation for a fixed longitude and the thick black graph denotes the average behaviour of all these graphs. (c) Temporal autocorrelation with time lags of station $26.05^{\circ}$ N, $88.05^{\circ}$E for 20 years. The thin graphs (from light blue to dark blue) denotes the temporal autocorrelation for different years and the thick black graph depicts the average behaviour of 20 years. The periodic ripples (indicated by the black arrows) appear at a mean interval of 24 hours. Inset shows a zoomed version of the mean graph with a red shaded area indicating mean time interval of occurrence of ripples.} \label{fig4}
\end{figure*}

\subsubsection{Spatial and Temporal Correlation}  \label{sec5}

rainfall is multiscale phenomena in which the event is correlated in the space as well as in the time. To probe the nature of these correlation next we move our focus in analyzing the spatial and temporal correlation of the rainfall time-series collected for the 700 stations of North East region of India of span of twenty years. Here we compute the correlation among different stations' rainfall pattern for both longitude and latitude spatial directions. Pearson correlation coefficient used for the spatial correlation between the rainfall time series of the neighbouring stations both in latitudinal and longitudinal direction is defines as
\begin{equation}
  \rho(A,B)= \frac{1}{N-1} \sum_{i=1}^N  \frac{(A_{i}-\Bar{A})}{\sigma_{A}} \frac{(B_{i}-\Bar{B})}{\sigma_{B}} \label{eq19}
\end{equation}
where $A$ and $B$ are the variable with two different time series, $\Bar{A}$ and $\Bar{B}$ are the mean and $\sigma_{A}$ and $\sigma_{B}$ are the standard deviation of the time series $A$ and $B$ respectively and $N$ is the total number of observations in each time series. To find out the correlation among different stations, we have first considered the stations varying in latitude direction keeping the longitude variation fixed and then we repeat the same calculation by keeping latitude fixed. 
In Fig.~\ref{fig4}(a) we plot the variation of correlation coefficients ($\rho(\theta)$) with longitude ($\theta$) of different stations having same latitude ($\phi$) values for the year 2005. The shaded lines (from light blue to dark blue) correspond to correlation coefficients variation for a fixed latitude and longitude varying from $88.05^{\circ}$ E to $94.95^{\circ}$ E. However the curves for different latitude values from $26.05^{\circ}$ N to $26.95^{\circ}$ N are denoted by different shades of blue [see adjacent colorbar]. The thick black graph exhibits the average behaviour of all these graphs. For the longitudinal direction, we find the mean spatial correlation length is to be 0.7350$^{\circ}$ (Max value 0.9267$^{\circ}$ to min value 0.5370$^{\circ}$) i.e. 73.50 km which is equivalent to almost seven neighbouring stations in longitude direction and the magnitude of this length has a gradual decreasing trend with increasing longitude value. In the inset of Fig.~\ref{fig4}(a), we fit the mean curve (thick continuous black curve) with an exponential function (dotted red curve). The thin dotted black line gives an estimate for the the correlation length the value at which correlation falls by  $1/e$. In Fig.~\ref{fig4}(b) we present the variation of the correlation coefficients with latitude ($\phi$) fixing longitude ($\theta$) values for the year 2005. The thin graphs (from light blue to dark blue) denote the corresponding correlation coefficients variation for a fixed longitude but latitude varying from $26.05^{\circ}$N to $26.95^{\circ}$N. The different longitude values from $88.05^{\circ}$ E to $94.95^{\circ}$ E are denoted by different shades of blue [see adjacent colorbar]. The thick black graph exhibits the average behaviour of all these graphs. After analysing those stations which are aligned latitude wise, we obtain the mean spatial correlation length to be $0.4362^{\circ}$ (max value 0.4581$^{\circ}$ to min value 0.4052$^{\circ}$) i.e. 43.62 km which is equivalent to almost four neighbouring stations in latitude direction. Using same kind of approach, some studies done on Malawi using the rainfall data of forty two stations for forty six years (1960-2006) has reported that there exists strong spatial correlation among the stations situated within 20 km range~\cite{ngongondo2011evaluation}. Another study done based on the rainfall data obtained from 391 stations over Korean peninsula from May to September for four years (1999-2002) reported that the spatial correlation length ranges from 50-100 km depending on month wise analysis~\cite{ha2007spatial}.

Next to understand the temporal correlation in different year's rainfall for any specific station we compute the autocorrelation coefficient defined as 
\begin{equation}
 C(\tau)= \frac{<(X(t+\tau)-\Bar{X})(X(t)-\Bar{X})>_{t}}{\sigma^2_{X}} \label{eq20}
\end{equation}
where $\tau$ is the time lag, $\Bar{X}$ is the mean and $\sigma_{X}$ is the standard deviation of the time series and $<\cdot>_{t}$ denotes the average of the time series $X(t)$ over time. We have considered the temporal autocorrelation of different stations for 20 years (from 2001 to 2020). In Fig.~\ref{fig4}(c) we show the variation of the temporal autocorrelation (C($\tau$)) with time lags ($\tau$) up to 500 half hour of the station having latitude $26.05^{\circ}$N and longitude $88.05^{\circ}$E for 20 years. The thin graphs (from light blue to dark blue) correspond to the temporal autocorrelation for each year from 2001 to 2020 and the thick black graph denotes the average behaviour of these twenty curves. The periodic ripples (indicated by the black arrows) occur at a mean interval of $24$ hours. In the inset of Fig.~\ref{fig4}(c), we present the zoomed version of the mean graph. The light red shaded area indicates the width between two consecutive periodic ripples. The autocorrelation decreases rapidly with time lag and the mean correlation time (defined as the value where C($\tau$) falls to 1/e) comes out to be 4.5 half hour or 2.3 hour. One study done based on the rainfall data obtained from 391 stations over Korean peninsula from May to September for four years (1999-2002) reported that the temporal correlation length is quite short (1.34 to 1.87 hours for different months) for the whole region while the rainfall events of the coastal part of the country exhibit comparatively long temporal correlation (1.52 to 2.45 hours for different months) ~\cite{ha2007spatial}.

Through these analysis we find the presence of typical spatial and temporal length scales associated with the rainfall in the Northeast region of India. Next to extract the presence of characteristic frequency modes we present the spectral analysis of the rainfall.

\subsection{Dominant temporal frequency modes in the rainfall event}  \label{sec6}
\subsubsection{Global Spectral Analysis}
\begin{figure} [!htp]   
\centering
\includegraphics[width=0.48\textwidth]{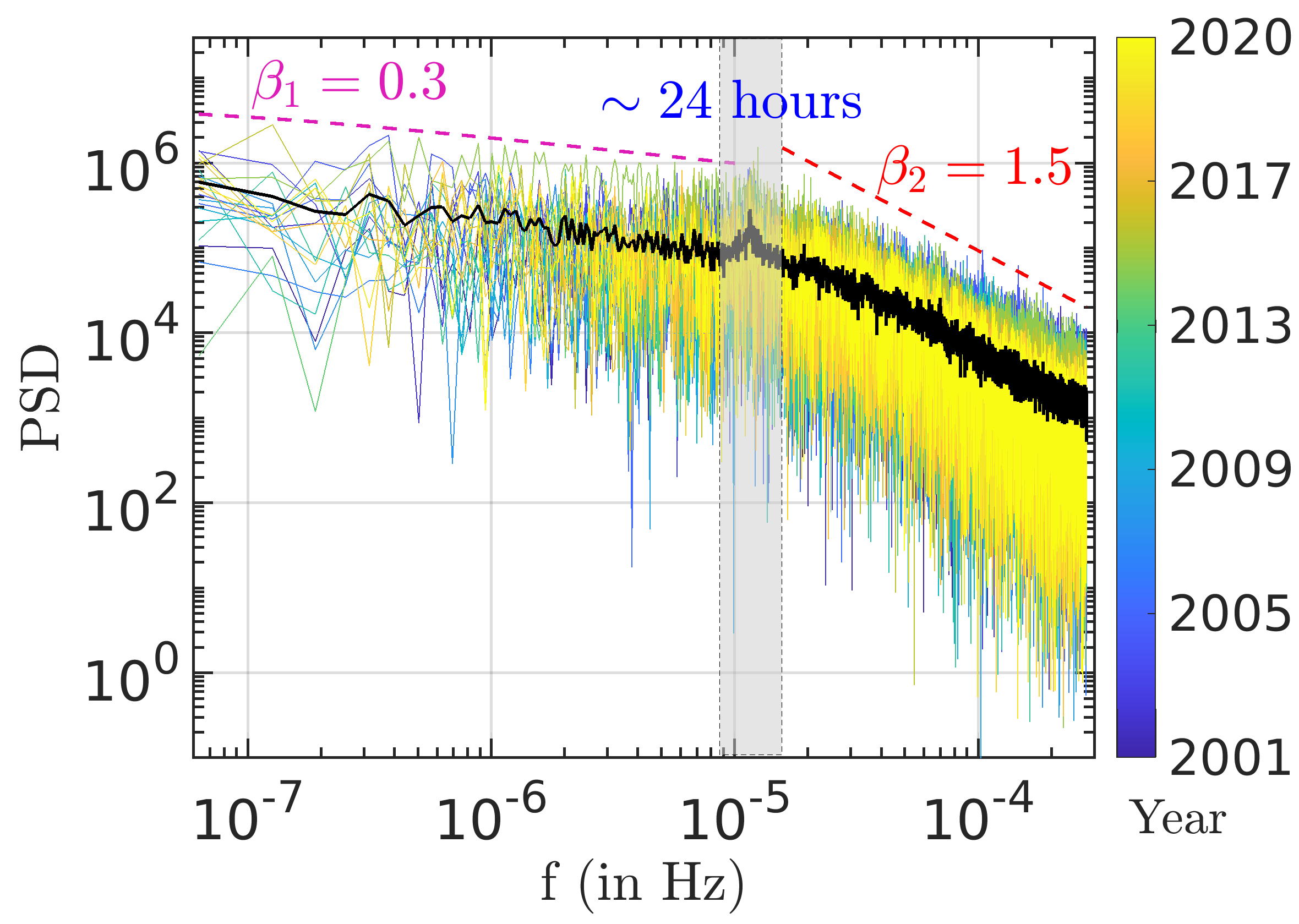}
\caption{PSD of the rainfall time series of the station $26.15^{\circ}N,91.05^{\circ}E$ for 20 years in log-log scale. The thin lines (varying from blue to yellow) corresponds to different years while the thick black line denotes the temporal average. At higher frequency range, it shows power law behaviour (red dotted line) with the exponent value ($\beta_{2}$) 1.5 whereas in the low frequency range, it follows power law behaviour (pink dotted line) with an exponent ($\beta_{1}$) 0.3. The shaded gray rectangle highlights the presence of the distinct peak at almost 24 hours.} \label{fig5}
\end{figure}

Power spectral density (PSD) analysis is an essential tool to characterize the presence multicale phenomena. In general many of the complex phenomena , such as critical phenomena, preferential processes, self organised criticality (SOC), multiplicative processes with constraints, optimisation etc.~\cite{holovatch2017complex} show the power law nature of PSD.   



In Fig.~\ref{fig5} we show the PSD of the rainfall time series of the station with latitude $26.15^{\circ}$N and longitude $91.05^{\circ}$E for 20 years (varying from blue to yellow as depicted by the adjacent colorbar) in log-log scale. The thick black line denotes the average behaviour over these years. The dashed red and pink line indicate the power law behaviour in the higher and lower frequency range respectively. Here we perform the power spectrum analysis for all the stations over 20 years (from 2001 to 2020) to estimate the dominant frequencies. We find the presence of a distinct kink in all the spectres at an average frequency of 1.16 $\times 10^{-5}$ Hz ($\sim 24$ hour in time scale) that suggests the presence of a dominate period of rainfall nearly equal to 24 Hrs (one day). The spectrum exhibits power law behaviour in both the higher and lower frequency range divided by the one day peak with two different exponents $\beta_{1}$ (for low frequency region) and $\beta_{2}$ (for high frequency region). To obtain the exponent values, we employ linear regression method and find that $\beta_{1}$ comes out to be 0.3 approximately while $\beta_{2}$ is found to be almost 1.5. As such there is no theory which can fully explain this scaling behaviour of rainfall. However the rain within one day frequency or less can be considered to be mainly driven by mean circulation of the wind. According to the theory of turbulent field, the spectrum of wind follows the Kolmogorov scaling with the PSD falls with frequency with an exponent $-5/3$~\cite{kolmogorov1962refinement}. Our obtained exponent in high frequency range (i.e. $\beta_{2}$) is closer to this which indicate that the rainfall pattern within a day is mainly dictated by the mean-wind velocity field. Similar kind of power law behaviour of PSD of rainfall with an exponent  $\sim 0.7$ within a day time interval (3-24Hrs) has also been reported from a X-band radar situated in the southeast part of France during the summer season. However,  for the time window of one to ten days, it exhibits power law behaviour with an exponent almost $\sim 1.2$~\cite{rysman2013space}.

PSD essentially brings out the most dominant frequencies present in the system. To understand the different ranges of frequencies present in the rainfall process, it is pertinent to investigate  the local trends embedded in the system. For that we employ two non-linear techniques, namely Wavelet analysis and Hilbert-Huang Transformation to investigate the rainfall time series. There we successfully obtain the two power law exponents ($\beta_{1}$ and $\beta_{2}$) along with other local features.

\subsubsection{Wavelet Analysis}
\begin{figure}  [!htp]      
\centering
\includegraphics[width=0.48\textwidth]{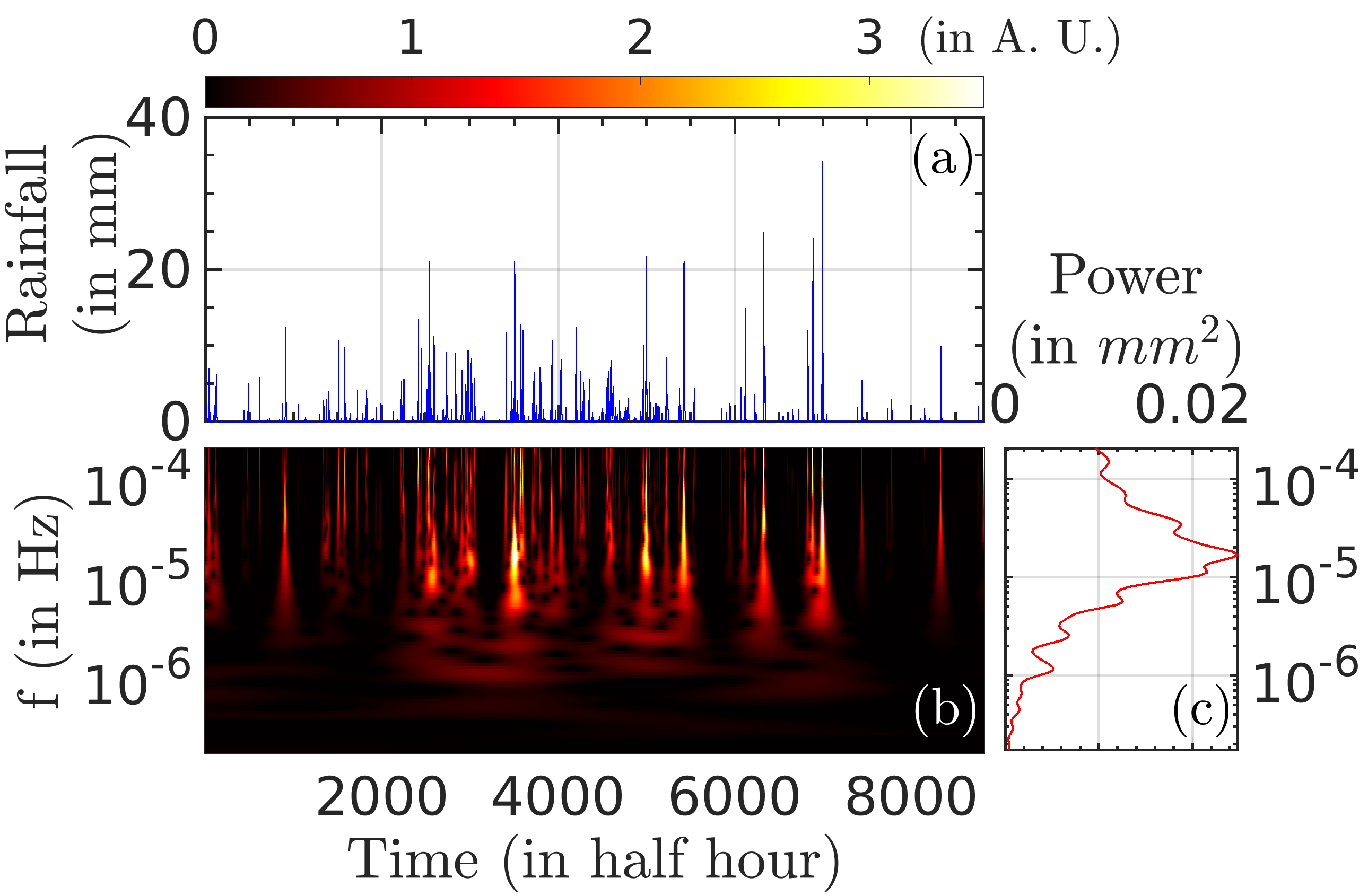}
\caption{Wavelet analysis. (a) Temporal distribution of rainfall of the station $26.15^{\circ}N,91.05^{\circ}E$ for the year 2010, (b) Wavelet spectrum of the rainfall data shown in (a). The colorbar shows the power variable in arbitrary unit (c) The global wavelet power spectrum. It reveals the presence of local peaks in the frequency domain.}\label{fig6}
\end{figure}
Fourier analysis can only provide the global behaviour of the time series data over that time span. But rainfall data is non-linear and non-stationary in nature. To get more regional and local insight about the process, we analyse our data using wavelet techniques~\cite{torrence1998practical} and Hilbert-Huang Transformation (HHT)~\cite{huang1998empirical}. Wavelet analysis can extract the multi-frequency, multi-scale features hidden in the non-stationary data like rainfall. It takes wavelets instead of continuous sine or cosine wave as its basis and select the local characteristics of the time series which finally yields the representation of how the signal's frequency and amplitude evolve with time. In wavelet analysis, the signal not only gets decomposed into its spectral components, but the spatial information i.e. where it was oscillating also gets retained. This analysis has been used quite widely in different multiscale phenomena such as, nanoscale thermal transportation in molecular dynamics simulations~\cite{PhysRevB.86.104306}, detecting the scaling behaviour in noisy experimental data~\cite{contoyiannis2020wavelet}, earth's atmosphere and climate~\cite{paluvs2019coupling}.


Wavelet Transformation (WT) is a comparatively new time series analysing tool which provides the instantaneous temporal behaviour of the magnitude of the amplitude of the time series along with its behaviour in the frequency domain. Thus the signal's behaviour in both time and frequency space can be readily get retaining all its local variations. In WT, wavelets from different families are used as the basis of the analysis instead of the continuous sine and cosine functions used as the basis in Fourier transformation. The standard wavelet is called the mother wavelet and the different sections of the time series under inspection is matched with different scaled and shifted versions of that mother wavelet to extract the local frequency and temporal behaviour. Among different types of mother wavelet, Morlet wavelet is used here to analyse the rainfall time series because of its many oscillations resembles the nature of the pattern of the rainfall time series. The two dimensional plot of the wavelet spectrum displays the time scale in one axis, frequency scale in another scale and shows the magnitude of the amplitude of rain intensity using color. We  obtain the global wavelet power spectrum by averaging over the time scale. The main feature of this spectrum is that it retains the local behaviour of the time series. For the completeness of the paper, we discuss briefly about the methodology of WT in the appendix~\ref{app:b}. More elaborated analysis can be found in~\cite{koornwinder1993wavelets,daubechies1992ten,simonsen1998determination,torrence1998practical}.

\begin{table}  [!htp] 
\caption{The prominent time scales obtained from the global wavelet power spectrum and the associated physical events related to monsoon dynamics~\cite{rajeevan2010active}.} \label{table3}
\centering
\begin{tabular}{p {3 cm} p {4.5 cm}}
\hline \\
Time scale procured from WT & Time scale of the attributed physical events\\ \\
\hline\hline \\
21 days & 10-20 days: The intraseasonal variation of monsoon circulation \\ \\
4 days & 3-4 days: The mean life span of active and break events and the typical longevity of the synoptic systems (e.g. depressions formed in low pressured region)\\ \\ 
1 day & Daily periodicity \\ \\
$\frac{1}{2}$ day & Harmonic of the daily periodicity\\ \\
\hline
\end{tabular}
\end{table}

In Fig.~\ref{fig6}, we present the wavelet analysis of our rainfall data. In Fig.~\ref{fig6}(a) we show the temporal evolution of rainfall of the station $26.15^{\circ}N,91.05^{\circ}E$ for the year 2010. In Fig.~\ref{fig6}(b) and Fig.~\ref{fig6}(c) we show the wavelet spectrum  and global wavelet power spectrum of the rainfall time series shown in Fig.~\ref{fig6}(a) respectively. After computing for all the stations over 20 years, we find the presence of three major bands of frequency where the rainfall data shows significant local peaks in intensity in the global wavelet power spectrum. The characteristics time scale embedded in the rainfall time series have been listed in the table~\ref{table3}. In the first band of frequency ranging from $10^{-6}$ Hz to $10^{-7}$ Hz, the mean frequency comes out to be 5.4 $\times 10^{-7}$ Hz i.e. almost 21 days. This result can be associated with the intraseasonal variation of monsoon circulation which is found to be around 10 to 20 days for the core monsoon zone of India~\cite{rajeevan2010active,gadgil2003breaks}. In the second band ranging from $10^{-5}$ Hz to $10^{-6}$ Hz, the mean frequency of the existing local peaks is found to be 2.7 $\times 10^{-6}$ Hz i.e. almost 4 days which can be attributed to the mean life span of active and break spell events consistent with observation reported in~\cite{rajeevan2010active}. The main driving force of these events are the lows and depressions formed in Bay of Bengal and these synoptic systems also possess an average longevity of 3 to 4 days. The last and third frequency band is extended from $10^{-4}$ Hz to $10^{-5}$ Hz and the mean peak position of this range is found out to be 2.4 $\times 10^{-5}$ Hz i.e. almost 12 hours. In between $10^{-5}$ Hz to $2.15 \times 10^{-5}$ Hz, the presence of a peak at almost 24 hours in each dataset complement the prominent peak observed in PSD of rainfall data (See Fig.~\ref{fig5}). The almost 12 hours peak can be attributed as the harmonic of this globally dominated one day peak. These unique local behaviours stem from the presence of the quasi-rhythmic intraseasonal oscillations of Indian summer monsoon rising the quasi-periodic emergence of active and break spell in the monsoon season. Along with this, the seasonal migration of the intertropical convergence zone toward north during monsoon season contributes to these signature features of Indian summer monsoon~\cite{gadgil2003indian}. After obtaining the characteristics frequencies presence in the rainfall locally at a given time of instant now we seek to disentangle the embedded noise (fluctuating at short time scale) from the mean oscillating component of the rainfall time series using the Hilbert Huang Transformation.

\subsubsection{Characterizing the noise embedded in rainfall event}
\begin{figure*}  [!htp] 
\centering
\includegraphics[width=1\textwidth]{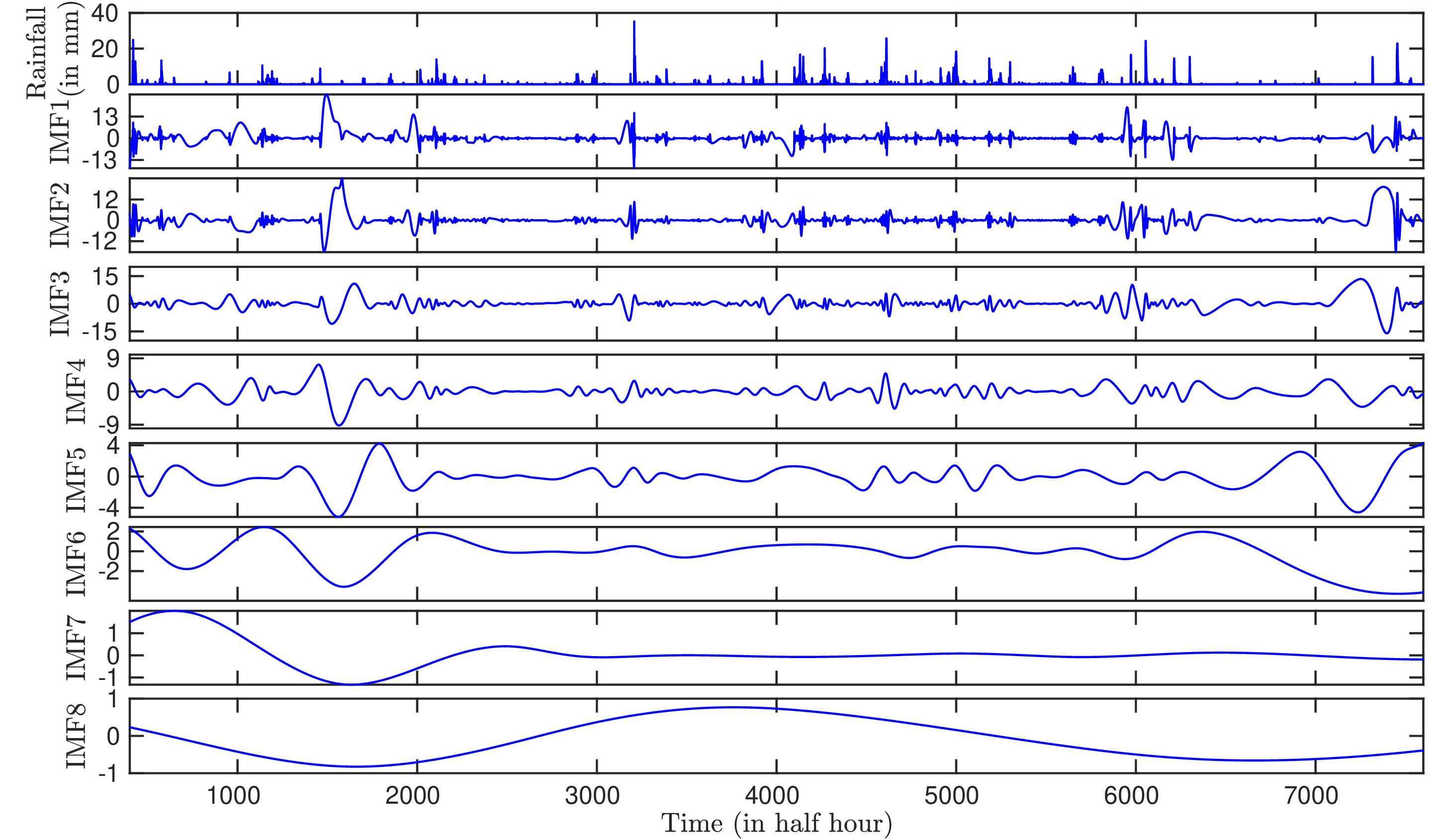}
\caption{IMFs obtained after EMD in HHT analysis of the station $26.15^{\circ}N,91.05^{\circ}E$ for the year 2009. The top most panel shows the original rainfall signal and the rest of the panels contain the IMFs obtained after the decomposition. IMF1 contains the highest frequency present in the rainfall time series and subsequently other IMFs contain the next characteristic frequencies from high to low. } \label{fig7}
\end{figure*}

Unlike Fourier analysis and Wavelet analysis, Hilbert-Huang Transformation (HHT) does not depend on the preselected basis functions and it is suitable for both non-linear and non-stationary data as it decomposes the mother signal into several intrinsic mode functions (IMF) using empirical mode decomposition (EMD) technique. Thus it is a quite useful technique to extract different significant patterns present in different frequency scales. The formalism of HHT has been successfully applied to analysis the gravitational wave data from ground and space based gravitational wave detectors, Laser Interferometer Gravitational Wave Observatory (LIGO) and Laser Interferometer Space Antenna (LISA)~\cite{camp2007application}. Further this technique has been extended to characterise a simulated gravitational wave signal from a core-collapse supernova, which found to acquire quite promising results~\cite{takeda2021application}. This technique has also been applied to resolve the non-stationary wind speed time series to obtain the spectral information~\cite{vincent2010resolving}. The capability of HHT in detecting and characterising the embedded signals in a noisy data using its adaptive and local analysis are reported in~\cite{stroeer2009methods}. In our case, we utilize HHT to analyze real time rainfall data which also encompasses complex and noisy mechanisms in it like different systems mentioned earlier. 
\begin{figure}  [!htp] 
\centering
\includegraphics[width=0.48\textwidth]{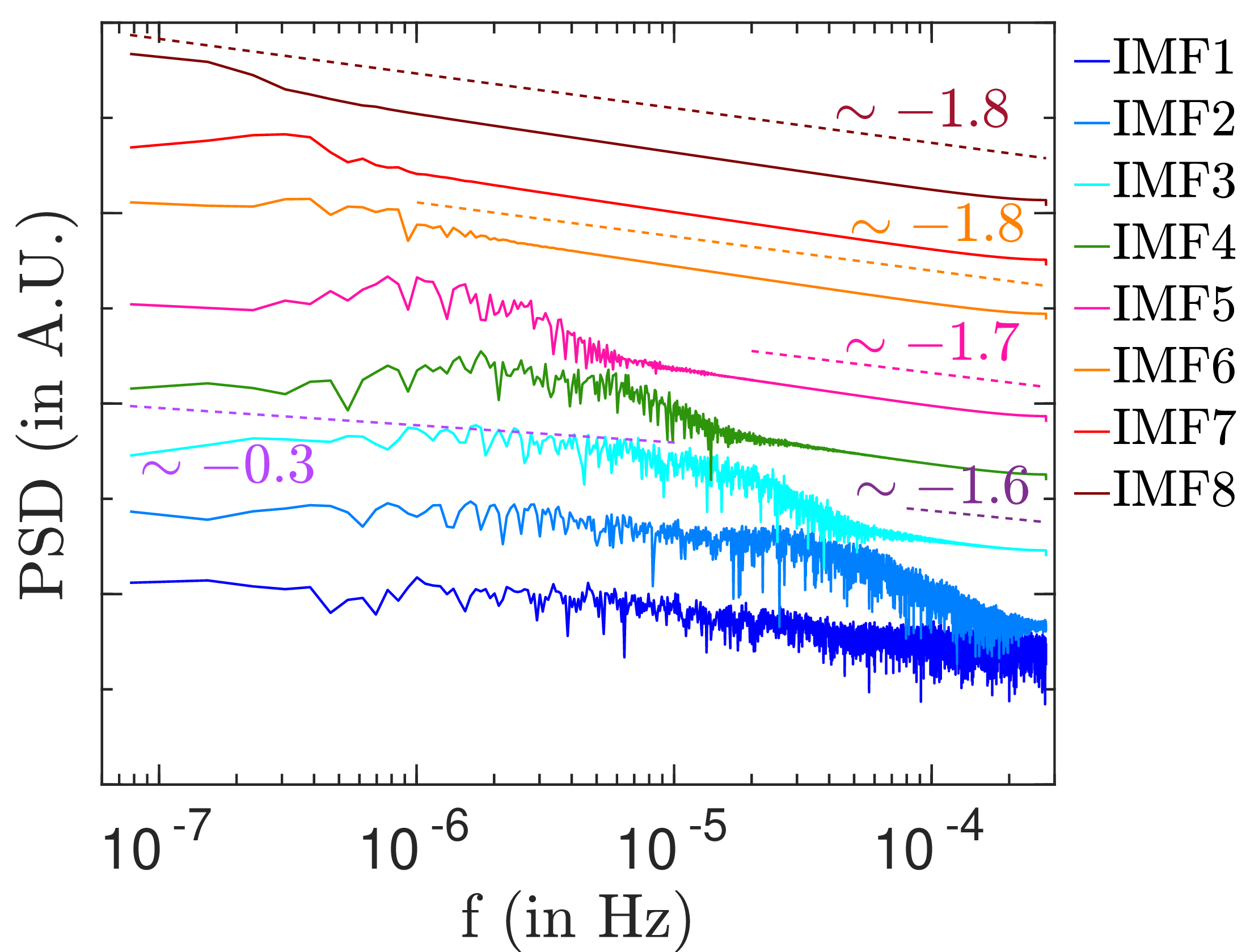}
\caption{PSD of the individual IMFs shown in Fig.~\ref{fig7}. The dotted lines indicate the power law behaviour of that particular region. This depicts how the IMFs evolve from the noisy to quasi-periodic behaviour.} \label{fig8}
\end{figure}
In general HHT is an empirical approach which consists of two components namely empirical mode decomposition (EMD) and Hilbert spectral analysis. EMD disintegrates the original signal, in our case here rainfall time series, into several sub-signals each with a specific frequency. These sub-signals known as intrinsic mode functions (IMF) and they are derived empirically from the original signal without any user specified filter. The instantaneous frequency accounts for the signal's frequency at every time instance and it is calculated as the rate of change of the phase angle at the analysis time instance. All the IMFs are real valued signal and analytic signal method is used to find out their instantaneous frequency. The overall effect of instantaneous frequency for all the IMFs construct the temporal variation with frequency in addition to the variation of the magnitude of the amplitude of the signal with frequency. Hilbert spectrum is the final representation of this analysis which comprises the joint distribution of the magnitude of the amplitude as well as the time with frequency simultaneously. Without considering the final residue, the original signal is expressed after performing the Hilbert transformation on each of the IMFs. The three dimensional Hilbert spectrum is constructed using those and the amplitude and instantaneous frequency is represented as a function of time in the Hilbert spectrum. Detailed methodology can be obtained from~\cite{huang1996mechanism,huang1999new}. A brief discussion of the construction of IMFs and Hilbert spectrum is included in the appendix~\ref{app:c}.
\begin{figure}  [!htp] 
\centering
\includegraphics[width=0.48\textwidth]{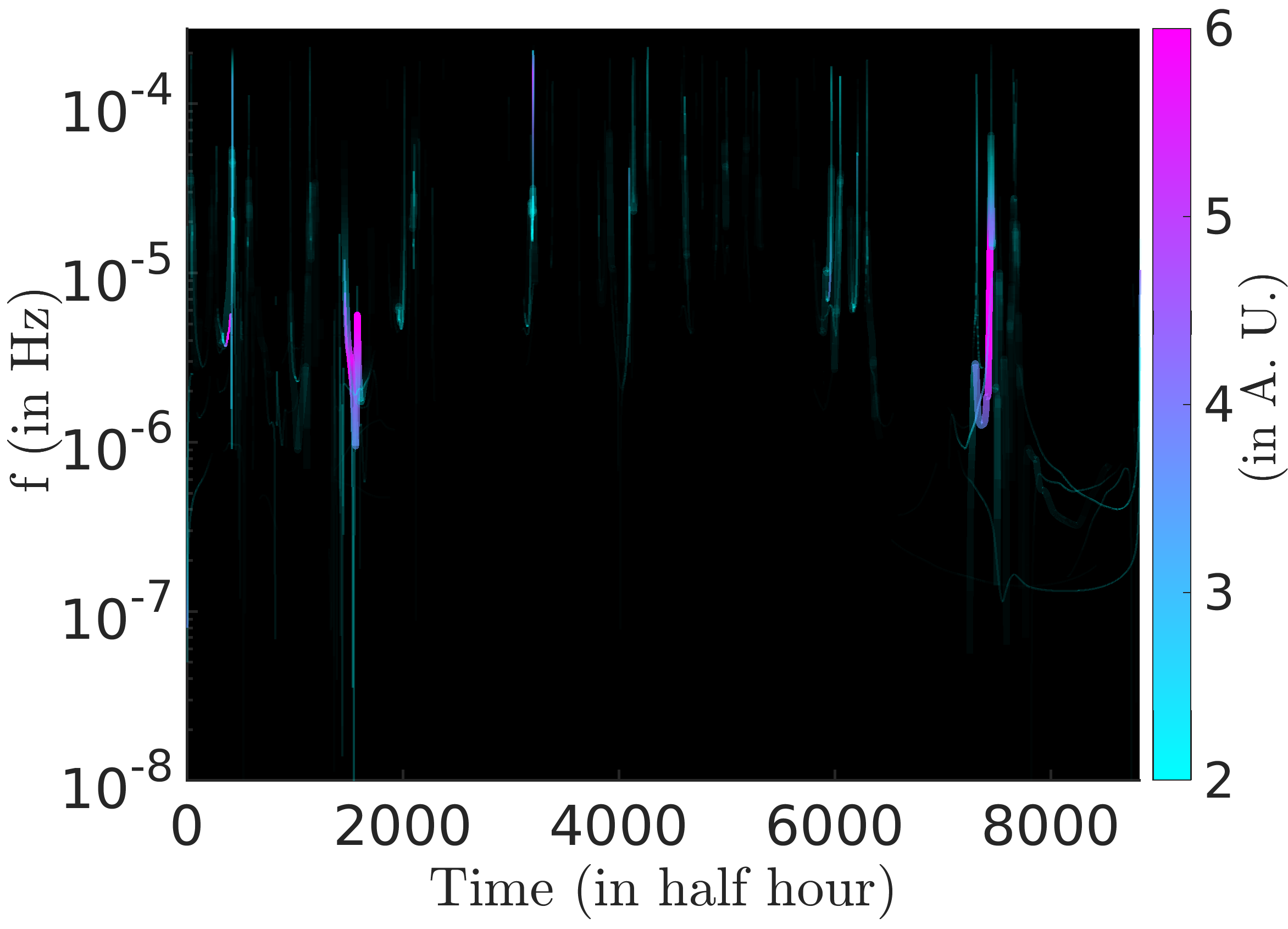}
\caption{HHT spectrum of the station $26.15^{\circ}N,91.05^{\circ}E$ for the year 2009. Both frequency and intensity of rainfall is plotted as a function of time. The colorbar denotes the value of rain intensity in arbitrary unit. It distinctly reveals the presence of intense rain events over different temporal and frequency scales.} \label{fig9}
\end{figure}

In Fig.~\ref{fig7} we present the IMFs obtained from the original rainfall signal (first upper panel of Fig.~\ref{fig7}) of the station $26.15^{\circ}N,91.05^{\circ}E$ for the year 2010 following the procedure explained briefly in the appendix~\ref{app:c} as EMD. The 1st IMF contains the highest frequency present in the original signal and the 2nd IMF represents the second highest frequency of the mother signal and so on. In Fig.~\ref{fig7} we can clearly see that first few IMFs contain the higher fluctuations in them and gradually we can observe more periodic behaviour in the last few IMFs. In Fig.~\ref{fig8} we show the PSD of the IMFs. Here we can clearly see how the frequencies present in the original signal get distributed among several IMFs and subsequently we can identify which frequencies are contributing how much and what is the embedded mean frequency coupling with the noise. From Fig.~\ref{fig8}, we find that the IMFs gets evolved from the noisy behaviour to gradually quasi periodic behaviour. Here, the last Few IMFs (IMF8,IMF7 and IMF6) show a power law behaviour with an exponent $\sim -1.8$ almost over the entire range of frequency. Subsequently this behaviour gets shifted gradually towards the high frequency range and the value of the exponent also gradually decreases. In the IMF3 (cyan blue), we find  the presence of distinct separation between the modes where the high frequency region exhibits the power law behaviour with an exponent $\sim -1.6$ whereas in the low frequency region, the IMF shows the same behaviour with an exponent $\sim -0.3$. All these  exponents are calculated after analyzing all the stations over the whole range of durations (2001-2020). These analysis reveals the gradual emergence of different significant features, some of which we have also procured with power spectrum analysis in Fig.~\ref{fig5}. In Fig.~\ref{fig9}, we display the HHT spectrum of the same rainfall time series. In the spectrum, the time coordinate is displayed along the x-axis, frequency values are depicted along y-axis (in log scale) and the pseudo colors exhibit the strength of the amplitude in arbitrary unit. Using this spectrum we obtain the frequency modes present a particular instant of  time along with its strength. We find the presence of distinct frequency peaks corresponding to the intense rainfall event which point out the dominant frequencies present at that particular time window.

So far we find that the rainfall distribution is of log-normal in nature that implies the presence of extreme events. In addition the rainfall spectrum shows the presence of Kolmogorov like scaling confirming the presence of the characteristics of the turbulent features. All these features of the rainfall indicate towards the interesting behaviour of the fluctuations at different time scale. In order to analyze these features of the fluctuations also responsible for the extreme events at different scale next we perform the fractal analysis of the rainfall data.


\subsection{Evidence of multifractal nature of rainfall event}  \label{sec7}

In both equilibrium and non-equilibrium systems, the natural fluctuations often found to follow a power law scaling relation over several orders of magnitude of the scales. This kind of power law scaling relations indicate towards the similarity of the behaviour of the system at different scale. Fractal analysis is a suitable tool to understand and analyze complex systems like rainfall in our case. The fat tail of PDF of the wet half hourly rainfall data appears due to the presence of extreme events in the time series. This indicates the presence of fractality in the underlying time series~\cite{kantelhardt2002multifractal}. In addition the power law nature of the spectrum of the rainfall time series also carry a signature of the fractal nature. A fractal process basically means the occurrence of the same elementary action over different scales such that the behaviour of any particular part of the system mimics the general behaviour of the whole system. Rainfall also possesses this fractal nature like many other atmospheric, hydrological, climatic complex systems~\cite{monjo2024fractal}, seismic events~\cite{saylor2021multifractal}, complex biological processes related to the reproduction~\cite{pigolotti2020bifractal}, matter wave localization~\cite{PhysRevLett.102.106406}, etc. However, for highly non-linear system with long range correlation shows the signature of multifractal nature events\cite{kwapien2023genuine}. 

The analysis to ascertain the fractal nature of the event one needs to rely over several techniques. In this work we implement the multifractal detrended fluctuation analysis (MFDFA)~\cite{kantelhardt2002multifractal} for our rainfall data, one of the notable techniques among the other existing schemes to analysis the fractal nature of any time series data. The brief discussion of this method and the distinct features of mono and multifractality have been included in the appendix~\ref{app:d}. Briefly one can see that if a time series contains any fractal nature, the fluctuation function $F_{q}(s)$ [see appendix~\ref{app:d}] shows a power law behaviour with the segment length $s$ of the time series such that $F_{q}(s) \sim s^{H_{q}}$, generalized Hurst exponent, for any fixed value of fluctuation order $q$. Another scale parameter mass exponent ($\tau_{q}$) is used to show the scaling properties and can be defined using Hurst exponent as
\begin{equation}
    \tau_{q}= (q\times H_{q})-1     \label{eq35}
\end{equation}
To obtain the singularity spectrum, we need the singularity strength $h_{q}$ which is related to $\tau_{q}$ via Legendre transformation such that 
\begin{equation}
     h_{q}= \tau^{\prime}_{q}       \label{eq36}
\end{equation}
and the singularity spectrum ($D_{q}$) of $q$th order can be defined as
\begin{equation}
     D_{q}=(q\times h_{q}) - \tau_{q}       \label{eq37}
\end{equation}
Detailed methodology of MFDFA can be found in~\cite{cao2018multifractal,PhysRevE.74.016103,ihlen2012introductionMFDFA}.

\begin{figure}  [!htp]       
\centering
\includegraphics[width=0.48\textwidth]{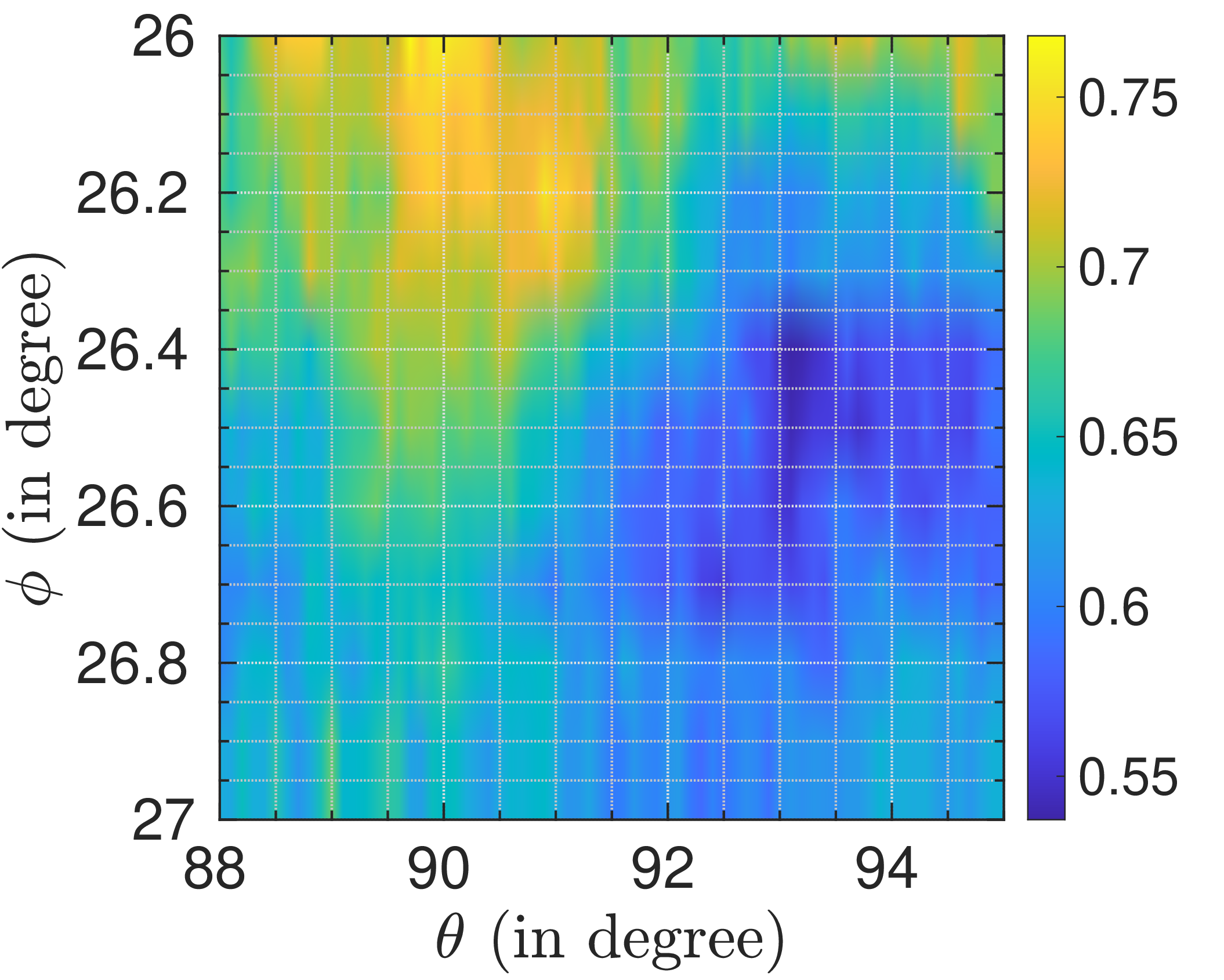}
\caption{The Hurst exponent values obtained for different stations averaging over twenty years using MFDFA plotted on the $\theta - \phi$ plane. The mean value of the Hurst exponent is indicated by the color at that particular station and it reveals that the rainfall time series is persistence in nature.} \label{fig10}
\end{figure}

In Fig.~\ref{fig10}, we show the mean value of the Hurst exponent ($H$) over 20 years for different stations on the $\theta - \phi$ plane. The mean value of hurst exponent is found to be of the order of $0.64$ indicating the time series of all the stations is persistent in nature and thus carry the predictable features for a shorter time period by knowing the initial state. This value is in consistence with the mean hurst value obtained using R/S method~\cite{hurst1951long}.
\begin{figure} [!htp]     
\centering
\includegraphics[width=0.48\textwidth]{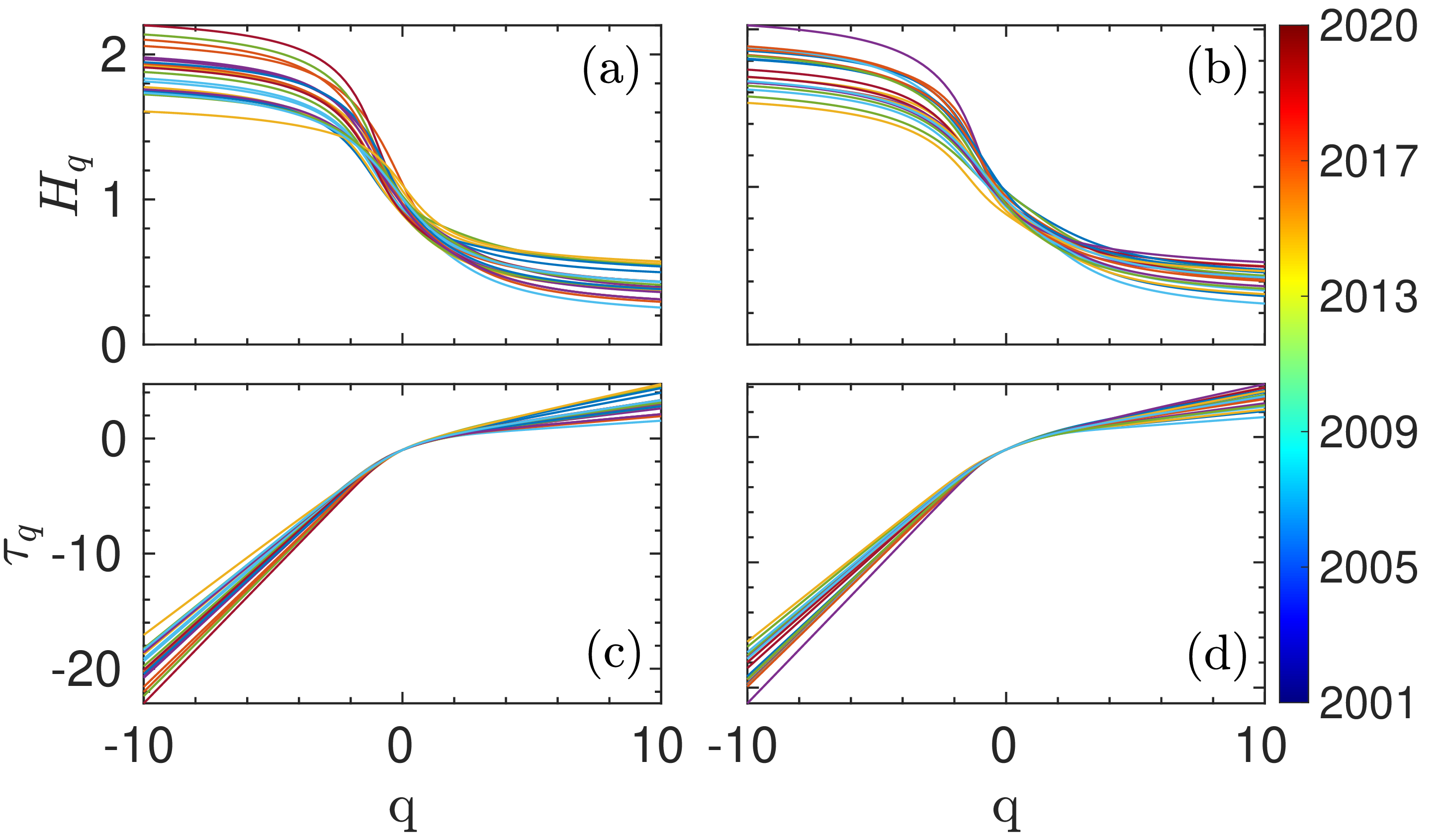}
\caption{Generalised hurst exponent ($H_{q}$) and mass exponent ($\tau_{q}$) variation with the order q using MFDFA method. (a) and (b) display the  generalised hurst exponent for two different stations $26.05^{\circ}N,88.05^{\circ}E$ and $26.95^{\circ}N,94.95^{\circ}E$ respectively. (c) and (d) show the mass exponent for those two different stations $26.05^{\circ}N,88.05^{\circ}E$ and $26.95^{\circ}N,94.95^{\circ}E$ respectively. The adjacent colorbar is showing different colors attributed to different year's dataset. The dependency of $H_{q}$ on q and the non-linear variation of $\tau_{q}$ with q reveals the multifractal nature of rainfall time series.} \label{fig11}
\end{figure}
Generalised Hurst exponent $H_{q}$ does not depend on the parameter $q$ for the monofractal time series but for the multifractal case it varies with q. In Fig.~\ref{fig11}(a) and (b), we plot $H_{q}$ as a function of $q$ for two different stations with coordinates $26.05^{\circ}N,88.05^{\circ}E$ and $26.95^{\circ}N,94.95^{\circ}E$ respectively, each for 20 years. Here we see that $H_{q}$ varies with different values of $q$ which is a signature of multifractality. The negative values of $q$, $H_{q}$ describes the behaviour of the segments with small fluctuations. On the other hand the positive values of $q$, $H_{q}$ depict the segments with the large fluctuations. Mass exponent $\tau_{q}$ can be obtained using the Eq.~(\ref{eq35}). We show the  variation of $\tau_{q}$ with $q$ in  Fig.~\ref{fig11}(c) and (d) for two different stations $26.05^{\circ}N,88.05^{\circ}E$ and $26.95^{\circ}N,94.95^{\circ}E$ respectively, at the span of 20 years. We find that  $\tau_{q}$ exhibits non-linear dependence on $q$ indicating the presence of multifractal nature of the rainfall event. Our detailed analysis performed on all the station shows the presence of same multifractality behaviour.
\begin{figure}  [!htp]    
\centering
\includegraphics[width=0.48\textwidth]{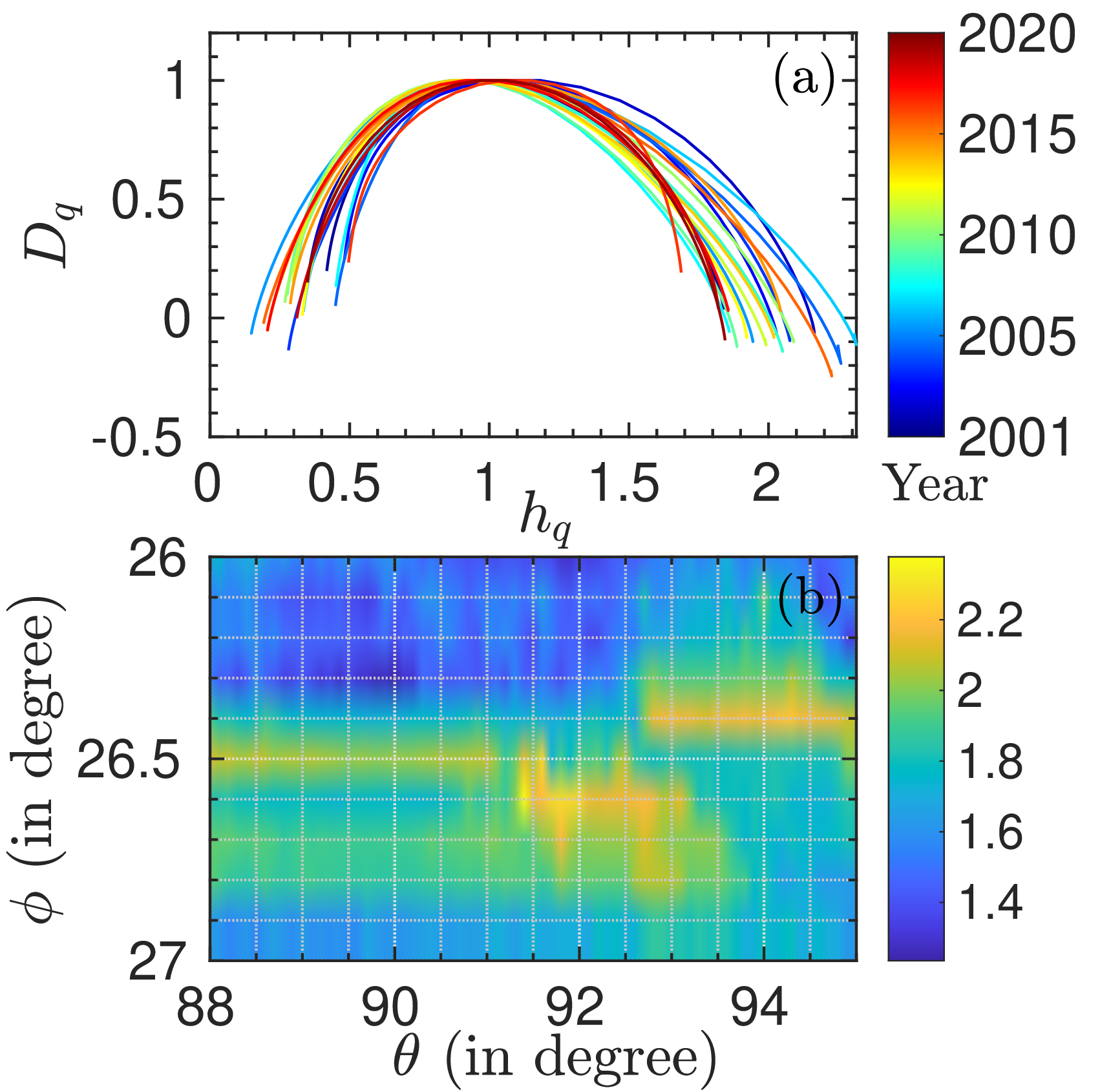}
\caption{Multifractal spectrum and its width distribution. (a) Multifractal spectrum of the station $26.15^{\circ}N,88.05^{\circ}E$ for all the 20 years. The adjacent colorbar displays different colors assigned to the datasets of different years. (b) Multifractal spectrum width values obtained for different stations averaging over twenty years using MFDFA method plotted on the $\theta - \phi$ plane. The color variation depicts the mean value of the width of the spectrum of that particular station.} \label{fig12}
\end{figure}
Next we compute the multifractal spectrum which carries the information about the distribution of the PSD scaling exponent of the multiscale system at different time scales. The multifractal spectrum can be obtained using Eq.~(\ref{eq37}). In Fig.~\ref{fig12}(a) we show the multifractral spectrum ($D_q$) for the station with coordinate $26.15^{\circ}N,88.05^{\circ}E$  spanning over 20 years (from 2001 to 2020). We notice that the spectrum with respect to $h_q$ exhibits a wide distribution a typical characteristic of multifractal nature of the rainfall events. To make it self content in the appendix we provide a detailed criteria to characterize the nature of different degrees of multifractality based general Hurst exponent, mass exponent, and spectrum behaviour[see appendix~\ref{app:d}]. In Fig.~\ref{fig12}(b), we report the mean of multifractal spectrum  width distribution averaged over twenty years for different stations as in the $\theta-\phi$ plane. We observe that multifractal spectrum width spread over a quite wide range ($\sim 1.2-2.4$) depict the strong evidence of multifractality in the rainfall events of North East part of India. In order make the analysis more concrete we have also used alternate scheme,  Wavelet Transform Modulus Maxima (WTMM)~\cite{PhysRevE.74.016103}, to calculate the multifractal spectrum and found the behaviour consistent with those obtained using MFDFA. 

\section{Summary and Conclusions} \label{sec8}


Using the half hourly rainfall data of summer-monsoon season (May-October month) in the span of two decades (2001-2020) we have performed a detailed statistical analysis of the rainfall data of North East part of India collected at 700 stations. The analysis reveals the presence of skewed distribution with heavy tails a typical features of extreme events for all the stations. The probability distribution of the rainfall event shows Log-Normal distribution which has been established by computing different characteristics quantities for goodness of fit like RMSE, MAE and RML, etc. Further we have computed the spatial as well temporal correlation of the rain fall time-series for all the stations that reveal seven  neighbouring stations are correlated in the longitudinal direction. However the temporal correlation of rainfall indicate the presence of short time correlation ($\sim 2.3 hours$) as well longtime correlation manifests as a periodic oscillation. The spectral analysis of the rainfall time series shows a signature of turbulent nature of the rainfall event which is quite evident through the power law behaviour of PSD of rainfall time series with an exponent of $\sim -1.5$ in the high frequency range (for the time interval less than 24Hrs).  The behaviour of the PSD in the high frequency region can be attributed to the Kolmogorov scaling of mean wind circulation power spectrum which shows an exponent $\sim -1.67$ (close to the exponent we get in our case i.e. $\sim -1.5$) and thus we find  that within one day time scale, the rainfall is mainly driven by the mean wind circulation field. This feature of PSD of rainfall is very much similar to the Kolmogorov like of exponent exhibited by the passive scalars such as temperature in the turbulent convective flow~\cite{Mishra2010}. On the other hand we obtain the exponent of the PSD of rainfall as $\sim-0.3$ for the low frequency range (time scale higher than 24 Hrs.). 

Further to ascertain the presence of different characteristics frequency at different time we have performed the wavelet analysis of the rainfall data for all the stations. Through this analysis we find the presence of one day periodicity as well as characteristic harmonics at the intervals of 12 hours. The wavelet analysis also shows the presence of dominant frequency corresponding to the 4 days which can be attributed to the mean life span of active and break spells as well as the mean longevity of the synoptic systems, a typical features of depressions formed in Bay of Bengal. Another dominant frequency in the wavelet analysis appears around 21 days which can be associated with the typical time scale of the intraseasonal variation of monsoon circulation in the core monsoon zone of India. Through the HHT analysis we have been able to characterize the noise present in the system. Further we have utilized the HHT spectrum to compute the high intense rain events present in this region. We have observed the multifractal nature of the rainfall time-series of North East part of India through the calculation of the Hurst exponent that suggest the persistent and predictable nature of the rainfall. 

\acknowledgments 
J. GhoshDastider acknowledges Ministry of Education (MOE), Government of India for providing financial support for her research work through the Prime Minister’s Research Fellowship (PMRF) May 2022 scheme. We thank S. B. Santra and Alex Hansen for the useful discussions and comments.
\onecolumngrid
\appendix \label{app} 
 
 \section{RMSE and MAE value} \label{app:a}
In this appendix we provide a detailed analysis of the goodness of fitting of the probability distribution function of time-series of rainfall data during Monsoon season of all the 700 station of North-East part of India. Following this we provide a detailed methodology used for the wavelet, Hilbert Huang transformation and fractal analysis. 

The required formula for calculating RMSE $(\mathcal{E}_{RMSE})$ and MAE ($\mathcal{E}_{MAE}$) is shown in the eqns. The normalised RMSE and MAE value can be calculated as $\mathcal{E}_{RMSE}/(max(A)-min(A))$ and $\mathcal{E}_{MAE}/(max(A)-min(A))$ respectively.

\begin{table} [hb!]
	\centering	
	\resizebox{1\textwidth}{!}{\begin{tabular}{c c c c c c }
			\hline \hline
			\multirow{2}{*}{Station} & \multirow{2}{*}{Year} &\multicolumn{2}{ c }{Normalised RMSE value (in $\%$)}&\multicolumn{2}{ c }{Normalised MAE value (in $\%$)}\\
			& &  Gamma & Log-Normal & Gamma & Log-Normal \\ 
			\hline \hline

    $26.05^{\circ}$ N, $88.05^{\circ}$ E & 2005  & 1.05 & 0.42  & 0.21 & 0.13  \\ 

    $26.05^{\circ}$ N, $88.05^{\circ}$ E & 2015  & 1.66 & 0.49  & 0.42 & 0.18  \\ 

    $26.55^{\circ}$ N, $91.65^{\circ}$ E & 2005  & 1.27 & 0.61  & 0.22 & 0.15  \\ 
   
    $26.55^{\circ}$ N, $91.65^{\circ}$ E & 2015  & 0.99 & 0.41  & 0.18 & 0.10  \\ \hline \hline

\end{tabular}}
	\caption{ The normalised $\mathcal{E}_{RMSE}$ and $\mathcal{E}_{MAE}$ values in percentage for two different station for two different years whose PDF fitting is shown in the Fig.~\ref{fig3}.  }\label{table4}
\end{table}

\begin{figure}  [!htp] 
\centering
        \subfloat{\includegraphics[width=0.48\textwidth]{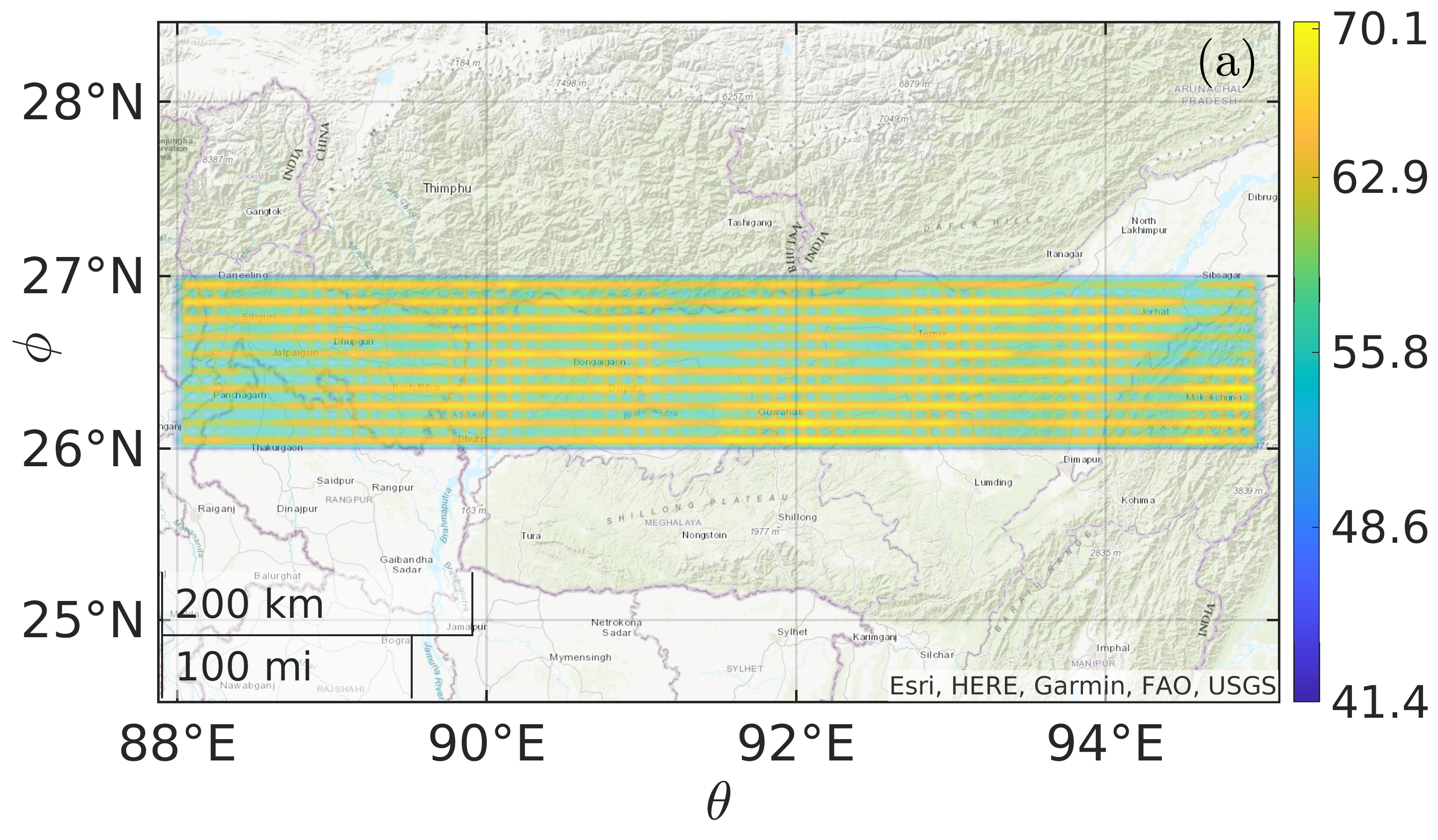}}
        \subfloat{\includegraphics[width=0.48\textwidth]{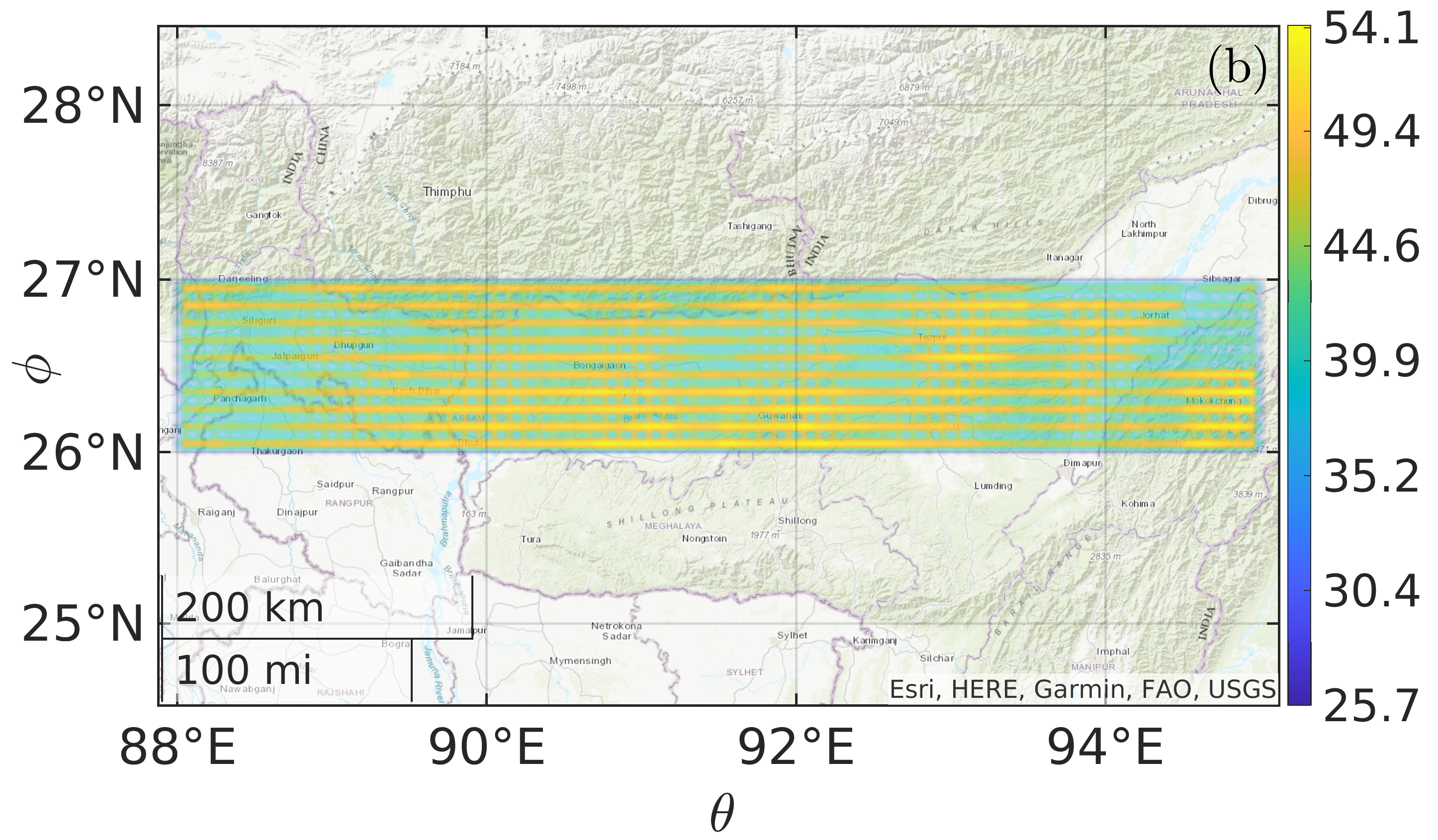}}
\caption{The difference of the error ((a) RMSE and (b) MAE) values (in percentage) of Gamma and Log Normal PDF fitting. The values are averaged over 20 years and the colorbar depicts the numerical values of the difference in error between the two fittings associated with the specific colors. (a) The difference of the RMSE value of the Gamma PDF and the RMSE value of the Log-Normal PDF is plotted and it comes out to be positive at all the stations which implies the Log Normal fitting has less error than the Gamma fitting for all the stations. (b) The same difference for MAE is shown and it also concludes that LOG-Normal PDF fits better than the Gamma PDF.}  \label{fig13}
\end{figure}

\begin{figure} [!htp]     
\centering
\includegraphics[width=0.48\textwidth]{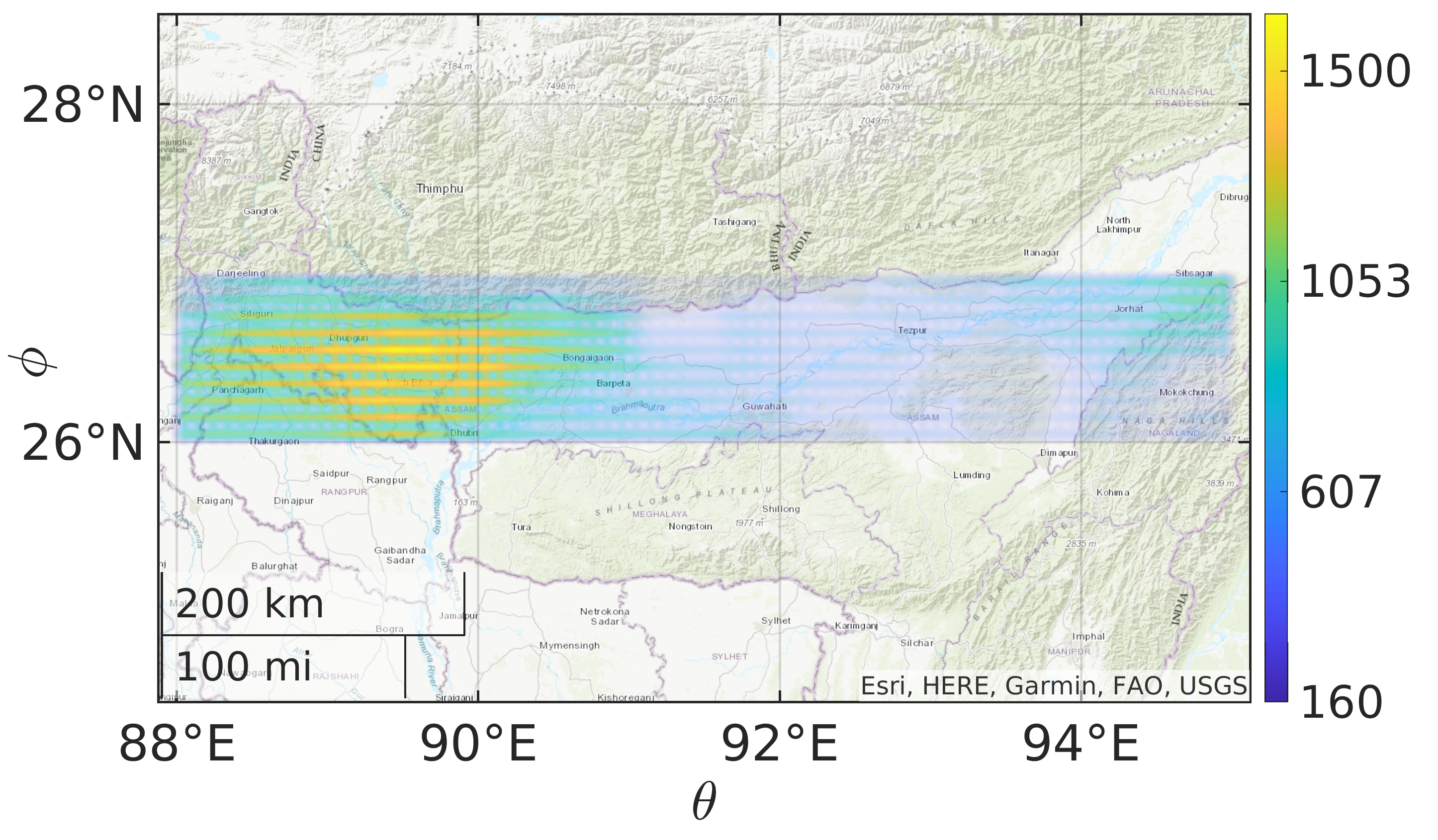}
\caption{The natural logarithm of RML value ($T$) is plotted on the $\theta - \phi$ plane. The values are averaged over 20 years. All values of the $T$ is positive which implies the Log-Normal PDF fits better than the Gamma PDF.}  \label{fig14}
\end{figure}

\section{Wavelet Analysis}  \label{app:b}

Morlet wavelet~\cite{torrence1998practical} can be expressed approximately using the following equation:
\begin{equation}
   \psi(\tau)=\pi^{-1/4} \exp(-i\omega_{0}\tau) \exp({-\frac{\tau^2}{2})} \label{eq22}
\end{equation}
where $\omega_{0}$ and $\tau$ is the non-dimensional frequency and non-dimensional time parameter respectively. Here continuous WT is used for analysing the rainfall time series as it provides a smooth transition of wavelet power between the scales or frequencies which are present in the time series. The continuous WT can be approximated for a discrete time series $x(t)$ is defined as the convolution of $x(t)$ with a scaled and translated version of ($\psi(\tau)$) such that the discrete wavelet transform ($W(b,a)$) can be obtained as:
\begin{equation}
   W(b,a)=\sum_{\tau=0}^{N-1} \frac{1}{a^{\frac{1}{2}}} \psi^*(\frac{\tau-b}{a}) x(t) \Delta t \label{eq23}
\end{equation}
where $\psi^*$ is the complex conjugate of the wavelet function $\psi$, $b$ is the localised time index of the mother wavelet and $a (a>0)$ is the scale or dilation parameter of the mother wavelet. For this analysis $\omega_{0}$ is set to be at 6 which gives a more convenient form the mother wavelet to carry out and interpret the results. The wavelet function  $\psi(\tau)$ is in general complex in nature and hence the wavelet transform is also complex. The Global wavelet power spectrum is obtained by averaging the wavelet spectrum in time over the entire time period of the time series and can be expressed mathematically as:
\begin{equation}
    \Bar{W}^2(a)= \frac{1}{N} \sum_{b=0}^{N-1} |{W(b,a)}|^2 \label{eq24}
\end{equation}
This spectrum gives an one dimensional plot of wavelet power vs frequency while retaining the traces of its localised time domain.

\section{Hilbert Huang Transformation (HHT)} \label{app:c}

HHT contains two steps namely empirical mode decomposition (EMD) and Hilbert spectral analysis. EMD disintegrates the original signal into several sub-signals each with a specific frequency. These sub-signals are called Intrinsic mode functions (IMF).
Each IMFs must satisfy the following conditions~\cite{huang1998empirical}: (a) the number of extrema and the number of zero crossing in the whole sub-signal must be equal or can differ at most by one and (b) at any point in the sub-signal, the mean value of the envelope defined by the local maxima and local minima is zero. EMD separates these sub-signals subsequently from its lower frequencies and trend. Here we are discussing the general algorithm to obtain these IMFs:
Let the time series be $x(t)$.
\begin{enumerate}
    \item  find out the extrema (both maxima and minima) of $x(t)$.
    \item  construct the upper and lower envelop $h(t)$ and $l(t)$ respectively, by connecting maxima and minima points in each case using cubic spline interpolation.
    \item calculate the local mean $m_{1}(t) = (h(t) + l(t))/2$
    \item as zero local mean is one of the requirement of being IMF, $m_{1}(t)$ is subtracted from $x(t)$: $g_{1}(t) = (x(t) - m_{1}(t))$
    \item examine whether $g_{1}(t)$ is an IMF or not
    \item repeat the steps from 1 to 5 to get the IMF $g_{1}(t)$
\end{enumerate}
That's how we obtain the 1st IMF $C_{1}(t)=g_{1}(t)$ which contains the finest temporal scale of the signal and the highest frequency present in the original signal. In order to proceed to find other IMFs, we have to get the 1st residue $r_{1}(t)$ as $r_{1}(t)=(x(t)-C_{1}(t))$. This residue $r_{1}(t)$ contains all other IMFs with higher temporal periods and we can extract them using the procedure explained earlier. So in each iteration we construct subsequent IMFs $C_{2}(t)$, $C_{3}(t)$ ... up to $C_{n}(t)$ with residues $r_{2}(t)=(r_{1}(t)-C_{2}(t))$, $r_{3}(t)=(r_{2}(t)-C_{3}(t))$ ... up to $r_{n}(t)=(r_{n-1}(t)-C_{n}(t))$ where $r_{n}(t)$ is a constant or monotonic function or a function left with only one extremum such that no other IMF can be derived. We can reconstruct the original signal as 
\begin{equation}
x(t)= \sum_{i=1}^{n}C_{i}(t)+r_{n}(t)  \label{eq25}
\end{equation}
The instantaneous frequency accounts for the signal's frequency at every time instance and it is calculated as the rate of change of the phase angle at the analysis time instance. All the IMFs are real valued signal and analytic signal method is used to find out their instantaneous frequency. The real valued IMFs need to express in the complex form and that is defined as 
\begin{equation}
z_{i}(t)= C_{i}(t) + jH[C_{i}(t)]=a_{i}(t)e^{j\theta_{i}(t)}  \label{eq26}
\end{equation}
Here, $a_{i}(t)$ and $\theta_{i}(t)$ are the instantaneous amplitude and phase respectively of the IMF $C_{i}(t)$ and $H[\cdot]$ is the Hilbert transform operator. For any arbitrary time series $X(t)$, the Hilbert transform $Y(t)$ can be obtained as
\begin{equation}
Y(t)=\frac{1}{\pi} P \int{\frac{X(t^{\prime})}{t-t^{\prime}}}dt^{\prime}    \label{eq27}
\end{equation}
where P is the Cauchy principal value. The Hilbert transform introduces a phase shift of $\pm\pi/2$ to all the frequency components. Following this formalism, $X(t)$ and $Y(t)$ become a complex conjugate and the analytical signal can be expressed as $Z(t)=X(t)+j Y(t)$. So the instantaneous frequency of the $i$th IMF can be obtained as 
\begin{equation}
\omega_{i}(t)=\frac{d\theta_{i}(t)}{dt}     \label{eq28}
\end{equation}
Thus the overall effect of instantaneous frequency for all the IMFs construct the temporal variation with frequency in addition to the variation of the magnitude of the amplitude of the signal with frequency. Hilbert spectrum is the final representation of this analysis which comprise the joint distribution of the magnitude of the amplitude as well as the time with frequency simultaneously. Without considering the final residue, the original signal can be expressed as after performing the Hilbert transformation on each of the IMFs as 
\begin{equation}
x(t)=\sum_{i=1}^{n}a_{i}(t)e^{j\int{\omega_{i}(t)dt}}       \label{eq29}
\end{equation}
This way the three dimensional Hilbert spectrum is constructed where the amplitude and instantaneous frequency is represented as a function of time. 

\section{Multifractal Detrended Fluctuation Analysis (MFDFA)} \label{app:d}

This technique~\cite{kantelhardt2002multifractal,ihlen2012introductionMFDFA} is briefly discussed here: 
Let $X_{k} (k=1,2,3....N)$ is any arbitrary time series and it is being analyzed by this MFDFA technique.
\begin{enumerate}
    \item determine the profile by subtracting the mean value ($<X>= \frac{1}{N}\sum_{k=1}^N X_{k}$) of the time series $X$ from each of its entries and then adding them resulting $1,2,...N$ no of profiles as
    \begin{equation}
    Y(i)=\sum_{k=1}^{i}[X_{k}-<X>],~~ i=1,2,...N        \label{eq30}
    \end{equation}
    \item the obtained profile $Y(i)$ is divided into $N_{s}=int(N/s)$ number of non overlapping segments of equal length s. Sometime, N is not a multiple number of s and the dividing process has to be started from the back end of the series to include the odd number of data in the last segment made earlier. Thus in total $2N_{s}$ number of segments of equal length is obtained following this process.
    \item in order to find out the local trend of each segments, polynomials of different degrees are fitted using least square fit and according to the degree of the fitted polynomial, the procedure is also named as MFDFA1 (for linear fit), MFDFA2 (for quadratic fit), MFDFA3 (for cubic fit), ..., MFDFAn (for nth order of polynomial fit). Lets take $Y_{p}$ as the best fitted polynomial to an arbitrary section p of the series, then its variance can be obtained as
    \begin{equation}
    F^{2}(p,s)=\frac{1}{s}\sum_{i=1}^{s}(Y[(p-1)s+i]-Y_{p}(i))^{2}, for~~ p=1,2,...,N_{s} \label{eq31}
    \end{equation} 

    \begin{equation}
    F^{2}(p,s)=\frac{1}{s}\sum_{i=1}^{s}(Y[N-(p-N_{s})s+i]-Y_{p}(i))^{2}, for~~ p=N_{s}+1,...2N_{s}   \label{eq32}
    \end{equation}

    \item at last the $q$th order MFDFA fluctuation is calculated using the following formulae
    \begin{equation}
     F_{q}(s)=\left[\frac{1}{2N_{s}}\sum_{p=1}^{2N_{s}}(F^{2}(p,s))^{q/2}\right]^{1/q}, \forall~~ q\neq0    \label{eq33}
    \end{equation}

    \begin{equation}
    F_{0}(s)=\exp \left[{\frac{1}{4N_{s}}\sum_{p=1}^{2N_{s}}\ln{(F^{2}(p,s))}}\right], for~~ q\rightarrow0      \label{eq34}
    \end{equation}
    \item if the time series contains any fractal nature, $F_{q}(s)$ shows a power law behaviour with $s$ such that $F_{q}(s) \sim s^{H_{q}}$ for any fixed value of q. Thus the generalised Hurst exponent ( $H_{q}$) and other scaling parameters like mass exponent ($\tau_{q}$), singularity strength $h_{q}$ and singularity spectrum ($D_{q}$) can be deduced from here.
    
\end{enumerate}
Using this procedure, we can find out if the time series under consideration possess any fractal behaviour and if it has then which type of fractality (monofractal or multifractal) is present in it. Based on different behaviours of Generalised Hurst exponent $H_{q}$ and Mass exponent $\tau_{q}$ with the scaling order $q$ and the width of the multifractal spectrum, we can divide the fractal nature as monofractal and multifractal. The key features of monofractals are: 
\begin{enumerate}
    \item the fluctuation function $F_{q}(s)$ is the same for all the segments of the original signal.
    \item the generalised Hurst exponent $H_{q}$ does not depend on the values of $q$.
    \item the mass exponent $\tau_{q}$ increases linearly with $q$.
    \item the singularity spectrum width ($\Delta h_{q}=(h_{q})_{max} - (h_{q})_{min}$) is very narrow, i.e. ($\Delta h_{q} \sim 0$).
\end{enumerate}
And the signature behaviour of the multifractals are:
\begin{enumerate}
    \item for the positive values of $q$, $F_{q}(s)$ is dominated by the segments having large deviation and for the negative values of $q$, $F_{q}(s)$'s behaviour is mostly controlled by the segments having small deviation.
    \item similarly the generalised Hurst exponent $H_{q}$ describes the effects of small fluctuations in the region of negative $q$ values whereas for positive $q$ values, $H_{q}$ shows the attributes of large fluctuations.
    \item the mass exponent $\tau_{q}$ increases non-linearly with $q$.
    \item the singularity spectrum is wide enough as compared to the monofractal case.
\end{enumerate}
So using these characteristic features, we can find out which kind of fractality is present in the our system.

\twocolumngrid
\bibliography{references}

\begin{thebibliography}{84}%
\makeatletter
\providecommand \@ifxundefined [1]{%
 \@ifx{#1\undefined}
}%
\providecommand \@ifnum [1]{%
 \ifnum #1\expandafter \@firstoftwo
 \else \expandafter \@secondoftwo
 \fi
}%
\providecommand \@ifx [1]{%
 \ifx #1\expandafter \@firstoftwo
 \else \expandafter \@secondoftwo
 \fi
}%
\providecommand \natexlab [1]{#1}%
\providecommand \enquote  [1]{``#1''}%
\providecommand \bibnamefont  [1]{#1}%
\providecommand \bibfnamefont [1]{#1}%
\providecommand \citenamefont [1]{#1}%
\providecommand \href@noop [0]{\@secondoftwo}%
\providecommand \href [0]{\begingroup \@sanitize@url \@href}%
\providecommand \@href[1]{\@@startlink{#1}\@@href}%
\providecommand \@@href[1]{\endgroup#1\@@endlink}%
\providecommand \@sanitize@url [0]{\catcode `\\12\catcode `\$12\catcode
  `\&12\catcode `\#12\catcode `\^12\catcode `\_12\catcode `\%12\relax}%
\providecommand \@@startlink[1]{}%
\providecommand \@@endlink[0]{}%
\providecommand \url  [0]{\begingroup\@sanitize@url \@url }%
\providecommand \@url [1]{\endgroup\@href {#1}{\urlprefix }}%
\providecommand \urlprefix  [0]{URL }%
\providecommand \Eprint [0]{\href }%
\providecommand \doibase [0]{https://doi.org/}%
\providecommand \selectlanguage [0]{\@gobble}%
\providecommand \bibinfo  [0]{\@secondoftwo}%
\providecommand \bibfield  [0]{\@secondoftwo}%
\providecommand \translation [1]{[#1]}%
\providecommand \BibitemOpen [0]{}%
\providecommand \bibitemStop [0]{}%
\providecommand \bibitemNoStop [0]{.\EOS\space}%
\providecommand \EOS [0]{\spacefactor3000\relax}%
\providecommand \BibitemShut  [1]{\csname bibitem#1\endcsname}%
\let\auto@bib@innerbib\@empty
\bibitem [{\citenamefont {Signoles}\ \emph {et~al.}(2021)\citenamefont
  {Signoles}, \citenamefont {Franz}, \citenamefont {Ferracini~Alves},
  \citenamefont {G\"arttner}, \citenamefont {Whitlock}, \citenamefont
  {Z\"urn},\ and\ \citenamefont {Weidem\"uller}}]{PhysRevX.11.011011}%
  \BibitemOpen
  \bibfield  {author} {\bibinfo {author} {\bibfnamefont {A.}~\bibnamefont
  {Signoles}}, \bibinfo {author} {\bibfnamefont {T.}~\bibnamefont {Franz}},
  \bibinfo {author} {\bibfnamefont {R.}~\bibnamefont {Ferracini~Alves}},
  \bibinfo {author} {\bibfnamefont {M.}~\bibnamefont {G\"arttner}}, \bibinfo
  {author} {\bibfnamefont {S.}~\bibnamefont {Whitlock}}, \bibinfo {author}
  {\bibfnamefont {G.}~\bibnamefont {Z\"urn}},\ and\ \bibinfo {author}
  {\bibfnamefont {M.}~\bibnamefont {Weidem\"uller}},\ }\bibfield  {title}
  {\bibinfo {title} {Glassy dynamics in a disordered heisenberg quantum spin
  system},\ }\href {https://doi.org/10.1103/PhysRevX.11.011011} {\bibfield
  {journal} {\bibinfo  {journal} {Phys. Rev. X}\ }\textbf {\bibinfo {volume}
  {11}},\ \bibinfo {pages} {011011} (\bibinfo {year} {2021})}\BibitemShut
  {NoStop}%
\bibitem [{\citenamefont {Corps}\ \emph {et~al.}(2023)\citenamefont {Corps},
  \citenamefont {Str\'ansk\'y},\ and\ \citenamefont
  {Cejnar}}]{PhysRevB.107.094307}%
  \BibitemOpen
  \bibfield  {author} {\bibinfo {author} {\bibfnamefont {A.~L.}\ \bibnamefont
  {Corps}}, \bibinfo {author} {\bibfnamefont {P.}~\bibnamefont
  {Str\'ansk\'y}},\ and\ \bibinfo {author} {\bibfnamefont {P.}~\bibnamefont
  {Cejnar}},\ }\bibfield  {title} {\bibinfo {title} {Mechanism of dynamical
  phase transitions: The complex-time survival amplitude},\ }\href
  {https://doi.org/10.1103/PhysRevB.107.094307} {\bibfield  {journal} {\bibinfo
   {journal} {Phys. Rev. B}\ }\textbf {\bibinfo {volume} {107}},\ \bibinfo
  {pages} {094307} (\bibinfo {year} {2023})}\BibitemShut {NoStop}%
\bibitem [{\citenamefont {Cantor}\ and\ \citenamefont
  {Wojtacki}(2022)}]{PhysRevApplied.17.064021}%
  \BibitemOpen
  \bibfield  {author} {\bibinfo {author} {\bibfnamefont {D.}~\bibnamefont
  {Cantor}}\ and\ \bibinfo {author} {\bibfnamefont {K.}~\bibnamefont
  {Wojtacki}},\ }\bibfield  {title} {\bibinfo {title} {Effects of friction and
  spacing on the collaborative behavior of domino toppling},\ }\href
  {https://doi.org/10.1103/PhysRevApplied.17.064021} {\bibfield  {journal}
  {\bibinfo  {journal} {Phys. Rev. Appl.}\ }\textbf {\bibinfo {volume} {17}},\
  \bibinfo {pages} {064021} (\bibinfo {year} {2022})}\BibitemShut {NoStop}%
\bibitem [{\citenamefont {Oberlack}\ \emph {et~al.}(2022)\citenamefont
  {Oberlack}, \citenamefont {Hoyas}, \citenamefont {Kraheberger}, \citenamefont
  {Alc\'antara-\'Avila},\ and\ \citenamefont {Laux}}]{PhysRevLett.128.024502}%
  \BibitemOpen
  \bibfield  {author} {\bibinfo {author} {\bibfnamefont {M.}~\bibnamefont
  {Oberlack}}, \bibinfo {author} {\bibfnamefont {S.}~\bibnamefont {Hoyas}},
  \bibinfo {author} {\bibfnamefont {S.~V.}\ \bibnamefont {Kraheberger}},
  \bibinfo {author} {\bibfnamefont {F.}~\bibnamefont {Alc\'antara-\'Avila}},\
  and\ \bibinfo {author} {\bibfnamefont {J.}~\bibnamefont {Laux}},\ }\bibfield
  {title} {\bibinfo {title} {Turbulence statistics of arbitrary moments of
  wall-bounded shear flows: A symmetry approach},\ }\href
  {https://doi.org/10.1103/PhysRevLett.128.024502} {\bibfield  {journal}
  {\bibinfo  {journal} {Phys. Rev. Lett.}\ }\textbf {\bibinfo {volume} {128}},\
  \bibinfo {pages} {024502} (\bibinfo {year} {2022})}\BibitemShut {NoStop}%
\bibitem [{\citenamefont {Peters}\ and\ \citenamefont
  {Christensen}(2002)}]{PhysRevE.66.036120}%
  \BibitemOpen
  \bibfield  {author} {\bibinfo {author} {\bibfnamefont {O.}~\bibnamefont
  {Peters}}\ and\ \bibinfo {author} {\bibfnamefont {K.}~\bibnamefont
  {Christensen}},\ }\bibfield  {title} {\bibinfo {title} {Rain: Relaxations in
  the sky},\ }\href {https://doi.org/10.1103/PhysRevE.66.036120} {\bibfield
  {journal} {\bibinfo  {journal} {Phys. Rev. E}\ }\textbf {\bibinfo {volume}
  {66}},\ \bibinfo {pages} {036120} (\bibinfo {year} {2002})}\BibitemShut
  {NoStop}%
\bibitem [{\citenamefont {Zhang}\ \emph {et~al.}(2020)\citenamefont {Zhang},
  \citenamefont {Fan}, \citenamefont {Marzocchi}, \citenamefont {Shapira},
  \citenamefont {Hofstetter}, \citenamefont {Havlin},\ and\ \citenamefont
  {Ashkenazy}}]{PhysRevResearch.2.013264}%
  \BibitemOpen
  \bibfield  {author} {\bibinfo {author} {\bibfnamefont {Y.}~\bibnamefont
  {Zhang}}, \bibinfo {author} {\bibfnamefont {J.}~\bibnamefont {Fan}}, \bibinfo
  {author} {\bibfnamefont {W.}~\bibnamefont {Marzocchi}}, \bibinfo {author}
  {\bibfnamefont {A.}~\bibnamefont {Shapira}}, \bibinfo {author} {\bibfnamefont
  {R.}~\bibnamefont {Hofstetter}}, \bibinfo {author} {\bibfnamefont
  {S.}~\bibnamefont {Havlin}},\ and\ \bibinfo {author} {\bibfnamefont
  {Y.}~\bibnamefont {Ashkenazy}},\ }\bibfield  {title} {\bibinfo {title}
  {Scaling laws in earthquake memory for interevent times and distances},\
  }\href {https://doi.org/10.1103/PhysRevResearch.2.013264} {\bibfield
  {journal} {\bibinfo  {journal} {Phys. Rev. Res.}\ }\textbf {\bibinfo {volume}
  {2}},\ \bibinfo {pages} {013264} (\bibinfo {year} {2020})}\BibitemShut
  {NoStop}%
\bibitem [{\citenamefont {Moustakis}\ \emph {et~al.}(2020)\citenamefont
  {Moustakis}, \citenamefont {Onof},\ and\ \citenamefont
  {Paschalis}}]{moustakis2020atmospheric}%
  \BibitemOpen
  \bibfield  {author} {\bibinfo {author} {\bibfnamefont {Y.}~\bibnamefont
  {Moustakis}}, \bibinfo {author} {\bibfnamefont {C.~J.}\ \bibnamefont
  {Onof}},\ and\ \bibinfo {author} {\bibfnamefont {A.}~\bibnamefont
  {Paschalis}},\ }\bibfield  {title} {\bibinfo {title} {Atmospheric convection,
  dynamics and topography shape the scaling pattern of hourly rainfall extremes
  with temperature globally},\ }\href@noop {} {\bibfield  {journal} {\bibinfo
  {journal} {Commun. Earth Environ.}\ }\textbf {\bibinfo {volume} {1}},\
  \bibinfo {pages} {11} (\bibinfo {year} {2020})}\BibitemShut {NoStop}%
\bibitem [{\citenamefont {Amarouchene}\ \emph {et~al.}(2001)\citenamefont
  {Amarouchene}, \citenamefont {Boudet},\ and\ \citenamefont
  {Kellay}}]{PhysRevLett.86.4286}%
  \BibitemOpen
  \bibfield  {author} {\bibinfo {author} {\bibfnamefont {Y.}~\bibnamefont
  {Amarouchene}}, \bibinfo {author} {\bibfnamefont {J.~F.}\ \bibnamefont
  {Boudet}},\ and\ \bibinfo {author} {\bibfnamefont {H.}~\bibnamefont
  {Kellay}},\ }\bibfield  {title} {\bibinfo {title} {Dynamic sand dunes},\
  }\href {https://doi.org/10.1103/PhysRevLett.86.4286} {\bibfield  {journal}
  {\bibinfo  {journal} {Phys. Rev. Lett.}\ }\textbf {\bibinfo {volume} {86}},\
  \bibinfo {pages} {4286} (\bibinfo {year} {2001})}\BibitemShut {NoStop}%
\bibitem [{\citenamefont {Parisi}(1999)}]{parisi1999complex}%
  \BibitemOpen
  \bibfield  {author} {\bibinfo {author} {\bibfnamefont {G.}~\bibnamefont
  {Parisi}},\ }\bibfield  {title} {\bibinfo {title} {Complex systems: a
  physicist's viewpoint},\ }\href
  {https://www.sciencedirect.com/science/article/pii/S037843719800524X}
  {\bibfield  {journal} {\bibinfo  {journal} {Phys. A: Stat. Mech. Appl.}\
  }\textbf {\bibinfo {volume} {1}},\ \bibinfo {pages} {557} (\bibinfo {year}
  {1999})}\BibitemShut {NoStop}%
\bibitem [{\citenamefont {Papo}\ and\ \citenamefont
  {Buld{\'u}}(2023)}]{papo2023does}%
  \BibitemOpen
  \bibfield  {author} {\bibinfo {author} {\bibfnamefont {D.}~\bibnamefont
  {Papo}}\ and\ \bibinfo {author} {\bibfnamefont {J.}~\bibnamefont
  {Buld{\'u}}},\ }\bibfield  {title} {\bibinfo {title} {Does the brain behave
  like a (complex) network? i. dynamics},\ }\href@noop {} {\bibfield  {journal}
  {\bibinfo  {journal} {Phys. Life Rev.}\ } (\bibinfo {year}
  {2023})}\BibitemShut {NoStop}%
\bibitem [{\citenamefont {Taitelbaum}\ \emph {et~al.}(2020)\citenamefont
  {Taitelbaum}, \citenamefont {West}, \citenamefont {Assaf},\ and\
  \citenamefont {Mobilia}}]{PhysRevLett.125.048105}%
  \BibitemOpen
  \bibfield  {author} {\bibinfo {author} {\bibfnamefont {A.}~\bibnamefont
  {Taitelbaum}}, \bibinfo {author} {\bibfnamefont {R.}~\bibnamefont {West}},
  \bibinfo {author} {\bibfnamefont {M.}~\bibnamefont {Assaf}},\ and\ \bibinfo
  {author} {\bibfnamefont {M.}~\bibnamefont {Mobilia}},\ }\bibfield  {title}
  {\bibinfo {title} {Population dynamics in a changing environment: Random
  versus periodic switching},\ }\href
  {https://doi.org/10.1103/PhysRevLett.125.048105} {\bibfield  {journal}
  {\bibinfo  {journal} {Phys. Rev. Lett.}\ }\textbf {\bibinfo {volume} {125}},\
  \bibinfo {pages} {048105} (\bibinfo {year} {2020})}\BibitemShut {NoStop}%
\bibitem [{\citenamefont {Iranzo}\ \emph {et~al.}(2020)\citenamefont {Iranzo},
  \citenamefont {Pablo-Mart\'{\i}},\ and\ \citenamefont
  {Aguirre}}]{PhysRevResearch.2.043352}%
  \BibitemOpen
  \bibfield  {author} {\bibinfo {author} {\bibfnamefont {J.}~\bibnamefont
  {Iranzo}}, \bibinfo {author} {\bibfnamefont {F.}~\bibnamefont
  {Pablo-Mart\'{\i}}},\ and\ \bibinfo {author} {\bibfnamefont {J.}~\bibnamefont
  {Aguirre}},\ }\bibfield  {title} {\bibinfo {title} {Emergence of complex
  socioeconomic networks driven by individual and collective interests},\
  }\href {https://doi.org/10.1103/PhysRevResearch.2.043352} {\bibfield
  {journal} {\bibinfo  {journal} {Phys. Rev. Res.}\ }\textbf {\bibinfo {volume}
  {2}},\ \bibinfo {pages} {043352} (\bibinfo {year} {2020})}\BibitemShut
  {NoStop}%
\bibitem [{\citenamefont {Neelin}\ \emph {et~al.}(2022)\citenamefont {Neelin},
  \citenamefont {Martinez-Villalobos}, \citenamefont {Stechmann}, \citenamefont
  {Ahmed}, \citenamefont {Chen}, \citenamefont {Norris}, \citenamefont {Kuo},\
  and\ \citenamefont {Lenderink}}]{neelin2022precipitation}%
  \BibitemOpen
  \bibfield  {author} {\bibinfo {author} {\bibfnamefont {J.~D.}\ \bibnamefont
  {Neelin}}, \bibinfo {author} {\bibfnamefont {C.}~\bibnamefont
  {Martinez-Villalobos}}, \bibinfo {author} {\bibfnamefont {S.~N.}\
  \bibnamefont {Stechmann}}, \bibinfo {author} {\bibfnamefont {F.}~\bibnamefont
  {Ahmed}}, \bibinfo {author} {\bibfnamefont {G.}~\bibnamefont {Chen}},
  \bibinfo {author} {\bibfnamefont {J.~M.}\ \bibnamefont {Norris}}, \bibinfo
  {author} {\bibfnamefont {Y.-H.}\ \bibnamefont {Kuo}},\ and\ \bibinfo {author}
  {\bibfnamefont {G.}~\bibnamefont {Lenderink}},\ }\bibfield  {title} {\bibinfo
  {title} {Precipitation extremes and water vapor: Relationships in current
  climate and implications for climate change},\ }\href@noop {} {\bibfield
  {journal} {\bibinfo  {journal} {Curr. Clim. Change Rep.}\ }\textbf {\bibinfo
  {volume} {8}},\ \bibinfo {pages} {17} (\bibinfo {year} {2022})}\BibitemShut
  {NoStop}%
\bibitem [{\citenamefont {Holovatch}\ \emph {et~al.}(2017)\citenamefont
  {Holovatch}, \citenamefont {Kenna},\ and\ \citenamefont
  {Thurner}}]{holovatch2017complex}%
  \BibitemOpen
  \bibfield  {author} {\bibinfo {author} {\bibfnamefont {Y.}~\bibnamefont
  {Holovatch}}, \bibinfo {author} {\bibfnamefont {R.}~\bibnamefont {Kenna}},\
  and\ \bibinfo {author} {\bibfnamefont {S.}~\bibnamefont {Thurner}},\
  }\bibfield  {title} {\bibinfo {title} {Complex systems: physics beyond
  physics},\ }\href@noop {} {\bibfield  {journal} {\bibinfo  {journal} {Eur. J.
  Phys.}\ }\textbf {\bibinfo {volume} {38}},\ \bibinfo {pages} {023002}
  (\bibinfo {year} {2017})}\BibitemShut {NoStop}%
\bibitem [{\citenamefont {Mandal}\ \emph {et~al.}(2019)\citenamefont {Mandal},
  \citenamefont {Arunkumar}, \citenamefont {Breach},\ and\ \citenamefont
  {Simonovic}}]{mandal2019reservoir}%
  \BibitemOpen
  \bibfield  {author} {\bibinfo {author} {\bibfnamefont {S.}~\bibnamefont
  {Mandal}}, \bibinfo {author} {\bibfnamefont {R.}~\bibnamefont {Arunkumar}},
  \bibinfo {author} {\bibfnamefont {P.~A.}\ \bibnamefont {Breach}},\ and\
  \bibinfo {author} {\bibfnamefont {S.~P.}\ \bibnamefont {Simonovic}},\
  }\bibfield  {title} {\bibinfo {title} {Reservoir operations under changing
  climate conditions: hydropower-production perspective},\ }\href@noop {}
  {\bibfield  {journal} {\bibinfo  {journal} {J. Water Resour. Plan. Manag.}\
  }\textbf {\bibinfo {volume} {145}},\ \bibinfo {pages} {04019016} (\bibinfo
  {year} {2019})}\BibitemShut {NoStop}%
\bibitem [{\citenamefont {Kurihara}(1970)}]{kurihara1970statistical}%
  \BibitemOpen
  \bibfield  {author} {\bibinfo {author} {\bibfnamefont {Y.}~\bibnamefont
  {Kurihara}},\ }\bibfield  {title} {\bibinfo {title} {A statistical-dynamical
  model of the general circulation of the atmosphere},\ }\href@noop {}
  {\bibfield  {journal} {\bibinfo  {journal} {J. Atmos. Sci.}\ }\textbf
  {\bibinfo {volume} {27}},\ \bibinfo {pages} {847} (\bibinfo {year}
  {1970})}\BibitemShut {NoStop}%
\bibitem [{\citenamefont {Palmer}(2019)}]{palmer2019stochastic}%
  \BibitemOpen
  \bibfield  {author} {\bibinfo {author} {\bibfnamefont {T.}~\bibnamefont
  {Palmer}},\ }\bibfield  {title} {\bibinfo {title} {Stochastic weather and
  climate models},\ }\href@noop {} {\bibfield  {journal} {\bibinfo  {journal}
  {Nat. Rev. Phys.}\ }\textbf {\bibinfo {volume} {1}},\ \bibinfo {pages} {463}
  (\bibinfo {year} {2019})}\BibitemShut {NoStop}%
\bibitem [{\citenamefont {Hasselmann}(1976)}]{hasselmann1976stochastic}%
  \BibitemOpen
  \bibfield  {author} {\bibinfo {author} {\bibfnamefont {K.}~\bibnamefont
  {Hasselmann}},\ }\bibfield  {title} {\bibinfo {title} {Stochastic climate
  models part i. theory},\ }\href@noop {} {\bibfield  {journal} {\bibinfo
  {journal} {tellus}\ }\textbf {\bibinfo {volume} {28}},\ \bibinfo {pages}
  {473} (\bibinfo {year} {1976})}\BibitemShut {NoStop}%
\bibitem [{\citenamefont {Frankignoul}\ and\ \citenamefont
  {Hasselmann}(1977)}]{frankignoul1977stochastic}%
  \BibitemOpen
  \bibfield  {author} {\bibinfo {author} {\bibfnamefont {C.}~\bibnamefont
  {Frankignoul}}\ and\ \bibinfo {author} {\bibfnamefont {K.}~\bibnamefont
  {Hasselmann}},\ }\bibfield  {title} {\bibinfo {title} {Stochastic climate
  models, part ii application to sea-surface temperature anomalies and
  thermocline variability},\ }\href@noop {} {\bibfield  {journal} {\bibinfo
  {journal} {Tellus}\ }\textbf {\bibinfo {volume} {29}},\ \bibinfo {pages}
  {289} (\bibinfo {year} {1977})}\BibitemShut {NoStop}%
\bibitem [{\citenamefont {Lemke}(1977)}]{lemke1977stochastic}%
  \BibitemOpen
  \bibfield  {author} {\bibinfo {author} {\bibfnamefont {P.}~\bibnamefont
  {Lemke}},\ }\bibfield  {title} {\bibinfo {title} {Stochastic climate models,
  part 3. application to zonally averaged energy models},\ }\href@noop {}
  {\bibfield  {journal} {\bibinfo  {journal} {Tellus}\ }\textbf {\bibinfo
  {volume} {29}},\ \bibinfo {pages} {385} (\bibinfo {year} {1977})}\BibitemShut
  {NoStop}%
\bibitem [{\citenamefont {Raju}\ and\ \citenamefont
  {Kumar}(2020)}]{raju2020review}%
  \BibitemOpen
  \bibfield  {author} {\bibinfo {author} {\bibfnamefont {K.~S.}\ \bibnamefont
  {Raju}}\ and\ \bibinfo {author} {\bibfnamefont {D.~N.}\ \bibnamefont
  {Kumar}},\ }\bibfield  {title} {\bibinfo {title} {Review of approaches for
  selection and ensembling of gcms},\ }\href@noop {} {\bibfield  {journal}
  {\bibinfo  {journal} {J. Water Clim. Change}\ }\textbf {\bibinfo {volume}
  {11}},\ \bibinfo {pages} {577} (\bibinfo {year} {2020})}\BibitemShut
  {NoStop}%
\bibitem [{\citenamefont {Vallis}(1982)}]{vallis1982statistical}%
  \BibitemOpen
  \bibfield  {author} {\bibinfo {author} {\bibfnamefont {G.~K.}\ \bibnamefont
  {Vallis}},\ }\bibfield  {title} {\bibinfo {title} {A statistical-dynamical
  climate model with a simple hydrology cycle},\ }\href@noop {} {\bibfield
  {journal} {\bibinfo  {journal} {Tellus}\ }\textbf {\bibinfo {volume}
  {\textbf{34}}},\ \bibinfo {pages} {211} (\bibinfo {year} {1982})}\BibitemShut
  {NoStop}%
\bibitem [{\citenamefont {Peters}\ and\ \citenamefont
  {Neelin}(2006)}]{peters2006critical}%
  \BibitemOpen
  \bibfield  {author} {\bibinfo {author} {\bibfnamefont {O.}~\bibnamefont
  {Peters}}\ and\ \bibinfo {author} {\bibfnamefont {J.~D.}\ \bibnamefont
  {Neelin}},\ }\bibfield  {title} {\bibinfo {title} {Critical phenomena in
  atmospheric precipitation},\ }\href@noop {} {\bibfield  {journal} {\bibinfo
  {journal} {Nat. Phys.}\ }\textbf {\bibinfo {volume} {2}},\ \bibinfo {pages}
  {393} (\bibinfo {year} {2006})}\BibitemShut {NoStop}%
\bibitem [{\citenamefont {Bak}\ \emph {et~al.}(1987)\citenamefont {Bak},
  \citenamefont {Tang},\ and\ \citenamefont {Wiesenfeld}}]{PhysRevLett.59.381}%
  \BibitemOpen
  \bibfield  {author} {\bibinfo {author} {\bibfnamefont {P.}~\bibnamefont
  {Bak}}, \bibinfo {author} {\bibfnamefont {C.}~\bibnamefont {Tang}},\ and\
  \bibinfo {author} {\bibfnamefont {K.}~\bibnamefont {Wiesenfeld}},\ }\bibfield
   {title} {\bibinfo {title} {Self-organized criticality: An explanation of the
  1/f noise},\ }\href {https://doi.org/10.1103/PhysRevLett.59.381} {\bibfield
  {journal} {\bibinfo  {journal} {Phys. Rev. Lett.}\ }\textbf {\bibinfo
  {volume} {59}},\ \bibinfo {pages} {381} (\bibinfo {year} {1987})}\BibitemShut
  {NoStop}%
\bibitem [{\citenamefont {Peters}\ \emph {et~al.}(2001)\citenamefont {Peters},
  \citenamefont {Hertlein},\ and\ \citenamefont
  {Christensen}}]{PhysRevLett.88.018701}%
  \BibitemOpen
  \bibfield  {author} {\bibinfo {author} {\bibfnamefont {O.}~\bibnamefont
  {Peters}}, \bibinfo {author} {\bibfnamefont {C.}~\bibnamefont {Hertlein}},\
  and\ \bibinfo {author} {\bibfnamefont {K.}~\bibnamefont {Christensen}},\
  }\bibfield  {title} {\bibinfo {title} {A complexity view of rainfall},\
  }\href {https://doi.org/10.1103/PhysRevLett.88.018701} {\bibfield  {journal}
  {\bibinfo  {journal} {Phys. Rev. Lett.}\ }\textbf {\bibinfo {volume} {88}},\
  \bibinfo {pages} {018701} (\bibinfo {year} {2001})}\BibitemShut {NoStop}%
\bibitem [{\citenamefont {Peters}\ \emph {et~al.}(2010)\citenamefont {Peters},
  \citenamefont {Deluca}, \citenamefont {Corral}, \citenamefont {Neelin},\ and\
  \citenamefont {Holloway}}]{peters2010universality}%
  \BibitemOpen
  \bibfield  {author} {\bibinfo {author} {\bibfnamefont {O.}~\bibnamefont
  {Peters}}, \bibinfo {author} {\bibfnamefont {A.}~\bibnamefont {Deluca}},
  \bibinfo {author} {\bibfnamefont {{\'A}.}~\bibnamefont {Corral}}, \bibinfo
  {author} {\bibfnamefont {J.~D.}\ \bibnamefont {Neelin}},\ and\ \bibinfo
  {author} {\bibfnamefont {C.~E.}\ \bibnamefont {Holloway}},\ }\bibfield
  {title} {\bibinfo {title} {Universality of rain event size distributions},\
  }\href@noop {} {\bibfield  {journal} {\bibinfo  {journal} {J. Stat.
  Mech.-Theory E.}\ }\textbf {\bibinfo {volume} {2010}},\ \bibinfo {pages}
  {P11030} (\bibinfo {year} {2010})}\BibitemShut {NoStop}%
\bibitem [{\citenamefont {Stechmann}\ and\ \citenamefont
  {Neelin}(2014)}]{stechmann2014first}%
  \BibitemOpen
  \bibfield  {author} {\bibinfo {author} {\bibfnamefont {S.~N.}\ \bibnamefont
  {Stechmann}}\ and\ \bibinfo {author} {\bibfnamefont {J.~D.}\ \bibnamefont
  {Neelin}},\ }\bibfield  {title} {\bibinfo {title} {First-passage-time
  prototypes for precipitation statistics},\ }\href@noop {} {\bibfield
  {journal} {\bibinfo  {journal} {J. Atmos. Sci.}\ }\textbf {\bibinfo {volume}
  {71}},\ \bibinfo {pages} {3269} (\bibinfo {year} {2014})}\BibitemShut
  {NoStop}%
\bibitem [{\citenamefont {Yalcin}\ \emph {et~al.}(2016)\citenamefont {Yalcin},
  \citenamefont {Rabassa},\ and\ \citenamefont {Beck}}]{yalcin2016extreme}%
  \BibitemOpen
  \bibfield  {author} {\bibinfo {author} {\bibfnamefont {G.~C.}\ \bibnamefont
  {Yalcin}}, \bibinfo {author} {\bibfnamefont {P.}~\bibnamefont {Rabassa}},\
  and\ \bibinfo {author} {\bibfnamefont {C.}~\bibnamefont {Beck}},\ }\bibfield
  {title} {\bibinfo {title} {Extreme event statistics of daily rainfall:
  Dynamical systems approach},\ }\href@noop {} {\bibfield  {journal} {\bibinfo
  {journal} {J. Phys. A: Math. Theor.}\ }\textbf {\bibinfo {volume} {49}},\
  \bibinfo {pages} {154001} (\bibinfo {year} {2016})}\BibitemShut {NoStop}%
\bibitem [{\citenamefont {Ibebuchi}\ and\ \citenamefont
  {Abu}(2023)}]{ibebuchi2023rainfall}%
  \BibitemOpen
  \bibfield  {author} {\bibinfo {author} {\bibfnamefont {C.~C.}\ \bibnamefont
  {Ibebuchi}}\ and\ \bibinfo {author} {\bibfnamefont {I.-O.}\ \bibnamefont
  {Abu}},\ }\bibfield  {title} {\bibinfo {title} {Rainfall variability patterns
  in nigeria during the rainy season},\ }\href@noop {} {\bibfield  {journal}
  {\bibinfo  {journal} {Sci. Rep.}\ }\textbf {\bibinfo {volume} {13}},\
  \bibinfo {pages} {7888} (\bibinfo {year} {2023})}\BibitemShut {NoStop}%
\bibitem [{\citenamefont {Kurniawan}(2019)}]{kurniawan2019distribution}%
  \BibitemOpen
  \bibfield  {author} {\bibinfo {author} {\bibfnamefont {V.}~\bibnamefont
  {Kurniawan}},\ }\bibfield  {title} {\bibinfo {title} {Distribution fitting on
  rainfall data in jakarta},\ }in\ \href@noop {} {\emph {\bibinfo {booktitle}
  {IOP Conf. Ser. Mater. Sci. Eng.}}},\ Vol.\ \bibinfo {volume} {650}\
  (\bibinfo {organization} {IOP Publishing},\ \bibinfo {year} {2019})\ p.\
  \bibinfo {pages} {012060}\BibitemShut {NoStop}%
\bibitem [{\citenamefont {Cho}\ \emph {et~al.}(2004)\citenamefont {Cho},
  \citenamefont {Bowman},\ and\ \citenamefont {North}}]{cho2004comparison}%
  \BibitemOpen
  \bibfield  {author} {\bibinfo {author} {\bibfnamefont {H.-K.}\ \bibnamefont
  {Cho}}, \bibinfo {author} {\bibfnamefont {K.~P.}\ \bibnamefont {Bowman}},\
  and\ \bibinfo {author} {\bibfnamefont {G.~R.}\ \bibnamefont {North}},\
  }\bibfield  {title} {\bibinfo {title} {A comparison of gamma and lognormal
  distributions for characterizing satellite rain rates from the tropical
  rainfall measuring mission},\ }\href@noop {} {\bibfield  {journal} {\bibinfo
  {journal} {J. Appl. Meteorol}\ }\textbf {\bibinfo {volume} {43}},\ \bibinfo
  {pages} {1586} (\bibinfo {year} {2004})}\BibitemShut {NoStop}%
\bibitem [{\citenamefont {Rajeevan}\ \emph {et~al.}(2010)\citenamefont
  {Rajeevan}, \citenamefont {Gadgil},\ and\ \citenamefont
  {Bhate}}]{rajeevan2010active}%
  \BibitemOpen
  \bibfield  {author} {\bibinfo {author} {\bibfnamefont {M.}~\bibnamefont
  {Rajeevan}}, \bibinfo {author} {\bibfnamefont {S.}~\bibnamefont {Gadgil}},\
  and\ \bibinfo {author} {\bibfnamefont {J.}~\bibnamefont {Bhate}},\ }\bibfield
   {title} {\bibinfo {title} {Active and break spells of the indian summer
  monsoon},\ }\href@noop {} {\bibfield  {journal} {\bibinfo  {journal} {J.
  Earth Sys. Sci.}\ }\textbf {\bibinfo {volume} {119}},\ \bibinfo {pages} {229}
  (\bibinfo {year} {2010})}\BibitemShut {NoStop}%
\bibitem [{\citenamefont {Gadgil}(2003)}]{gadgil2003indian}%
  \BibitemOpen
  \bibfield  {author} {\bibinfo {author} {\bibfnamefont {S.}~\bibnamefont
  {Gadgil}},\ }\bibfield  {title} {\bibinfo {title} {The indian monsoon and its
  variability},\ }\href@noop {} {\bibfield  {journal} {\bibinfo  {journal}
  {Annu. Rev. Earth Planet. Sci.}\ }\textbf {\bibinfo {volume} {31}},\ \bibinfo
  {pages} {429} (\bibinfo {year} {2003})}\BibitemShut {NoStop}%
\bibitem [{\citenamefont {Mitra}\ \emph {et~al.}(2018)\citenamefont {Mitra},
  \citenamefont {Apte}, \citenamefont {Govindarajan}, \citenamefont {Vasan},\
  and\ \citenamefont {Vadlamani}}]{mitra2018spatio}%
  \BibitemOpen
  \bibfield  {author} {\bibinfo {author} {\bibfnamefont {A.}~\bibnamefont
  {Mitra}}, \bibinfo {author} {\bibfnamefont {A.}~\bibnamefont {Apte}},
  \bibinfo {author} {\bibfnamefont {R.}~\bibnamefont {Govindarajan}}, \bibinfo
  {author} {\bibfnamefont {V.}~\bibnamefont {Vasan}},\ and\ \bibinfo {author}
  {\bibfnamefont {S.}~\bibnamefont {Vadlamani}},\ }\bibfield  {title} {\bibinfo
  {title} {Spatio-temporal patterns of daily indian summer monsoon rainfall},\
  }\href@noop {} {\bibfield  {journal} {\bibinfo  {journal} {Dyn. and Stat.
  Clim. Sys.}\ }\textbf {\bibinfo {volume} {3}},\ \bibinfo {pages} {dzy010}
  (\bibinfo {year} {2018})}\BibitemShut {NoStop}%
\bibitem [{\citenamefont {Karmakar}\ \emph {et~al.}(2017)\citenamefont
  {Karmakar}, \citenamefont {Chakraborty},\ and\ \citenamefont
  {Nanjundiah}}]{karmakar2017increased}%
  \BibitemOpen
  \bibfield  {author} {\bibinfo {author} {\bibfnamefont {N.}~\bibnamefont
  {Karmakar}}, \bibinfo {author} {\bibfnamefont {A.}~\bibnamefont
  {Chakraborty}},\ and\ \bibinfo {author} {\bibfnamefont {R.~S.}\ \bibnamefont
  {Nanjundiah}},\ }\bibfield  {title} {\bibinfo {title} {Increased sporadic
  extremes decrease the intraseasonal variability in the indian summer monsoon
  rainfall},\ }\href@noop {} {\bibfield  {journal} {\bibinfo  {journal} {Sci.
  Rep.}\ }\textbf {\bibinfo {volume} {7}},\ \bibinfo {pages} {7824} (\bibinfo
  {year} {2017})}\BibitemShut {NoStop}%
\bibitem [{\citenamefont {Goswami}\ \emph {et~al.}(2010)\citenamefont
  {Goswami}, \citenamefont {Mukhopadhyay}, \citenamefont {Mahanta},\ and\
  \citenamefont {Goswami}}]{goswami2010multiscale}%
  \BibitemOpen
  \bibfield  {author} {\bibinfo {author} {\bibfnamefont {B.~B.}\ \bibnamefont
  {Goswami}}, \bibinfo {author} {\bibfnamefont {P.}~\bibnamefont
  {Mukhopadhyay}}, \bibinfo {author} {\bibfnamefont {R.}~\bibnamefont
  {Mahanta}},\ and\ \bibinfo {author} {\bibfnamefont {B.}~\bibnamefont
  {Goswami}},\ }\bibfield  {title} {\bibinfo {title} {Multiscale interaction
  with topography and extreme rainfall events in the northeast indian region},\
  }\href@noop {} {\bibfield  {journal} {\bibinfo  {journal} {J. Geophys. Res.
  Atmos.}\ }\textbf {\bibinfo {volume} {115}} (\bibinfo {year}
  {2010})}\BibitemShut {NoStop}%
\bibitem [{\citenamefont {Jain}\ \emph {et~al.}(2013)\citenamefont {Jain},
  \citenamefont {Kumar},\ and\ \citenamefont {Saharia}}]{jain2013analysis}%
  \BibitemOpen
  \bibfield  {author} {\bibinfo {author} {\bibfnamefont {S.~K.}\ \bibnamefont
  {Jain}}, \bibinfo {author} {\bibfnamefont {V.}~\bibnamefont {Kumar}},\ and\
  \bibinfo {author} {\bibfnamefont {M.}~\bibnamefont {Saharia}},\ }\bibfield
  {title} {\bibinfo {title} {Analysis of rainfall and temperature trends in
  northeast india},\ }\href@noop {} {\bibfield  {journal} {\bibinfo  {journal}
  {Int. J. Climatol.}\ }\textbf {\bibinfo {volume} {33}},\ \bibinfo {pages}
  {968} (\bibinfo {year} {2013})}\BibitemShut {NoStop}%
\bibitem [{\citenamefont {Sharma}\ \emph {et~al.}(2023)\citenamefont {Sharma},
  \citenamefont {Das},\ and\ \citenamefont {Goswami}}]{sharma2023variability}%
  \BibitemOpen
  \bibfield  {author} {\bibinfo {author} {\bibfnamefont {D.}~\bibnamefont
  {Sharma}}, \bibinfo {author} {\bibfnamefont {S.}~\bibnamefont {Das}},\ and\
  \bibinfo {author} {\bibfnamefont {B.}~\bibnamefont {Goswami}},\ }\bibfield
  {title} {\bibinfo {title} {Variability and predictability of the northeast
  india summer monsoon rainfall},\ }\href@noop {} {\bibfield  {journal}
  {\bibinfo  {journal} {Int. J. Climatol.}\ }\textbf {\bibinfo {volume} {43}},\
  \bibinfo {pages} {5248} (\bibinfo {year} {2023})}\BibitemShut {NoStop}%
\bibitem [{dat(2022)}]{datasets}%
  \BibitemOpen
  \href {https://disc.gsfc.nasa.gov} {\bibinfo {title}
  {https://disc.gsfc.nasa.gov}} (\bibinfo {year} {2022})\BibitemShut {NoStop}%
\bibitem [{\citenamefont {Samui}\ \emph {et~al.}(2019)\citenamefont {Samui},
  \citenamefont {Bui}, \citenamefont {Chakraborty},\ and\ \citenamefont
  {Deo}}]{samui2019handbook}%
  \BibitemOpen
  \bibfield  {author} {\bibinfo {author} {\bibfnamefont {P.}~\bibnamefont
  {Samui}}, \bibinfo {author} {\bibfnamefont {D.~T.}\ \bibnamefont {Bui}},
  \bibinfo {author} {\bibfnamefont {S.}~\bibnamefont {Chakraborty}},\ and\
  \bibinfo {author} {\bibfnamefont {R.}~\bibnamefont {Deo}},\ }\href@noop {}
  {\emph {\bibinfo {title} {{Handbook of probabilistic models}}}}\ (\bibinfo
  {publisher} {Butterworth-Heinemann},\ \bibinfo {year} {2019})\BibitemShut
  {NoStop}%
\bibitem [{\citenamefont {Deka}\ \emph {et~al.}(2008)\citenamefont {Deka},
  \citenamefont {Kakaty},\ and\ \citenamefont {Borah}}]{deka2008use}%
  \BibitemOpen
  \bibfield  {author} {\bibinfo {author} {\bibfnamefont {S.}~\bibnamefont
  {Deka}}, \bibinfo {author} {\bibfnamefont {S.}~\bibnamefont {Kakaty}},\ and\
  \bibinfo {author} {\bibfnamefont {M.}~\bibnamefont {Borah}},\ }\bibfield
  {title} {\bibinfo {title} {Use of probability distributions for the analysis
  of daily rainfall data of north east india},\ }\href@noop {} {\bibfield
  {journal} {\bibinfo  {journal} {MAUSAM}\ }\textbf {\bibinfo {volume} {59}},\
  \bibinfo {pages} {518} (\bibinfo {year} {2008})}\BibitemShut {NoStop}%
\bibitem [{\citenamefont {Forbes}\ \emph {et~al.}(2011)\citenamefont {Forbes},
  \citenamefont {Evans}, \citenamefont {Hastings},\ and\ \citenamefont
  {Peacock}}]{forbes2011statistical}%
  \BibitemOpen
  \bibfield  {author} {\bibinfo {author} {\bibfnamefont {C.}~\bibnamefont
  {Forbes}}, \bibinfo {author} {\bibfnamefont {M.}~\bibnamefont {Evans}},
  \bibinfo {author} {\bibfnamefont {N.}~\bibnamefont {Hastings}},\ and\
  \bibinfo {author} {\bibfnamefont {B.}~\bibnamefont {Peacock}},\ }\href@noop
  {} {\emph {\bibinfo {title} {Statistical distributions}}}\ (\bibinfo
  {publisher} {John Wiley \& Sons},\ \bibinfo {year} {2011})\BibitemShut
  {NoStop}%
\bibitem [{\citenamefont {Sagar}\ \emph {et~al.}(2023)\citenamefont {Sagar},
  \citenamefont {Cheng}, \citenamefont {McKinley},\ and\ \citenamefont
  {Agterberg}}]{sagar2023encyclopedia}%
  \BibitemOpen
  \bibfield  {author} {\bibinfo {author} {\bibfnamefont {B.~D.}\ \bibnamefont
  {Sagar}}, \bibinfo {author} {\bibfnamefont {Q.}~\bibnamefont {Cheng}},
  \bibinfo {author} {\bibfnamefont {J.}~\bibnamefont {McKinley}},\ and\
  \bibinfo {author} {\bibfnamefont {F.}~\bibnamefont {Agterberg}},\ }\href@noop
  {} {\emph {\bibinfo {title} {Encyclopedia of mathematical geosciences}}}\
  (\bibinfo  {publisher} {Springer Nature},\ \bibinfo {year}
  {2023})\BibitemShut {NoStop}%
\bibitem [{\citenamefont {Sreedhar}(2019)}]{sreedhar2019fitting}%
  \BibitemOpen
  \bibfield  {author} {\bibinfo {author} {\bibfnamefont {B.}~\bibnamefont
  {Sreedhar}},\ }\bibfield  {title} {\bibinfo {title} {Fitting of probability
  distribution for analyzing the rainfall data in the state of andhra pradesh,
  india},\ }\href@noop {} {\bibfield  {journal} {\bibinfo  {journal} {Int J
  Appl Eng Res}\ }\textbf {\bibinfo {volume} {14}},\ \bibinfo {pages} {835}
  (\bibinfo {year} {2019})}\BibitemShut {NoStop}%
\bibitem [{\citenamefont {Kedem}\ and\ \citenamefont
  {Chiu}(1987)}]{kedem1987lognormality}%
  \BibitemOpen
  \bibfield  {author} {\bibinfo {author} {\bibfnamefont {B.}~\bibnamefont
  {Kedem}}\ and\ \bibinfo {author} {\bibfnamefont {L.~S.}\ \bibnamefont
  {Chiu}},\ }\bibfield  {title} {\bibinfo {title} {On the lognormality of rain
  rate},\ }\href@noop {} {\bibfield  {journal} {\bibinfo  {journal} {Proc.
  Natl. Acad. Sci.}\ }\textbf {\bibinfo {volume} {84}},\ \bibinfo {pages} {901}
  (\bibinfo {year} {1987})}\BibitemShut {NoStop}%
\bibitem [{\citenamefont {Foster}\ \emph {et~al.}(2006)\citenamefont {Foster},
  \citenamefont {Bevis},\ and\ \citenamefont
  {Raymond}}]{foster2006precipitable}%
  \BibitemOpen
  \bibfield  {author} {\bibinfo {author} {\bibfnamefont {J.}~\bibnamefont
  {Foster}}, \bibinfo {author} {\bibfnamefont {M.}~\bibnamefont {Bevis}},\ and\
  \bibinfo {author} {\bibfnamefont {W.}~\bibnamefont {Raymond}},\ }\bibfield
  {title} {\bibinfo {title} {Precipitable water and the lognormal
  distribution},\ }\href@noop {} {\bibfield  {journal} {\bibinfo  {journal} {J.
  Geophys. Res.-Atmos.}\ }\textbf {\bibinfo {volume} {111}} (\bibinfo {year}
  {2006})}\BibitemShut {NoStop}%
\bibitem [{\citenamefont {Atkinson}(1970)}]{atkinson1970method}%
  \BibitemOpen
  \bibfield  {author} {\bibinfo {author} {\bibfnamefont {A.~C.}\ \bibnamefont
  {Atkinson}},\ }\bibfield  {title} {\bibinfo {title} {A method for
  discriminating between models},\ }\href@noop {} {\bibfield  {journal}
  {\bibinfo  {journal} {J. R. Stat.}\ }\textbf {\bibinfo {volume} {32}},\
  \bibinfo {pages} {323} (\bibinfo {year} {1970})}\BibitemShut {NoStop}%
\bibitem [{\citenamefont {Rohde}\ \emph {et~al.}(2014)\citenamefont {Rohde}
  \emph {et~al.}}]{rohde2014introductory}%
  \BibitemOpen
  \bibfield  {author} {\bibinfo {author} {\bibfnamefont {C.~A.}\ \bibnamefont
  {Rohde}} \emph {et~al.},\ }\href@noop {} {\emph {\bibinfo {title}
  {Introductory statistical inference with the likelihood function}}}\
  (\bibinfo  {publisher} {Springer},\ \bibinfo {year} {2014})\BibitemShut
  {NoStop}%
\bibitem [{\citenamefont {Casiraghi}(2021)}]{casiraghi2021likelihood}%
  \BibitemOpen
  \bibfield  {author} {\bibinfo {author} {\bibfnamefont {G.}~\bibnamefont
  {Casiraghi}},\ }\bibfield  {title} {\bibinfo {title} {The likelihood-ratio
  test for multi-edge network models},\ }\href@noop {} {\bibfield  {journal}
  {\bibinfo  {journal} {J. Phys.: Complex}\ }\textbf {\bibinfo {volume} {2}},\
  \bibinfo {pages} {035012} (\bibinfo {year} {2021})}\BibitemShut {NoStop}%
\bibitem [{\citenamefont {Pearson}\ and\ \citenamefont
  {Fox-Kemper}(2018)}]{pearson2018log}%
  \BibitemOpen
  \bibfield  {author} {\bibinfo {author} {\bibfnamefont {B.}~\bibnamefont
  {Pearson}}\ and\ \bibinfo {author} {\bibfnamefont {B.}~\bibnamefont
  {Fox-Kemper}},\ }\bibfield  {title} {\bibinfo {title} {Log-normal turbulence
  dissipation in global ocean models},\ }\href
  {https://doi.org/10.1103/PhysRevLett.120.094501} {\bibfield  {journal}
  {\bibinfo  {journal} {Phys. Rev. Lett.}\ }\textbf {\bibinfo {volume} {120}},\
  \bibinfo {pages} {094501} (\bibinfo {year} {2018})}\BibitemShut {NoStop}%
\bibitem [{\citenamefont {Slevin}\ and\ \citenamefont
  {Pendry}(1990)}]{slevin1990log}%
  \BibitemOpen
  \bibfield  {author} {\bibinfo {author} {\bibfnamefont {K.~M.}\ \bibnamefont
  {Slevin}}\ and\ \bibinfo {author} {\bibfnamefont {J.~B.}\ \bibnamefont
  {Pendry}},\ }\bibfield  {title} {\bibinfo {title} {Log-normal distribution as
  a description of fluctuations in one-dimensional disordered systems},\ }\href
  {https://doi.org/10.1103/PhysRevB.41.10240} {\bibfield  {journal} {\bibinfo
  {journal} {Phys. Rev. B}\ }\textbf {\bibinfo {volume} {41}},\ \bibinfo
  {pages} {10240} (\bibinfo {year} {1990})}\BibitemShut {NoStop}%
\bibitem [{\citenamefont {Tang}\ \emph {et~al.}(2020)\citenamefont {Tang},
  \citenamefont {Liu}, \citenamefont {Liu}, \citenamefont {Zhong},
  \citenamefont {Wang}, \citenamefont {Lu}, \citenamefont {Wang},\ and\
  \citenamefont {Shen}}]{tang2020lognormal}%
  \BibitemOpen
  \bibfield  {author} {\bibinfo {author} {\bibfnamefont {A.}~\bibnamefont
  {Tang}}, \bibinfo {author} {\bibfnamefont {H.}~\bibnamefont {Liu}}, \bibinfo
  {author} {\bibfnamefont {G.}~\bibnamefont {Liu}}, \bibinfo {author}
  {\bibfnamefont {Y.}~\bibnamefont {Zhong}}, \bibinfo {author} {\bibfnamefont
  {L.}~\bibnamefont {Wang}}, \bibinfo {author} {\bibfnamefont {Q.}~\bibnamefont
  {Lu}}, \bibinfo {author} {\bibfnamefont {J.}~\bibnamefont {Wang}},\ and\
  \bibinfo {author} {\bibfnamefont {Y.}~\bibnamefont {Shen}},\ }\bibfield
  {title} {\bibinfo {title} {Lognormal distribution of local strain: A
  universal law of plastic deformation in material},\ }\href
  {https://doi.org/10.1103/PhysRevLett.124.155501} {\bibfield  {journal}
  {\bibinfo  {journal} {Phys. Rev. Lett.}\ }\textbf {\bibinfo {volume} {124}},\
  \bibinfo {pages} {155501} (\bibinfo {year} {2020})}\BibitemShut {NoStop}%
\bibitem [{\citenamefont {Hosoda}\ \emph {et~al.}(2011)\citenamefont {Hosoda},
  \citenamefont {Matsuura}, \citenamefont {Suzuki},\ and\ \citenamefont
  {Yomo}}]{hosoda2011origin}%
  \BibitemOpen
  \bibfield  {author} {\bibinfo {author} {\bibfnamefont {K.}~\bibnamefont
  {Hosoda}}, \bibinfo {author} {\bibfnamefont {T.}~\bibnamefont {Matsuura}},
  \bibinfo {author} {\bibfnamefont {H.}~\bibnamefont {Suzuki}},\ and\ \bibinfo
  {author} {\bibfnamefont {T.}~\bibnamefont {Yomo}},\ }\bibfield  {title}
  {\bibinfo {title} {Origin of lognormal-like distributions with a common width
  in a growth and division process},\ }\href
  {https://doi.org/10.1103/PhysRevE.83.031118} {\bibfield  {journal} {\bibinfo
  {journal} {Phys. Rev. E}\ }\textbf {\bibinfo {volume} {83}},\ \bibinfo
  {pages} {031118} (\bibinfo {year} {2011})}\BibitemShut {NoStop}%
\bibitem [{\citenamefont {Manola}\ \emph {et~al.}(2018)\citenamefont {Manola},
  \citenamefont {Van Den~Hurk}, \citenamefont {De~Moel},\ and\ \citenamefont
  {Aerts}}]{manola2018future}%
  \BibitemOpen
  \bibfield  {author} {\bibinfo {author} {\bibfnamefont {I.}~\bibnamefont
  {Manola}}, \bibinfo {author} {\bibfnamefont {B.}~\bibnamefont {Van
  Den~Hurk}}, \bibinfo {author} {\bibfnamefont {H.}~\bibnamefont {De~Moel}},\
  and\ \bibinfo {author} {\bibfnamefont {J.~C.}\ \bibnamefont {Aerts}},\
  }\bibfield  {title} {\bibinfo {title} {Future extreme precipitation
  intensities based on a historic event},\ }\href@noop {} {\bibfield  {journal}
  {\bibinfo  {journal} {Hydrol. Earth Syst. Sci.}\ }\textbf {\bibinfo {volume}
  {22}},\ \bibinfo {pages} {3777} (\bibinfo {year} {2018})}\BibitemShut
  {NoStop}%
\bibitem [{\citenamefont {Ngongondo}\ \emph {et~al.}(2011)\citenamefont
  {Ngongondo}, \citenamefont {Xu}, \citenamefont {Gottschalk},\ and\
  \citenamefont {Alemaw}}]{ngongondo2011evaluation}%
  \BibitemOpen
  \bibfield  {author} {\bibinfo {author} {\bibfnamefont {C.}~\bibnamefont
  {Ngongondo}}, \bibinfo {author} {\bibfnamefont {C.-Y.}\ \bibnamefont {Xu}},
  \bibinfo {author} {\bibfnamefont {L.}~\bibnamefont {Gottschalk}},\ and\
  \bibinfo {author} {\bibfnamefont {B.}~\bibnamefont {Alemaw}},\ }\bibfield
  {title} {\bibinfo {title} {Evaluation of spatial and temporal characteristics
  of rainfall in malawi: a case of data scarce region},\ }\href@noop {}
  {\bibfield  {journal} {\bibinfo  {journal} {Theor. Appl. Climatol.}\ }\textbf
  {\bibinfo {volume} {106}},\ \bibinfo {pages} {79} (\bibinfo {year}
  {2011})}\BibitemShut {NoStop}%
\bibitem [{\citenamefont {Ha}\ \emph {et~al.}(2007)\citenamefont {Ha},
  \citenamefont {Jeon},\ and\ \citenamefont {Oh}}]{ha2007spatial}%
  \BibitemOpen
  \bibfield  {author} {\bibinfo {author} {\bibfnamefont {K.-J.}\ \bibnamefont
  {Ha}}, \bibinfo {author} {\bibfnamefont {E.-H.}\ \bibnamefont {Jeon}},\ and\
  \bibinfo {author} {\bibfnamefont {H.-M.}\ \bibnamefont {Oh}},\ }\bibfield
  {title} {\bibinfo {title} {Spatial and temporal characteristics of
  precipitation using an extensive network of ground gauge in the korean
  peninsula},\ }\href@noop {} {\bibfield  {journal} {\bibinfo  {journal}
  {Atmos. Res.}\ }\textbf {\bibinfo {volume} {86}},\ \bibinfo {pages} {330}
  (\bibinfo {year} {2007})}\BibitemShut {NoStop}%
\bibitem [{\citenamefont {Kolmogorov}(1962)}]{kolmogorov1962refinement}%
  \BibitemOpen
  \bibfield  {author} {\bibinfo {author} {\bibfnamefont {A.~N.}\ \bibnamefont
  {Kolmogorov}},\ }\bibfield  {title} {\bibinfo {title} {A refinement of
  previous hypotheses concerning the local structure of turbulence in a viscous
  incompressible fluid at high reynolds number},\ }\href@noop {} {\bibfield
  {journal} {\bibinfo  {journal} {J. Fluid Mech.}\ }\textbf {\bibinfo {volume}
  {13}},\ \bibinfo {pages} {82} (\bibinfo {year} {1962})}\BibitemShut {NoStop}%
\bibitem [{\citenamefont {Rysman}\ \emph {et~al.}(2013)\citenamefont {Rysman},
  \citenamefont {Verrier}, \citenamefont {Lema{\^\i}tre},\ and\ \citenamefont
  {Moreau}}]{rysman2013space}%
  \BibitemOpen
  \bibfield  {author} {\bibinfo {author} {\bibfnamefont {J.-F.}\ \bibnamefont
  {Rysman}}, \bibinfo {author} {\bibfnamefont {S.}~\bibnamefont {Verrier}},
  \bibinfo {author} {\bibfnamefont {Y.}~\bibnamefont {Lema{\^\i}tre}},\ and\
  \bibinfo {author} {\bibfnamefont {E.}~\bibnamefont {Moreau}},\ }\bibfield
  {title} {\bibinfo {title} {Space-time variability of the rainfall over the
  western mediterranean region: A statistical analysis},\ }\href@noop {}
  {\bibfield  {journal} {\bibinfo  {journal} {J. Geophys. Res.-Atmos.}\
  }\textbf {\bibinfo {volume} {118}},\ \bibinfo {pages} {8448} (\bibinfo {year}
  {2013})}\BibitemShut {NoStop}%
\bibitem [{\citenamefont {Torrence}\ and\ \citenamefont
  {Compo}(1998)}]{torrence1998practical}%
  \BibitemOpen
  \bibfield  {author} {\bibinfo {author} {\bibfnamefont {C.}~\bibnamefont
  {Torrence}}\ and\ \bibinfo {author} {\bibfnamefont {G.~P.}\ \bibnamefont
  {Compo}},\ }\bibfield  {title} {\bibinfo {title} {A practical guide to
  wavelet analysis},\ }\href@noop {} {\bibfield  {journal} {\bibinfo  {journal}
  {Bull. Am. Meteorol. Soc.}\ }\textbf {\bibinfo {volume} {79}},\ \bibinfo
  {pages} {61} (\bibinfo {year} {1998})}\BibitemShut {NoStop}%
\bibitem [{\citenamefont {Huang}\ \emph {et~al.}(1998)\citenamefont {Huang},
  \citenamefont {Shen}, \citenamefont {Long}, \citenamefont {Wu}, \citenamefont
  {Shih}, \citenamefont {Zheng}, \citenamefont {Yen}, \citenamefont {Tung},\
  and\ \citenamefont {Liu}}]{huang1998empirical}%
  \BibitemOpen
  \bibfield  {author} {\bibinfo {author} {\bibfnamefont {N.~E.}\ \bibnamefont
  {Huang}}, \bibinfo {author} {\bibfnamefont {Z.}~\bibnamefont {Shen}},
  \bibinfo {author} {\bibfnamefont {S.~R.}\ \bibnamefont {Long}}, \bibinfo
  {author} {\bibfnamefont {M.~C.}\ \bibnamefont {Wu}}, \bibinfo {author}
  {\bibfnamefont {H.~H.}\ \bibnamefont {Shih}}, \bibinfo {author}
  {\bibfnamefont {Q.}~\bibnamefont {Zheng}}, \bibinfo {author} {\bibfnamefont
  {N.-C.}\ \bibnamefont {Yen}}, \bibinfo {author} {\bibfnamefont {C.~C.}\
  \bibnamefont {Tung}},\ and\ \bibinfo {author} {\bibfnamefont {H.~H.}\
  \bibnamefont {Liu}},\ }\bibfield  {title} {\bibinfo {title} {The empirical
  mode decomposition and the hilbert spectrum for nonlinear and non-stationary
  time series analysis},\ }\href@noop {} {\bibfield  {journal} {\bibinfo
  {journal} {Proc. R. Soc. A: Math. Phys. Eng.}\ }\textbf {\bibinfo {volume}
  {454}},\ \bibinfo {pages} {903} (\bibinfo {year} {1998})}\BibitemShut
  {NoStop}%
\bibitem [{\citenamefont {Baker}\ \emph {et~al.}(2012)\citenamefont {Baker},
  \citenamefont {Jordan},\ and\ \citenamefont {Norris}}]{PhysRevB.86.104306}%
  \BibitemOpen
  \bibfield  {author} {\bibinfo {author} {\bibfnamefont {C.~H.}\ \bibnamefont
  {Baker}}, \bibinfo {author} {\bibfnamefont {D.~A.}\ \bibnamefont {Jordan}},\
  and\ \bibinfo {author} {\bibfnamefont {P.~M.}\ \bibnamefont {Norris}},\
  }\bibfield  {title} {\bibinfo {title} {Application of the wavelet transform
  to nanoscale thermal transport},\ }\href
  {https://doi.org/10.1103/PhysRevB.86.104306} {\bibfield  {journal} {\bibinfo
  {journal} {Phys. Rev. B}\ }\textbf {\bibinfo {volume} {86}},\ \bibinfo
  {pages} {104306} (\bibinfo {year} {2012})}\BibitemShut {NoStop}%
\bibitem [{\citenamefont {Contoyiannis}\ \emph {et~al.}(2020)\citenamefont
  {Contoyiannis}, \citenamefont {Potirakis},\ and\ \citenamefont
  {Diakonos}}]{contoyiannis2020wavelet}%
  \BibitemOpen
  \bibfield  {author} {\bibinfo {author} {\bibfnamefont {Y.~F.}\ \bibnamefont
  {Contoyiannis}}, \bibinfo {author} {\bibfnamefont {S.~M.}\ \bibnamefont
  {Potirakis}},\ and\ \bibinfo {author} {\bibfnamefont {F.~K.}\ \bibnamefont
  {Diakonos}},\ }\bibfield  {title} {\bibinfo {title} {Wavelet-based detection
  of scaling behavior in noisy experimental data},\ }\href
  {https://doi.org/10.1103/PhysRevE.101.052104} {\bibfield  {journal} {\bibinfo
   {journal} {Phys. Rev. E}\ }\textbf {\bibinfo {volume} {101}},\ \bibinfo
  {pages} {052104} (\bibinfo {year} {2020})}\BibitemShut {NoStop}%
\bibitem [{\citenamefont {Palu{\v{s}}}(2019)}]{paluvs2019coupling}%
  \BibitemOpen
  \bibfield  {author} {\bibinfo {author} {\bibfnamefont {M.}~\bibnamefont
  {Palu{\v{s}}}},\ }\bibfield  {title} {\bibinfo {title} {Coupling in complex
  systems as information transfer across time scales},\ }\href@noop {}
  {\bibfield  {journal} {\bibinfo  {journal} {Phil. Trans. R. Soc.}\ }\textbf
  {\bibinfo {volume} {377}},\ \bibinfo {pages} {20190094} (\bibinfo {year}
  {2019})}\BibitemShut {NoStop}%
\bibitem [{\citenamefont {Koornwinder}(1993)}]{koornwinder1993wavelets}%
  \BibitemOpen
  \bibfield  {author} {\bibinfo {author} {\bibfnamefont {T.~H.}\ \bibnamefont
  {Koornwinder}},\ }\href@noop {} {\emph {\bibinfo {title} {{Wavelets: an
  elementary treatment of theory and applications}}}},\ Vol.~\bibinfo {volume}
  {1}\ (\bibinfo  {publisher} {World Scientific},\ \bibinfo {year}
  {1993})\BibitemShut {NoStop}%
\bibitem [{\citenamefont {Daubechies}(1992)}]{daubechies1992ten}%
  \BibitemOpen
  \bibfield  {author} {\bibinfo {author} {\bibfnamefont {I.}~\bibnamefont
  {Daubechies}},\ }\href@noop {} {\emph {\bibinfo {title} {Ten lectures on
  wavelets}}}\ (\bibinfo  {publisher} {SIAM},\ \bibinfo {year}
  {1992})\BibitemShut {NoStop}%
\bibitem [{\citenamefont {Simonsen}\ \emph {et~al.}(1998)\citenamefont
  {Simonsen}, \citenamefont {Hansen},\ and\ \citenamefont
  {Nes}}]{simonsen1998determination}%
  \BibitemOpen
  \bibfield  {author} {\bibinfo {author} {\bibfnamefont {I.}~\bibnamefont
  {Simonsen}}, \bibinfo {author} {\bibfnamefont {A.}~\bibnamefont {Hansen}},\
  and\ \bibinfo {author} {\bibfnamefont {O.~M.}\ \bibnamefont {Nes}},\
  }\bibfield  {title} {\bibinfo {title} {Determination of the hurst exponent by
  use of wavelet transforms},\ }\href
  {https://doi.org/10.1103/PhysRevE.58.2779} {\bibfield  {journal} {\bibinfo
  {journal} {Phys. Rev. E}\ }\textbf {\bibinfo {volume} {58}},\ \bibinfo
  {pages} {2779} (\bibinfo {year} {1998})}\BibitemShut {NoStop}%
\bibitem [{\citenamefont {Gadgil}\ and\ \citenamefont
  {Joseph}(2003)}]{gadgil2003breaks}%
  \BibitemOpen
  \bibfield  {author} {\bibinfo {author} {\bibfnamefont {S.}~\bibnamefont
  {Gadgil}}\ and\ \bibinfo {author} {\bibfnamefont {P.}~\bibnamefont
  {Joseph}},\ }\bibfield  {title} {\bibinfo {title} {On breaks of the indian
  monsoon},\ }\href@noop {} {\bibfield  {journal} {\bibinfo  {journal} {J.
  Earth Sys. Sci.}\ }\textbf {\bibinfo {volume} {112}},\ \bibinfo {pages} {529}
  (\bibinfo {year} {2003})}\BibitemShut {NoStop}%
\bibitem [{\citenamefont {Camp}\ \emph {et~al.}(2007)\citenamefont {Camp},
  \citenamefont {Cannizzo},\ and\ \citenamefont
  {Numata}}]{camp2007application}%
  \BibitemOpen
  \bibfield  {author} {\bibinfo {author} {\bibfnamefont {J.~B.}\ \bibnamefont
  {Camp}}, \bibinfo {author} {\bibfnamefont {J.~K.}\ \bibnamefont {Cannizzo}},\
  and\ \bibinfo {author} {\bibfnamefont {K.}~\bibnamefont {Numata}},\
  }\bibfield  {title} {\bibinfo {title} {Application of the hilbert-huang
  transform to the search for gravitational waves},\ }\href
  {https://doi.org/10.1103/PhysRevD.75.061101} {\bibfield  {journal} {\bibinfo
  {journal} {Phys. Rev. D}\ }\textbf {\bibinfo {volume} {75}},\ \bibinfo
  {pages} {061101} (\bibinfo {year} {2007})}\BibitemShut {NoStop}%
\bibitem [{\citenamefont {Takeda}\ \emph {et~al.}(2021)\citenamefont {Takeda},
  \citenamefont {Hiranuma}, \citenamefont {Kanda}, \citenamefont {Kotake},
  \citenamefont {Kuroda}, \citenamefont {Negishi}, \citenamefont {Oohara},
  \citenamefont {Sakai}, \citenamefont {Sakai}, \citenamefont {Sawada},
  \citenamefont {Takahashi}, \citenamefont {Tsuchida}, \citenamefont
  {Watanabe},\ and\ \citenamefont {Yokozawa}}]{takeda2021application}%
  \BibitemOpen
  \bibfield  {author} {\bibinfo {author} {\bibfnamefont {M.}~\bibnamefont
  {Takeda}}, \bibinfo {author} {\bibfnamefont {Y.}~\bibnamefont {Hiranuma}},
  \bibinfo {author} {\bibfnamefont {N.}~\bibnamefont {Kanda}}, \bibinfo
  {author} {\bibfnamefont {K.}~\bibnamefont {Kotake}}, \bibinfo {author}
  {\bibfnamefont {T.}~\bibnamefont {Kuroda}}, \bibinfo {author} {\bibfnamefont
  {R.}~\bibnamefont {Negishi}}, \bibinfo {author} {\bibfnamefont
  {K.}~\bibnamefont {Oohara}}, \bibinfo {author} {\bibfnamefont
  {K.}~\bibnamefont {Sakai}}, \bibinfo {author} {\bibfnamefont
  {Y.}~\bibnamefont {Sakai}}, \bibinfo {author} {\bibfnamefont
  {T.}~\bibnamefont {Sawada}}, \bibinfo {author} {\bibfnamefont
  {H.}~\bibnamefont {Takahashi}}, \bibinfo {author} {\bibfnamefont
  {S.}~\bibnamefont {Tsuchida}}, \bibinfo {author} {\bibfnamefont
  {Y.}~\bibnamefont {Watanabe}},\ and\ \bibinfo {author} {\bibfnamefont
  {T.}~\bibnamefont {Yokozawa}},\ }\bibfield  {title} {\bibinfo {title}
  {Application of the hilbert-huang transform for analyzing
  standing-accretion-shock-instability induced gravitational waves in a
  core-collapse supernova},\ }\href
  {https://doi.org/10.1103/PhysRevD.104.084063} {\bibfield  {journal} {\bibinfo
   {journal} {Phys. Rev. D}\ }\textbf {\bibinfo {volume} {104}},\ \bibinfo
  {pages} {084063} (\bibinfo {year} {2021})}\BibitemShut {NoStop}%
\bibitem [{\citenamefont {Vincent}\ \emph {et~al.}(2010)\citenamefont
  {Vincent}, \citenamefont {Giebel}, \citenamefont {Pinson},\ and\
  \citenamefont {Madsen}}]{vincent2010resolving}%
  \BibitemOpen
  \bibfield  {author} {\bibinfo {author} {\bibfnamefont {C.}~\bibnamefont
  {Vincent}}, \bibinfo {author} {\bibfnamefont {G.}~\bibnamefont {Giebel}},
  \bibinfo {author} {\bibfnamefont {P.}~\bibnamefont {Pinson}},\ and\ \bibinfo
  {author} {\bibfnamefont {H.}~\bibnamefont {Madsen}},\ }\bibfield  {title}
  {\bibinfo {title} {Resolving nonstationary spectral information in wind speed
  time series using the hilbert--huang transform},\ }\href@noop {} {\bibfield
  {journal} {\bibinfo  {journal} {J. Appl. Meteorol. Climatol.}\ }\textbf
  {\bibinfo {volume} {49}},\ \bibinfo {pages} {253} (\bibinfo {year}
  {2010})}\BibitemShut {NoStop}%
\bibitem [{\citenamefont {Stroeer}\ \emph {et~al.}(2009)\citenamefont
  {Stroeer}, \citenamefont {Cannizzo}, \citenamefont {Camp},\ and\
  \citenamefont {Gagarin}}]{stroeer2009methods}%
  \BibitemOpen
  \bibfield  {author} {\bibinfo {author} {\bibfnamefont {A.}~\bibnamefont
  {Stroeer}}, \bibinfo {author} {\bibfnamefont {J.~K.}\ \bibnamefont
  {Cannizzo}}, \bibinfo {author} {\bibfnamefont {J.~B.}\ \bibnamefont {Camp}},\
  and\ \bibinfo {author} {\bibfnamefont {N.}~\bibnamefont {Gagarin}},\
  }\bibfield  {title} {\bibinfo {title} {Methods for detection and
  characterization of signals in noisy data with the hilbert-huang transform},\
  }\href {https://doi.org/10.1103/PhysRevD.79.124022} {\bibfield  {journal}
  {\bibinfo  {journal} {Phys. Rev. D}\ }\textbf {\bibinfo {volume} {79}},\
  \bibinfo {pages} {124022} (\bibinfo {year} {2009})}\BibitemShut {NoStop}%
\bibitem [{\citenamefont {Huang}\ \emph {et~al.}(1996)\citenamefont {Huang},
  \citenamefont {Long},\ and\ \citenamefont {Shen}}]{huang1996mechanism}%
  \BibitemOpen
  \bibfield  {author} {\bibinfo {author} {\bibfnamefont {N.~E.}\ \bibnamefont
  {Huang}}, \bibinfo {author} {\bibfnamefont {S.~R.}\ \bibnamefont {Long}},\
  and\ \bibinfo {author} {\bibfnamefont {Z.}~\bibnamefont {Shen}},\ }\bibfield
  {title} {\bibinfo {title} {The mechanism for frequency downshift in nonlinear
  wave evolution},\ }\href@noop {} {\bibfield  {journal} {\bibinfo  {journal}
  {Adv. Appl. Mech.}\ }\textbf {\bibinfo {volume} {32}},\ \bibinfo {pages} {59}
  (\bibinfo {year} {1996})}\BibitemShut {NoStop}%
\bibitem [{\citenamefont {Huang}\ \emph {et~al.}(1999)\citenamefont {Huang},
  \citenamefont {Shen},\ and\ \citenamefont {Long}}]{huang1999new}%
  \BibitemOpen
  \bibfield  {author} {\bibinfo {author} {\bibfnamefont {N.~E.}\ \bibnamefont
  {Huang}}, \bibinfo {author} {\bibfnamefont {Z.}~\bibnamefont {Shen}},\ and\
  \bibinfo {author} {\bibfnamefont {S.~R.}\ \bibnamefont {Long}},\ }\bibfield
  {title} {\bibinfo {title} {A new view of nonlinear water waves: the hilbert
  spectrum},\ }\href@noop {} {\bibfield  {journal} {\bibinfo  {journal} {Ann.
  Rev. Fluid Mech.}\ }\textbf {\bibinfo {volume} {31}},\ \bibinfo {pages} {417}
  (\bibinfo {year} {1999})}\BibitemShut {NoStop}%
\bibitem [{\citenamefont {Kantelhardt}\ \emph {et~al.}(2002)\citenamefont
  {Kantelhardt}, \citenamefont {Zschiegner}, \citenamefont {Koscielny-Bunde},
  \citenamefont {Havlin}, \citenamefont {Bunde},\ and\ \citenamefont
  {Stanley}}]{kantelhardt2002multifractal}%
  \BibitemOpen
  \bibfield  {author} {\bibinfo {author} {\bibfnamefont {J.~W.}\ \bibnamefont
  {Kantelhardt}}, \bibinfo {author} {\bibfnamefont {S.~A.}\ \bibnamefont
  {Zschiegner}}, \bibinfo {author} {\bibfnamefont {E.}~\bibnamefont
  {Koscielny-Bunde}}, \bibinfo {author} {\bibfnamefont {S.}~\bibnamefont
  {Havlin}}, \bibinfo {author} {\bibfnamefont {A.}~\bibnamefont {Bunde}},\ and\
  \bibinfo {author} {\bibfnamefont {H.~E.}\ \bibnamefont {Stanley}},\
  }\bibfield  {title} {\bibinfo {title} {Multifractal detrended fluctuation
  analysis of nonstationary time series},\ }\href@noop {} {\bibfield  {journal}
  {\bibinfo  {journal} {Phys. A: Stat. Mech. Appl.}\ }\textbf {\bibinfo
  {volume} {316}},\ \bibinfo {pages} {87} (\bibinfo {year} {2002})}\BibitemShut
  {NoStop}%
\bibitem [{\citenamefont {Monjo}\ and\ \citenamefont
  {Meseguer-Ruiz}(2024)}]{monjo2024fractal}%
  \BibitemOpen
  \bibfield  {author} {\bibinfo {author} {\bibfnamefont {R.}~\bibnamefont
  {Monjo}}\ and\ \bibinfo {author} {\bibfnamefont {O.}~\bibnamefont
  {Meseguer-Ruiz}},\ }\bibfield  {title} {\bibinfo {title} {Fractal geometry in
  precipitation},\ }\href@noop {} {\bibfield  {journal} {\bibinfo  {journal}
  {Atmos.}\ }\textbf {\bibinfo {volume} {15}},\ \bibinfo {pages} {135}
  (\bibinfo {year} {2024})}\BibitemShut {NoStop}%
\bibitem [{\citenamefont {Saylor}\ \emph {et~al.}(2021)\citenamefont {Saylor},
  \citenamefont {Rundle},\ and\ \citenamefont
  {Donnellan}}]{saylor2021multifractal}%
  \BibitemOpen
  \bibfield  {author} {\bibinfo {author} {\bibfnamefont {C.}~\bibnamefont
  {Saylor}}, \bibinfo {author} {\bibfnamefont {J.~B.}\ \bibnamefont {Rundle}},\
  and\ \bibinfo {author} {\bibfnamefont {A.}~\bibnamefont {Donnellan}},\
  }\bibfield  {title} {\bibinfo {title} {Multifractal analysis of a seismic
  moment distribution obtained from insar inversion},\ }\href@noop {}
  {\bibfield  {journal} {\bibinfo  {journal} {Earth Space Sci.}\ }\textbf
  {\bibinfo {volume} {8}},\ \bibinfo {pages} {e2020EA001433} (\bibinfo {year}
  {2021})}\BibitemShut {NoStop}%
\bibitem [{\citenamefont {Pigolotti}\ \emph {et~al.}(2020)\citenamefont
  {Pigolotti}, \citenamefont {Jensen}, \citenamefont {Zhan},\ and\
  \citenamefont {Tiana}}]{pigolotti2020bifractal}%
  \BibitemOpen
  \bibfield  {author} {\bibinfo {author} {\bibfnamefont {S.}~\bibnamefont
  {Pigolotti}}, \bibinfo {author} {\bibfnamefont {M.~H.}\ \bibnamefont
  {Jensen}}, \bibinfo {author} {\bibfnamefont {Y.}~\bibnamefont {Zhan}},\ and\
  \bibinfo {author} {\bibfnamefont {G.}~\bibnamefont {Tiana}},\ }\bibfield
  {title} {\bibinfo {title} {Bifractal nature of chromosome contact maps},\
  }\href {https://doi.org/10.1103/PhysRevResearch.2.043078} {\bibfield
  {journal} {\bibinfo  {journal} {Phys. Rev. Res.}\ }\textbf {\bibinfo {volume}
  {2}},\ \bibinfo {pages} {043078} (\bibinfo {year} {2020})}\BibitemShut
  {NoStop}%
\bibitem [{\citenamefont {Rodriguez}\ \emph {et~al.}(2009)\citenamefont
  {Rodriguez}, \citenamefont {Vasquez},\ and\ \citenamefont
  {R\"omer}}]{PhysRevLett.102.106406}%
  \BibitemOpen
  \bibfield  {author} {\bibinfo {author} {\bibfnamefont {A.}~\bibnamefont
  {Rodriguez}}, \bibinfo {author} {\bibfnamefont {L.~J.}\ \bibnamefont
  {Vasquez}},\ and\ \bibinfo {author} {\bibfnamefont {R.~A.}\ \bibnamefont
  {R\"omer}},\ }\bibfield  {title} {\bibinfo {title} {Multifractal analysis
  with the probability density function at the three-dimensional anderson
  transition},\ }\href {https://doi.org/10.1103/PhysRevLett.102.106406}
  {\bibfield  {journal} {\bibinfo  {journal} {Phys. Rev. Lett.}\ }\textbf
  {\bibinfo {volume} {102}},\ \bibinfo {pages} {106406} (\bibinfo {year}
  {2009})}\BibitemShut {NoStop}%
\bibitem [{\citenamefont {Kwapie\ifmmode~\acute{n}\else \'{n}\fi{}}\ \emph
  {et~al.}(2023)\citenamefont {Kwapie\ifmmode~\acute{n}\else \'{n}\fi{}},
  \citenamefont {Blasiak}, \citenamefont {Dro\ifmmode \dot{z}\else
  \.{z}\fi{}d\ifmmode~\dot{z}\else \.{z}\fi{}},\ and\ \citenamefont {O\ifmmode
  \acute{s}\else \'{s}\fi{}wi\ifmmode~\mbox{\k{e}}\else
  \k{e}\fi{}cimka}}]{kwapien2023genuine}%
  \BibitemOpen
  \bibfield  {author} {\bibinfo {author} {\bibfnamefont {J.}~\bibnamefont
  {Kwapie\ifmmode~\acute{n}\else \'{n}\fi{}}}, \bibinfo {author} {\bibfnamefont
  {P.}~\bibnamefont {Blasiak}}, \bibinfo {author} {\bibfnamefont
  {S.}~\bibnamefont {Dro\ifmmode \dot{z}\else \.{z}\fi{}d\ifmmode~\dot{z}\else
  \.{z}\fi{}}},\ and\ \bibinfo {author} {\bibfnamefont {P.}~\bibnamefont
  {O\ifmmode \acute{s}\else \'{s}\fi{}wi\ifmmode~\mbox{\k{e}}\else
  \k{e}\fi{}cimka}},\ }\bibfield  {title} {\bibinfo {title} {Genuine
  multifractality in time series is due to temporal correlations},\ }\href
  {https://doi.org/10.1103/PhysRevE.107.034139} {\bibfield  {journal} {\bibinfo
   {journal} {Phys. Rev. E}\ }\textbf {\bibinfo {volume} {107}},\ \bibinfo
  {pages} {034139} (\bibinfo {year} {2023})}\BibitemShut {NoStop}%
\bibitem [{\citenamefont {Cao}\ \emph {et~al.}(2018)\citenamefont {Cao},
  \citenamefont {He}, \citenamefont {Cao} \emph
  {et~al.}}]{cao2018multifractal}%
  \BibitemOpen
  \bibfield  {author} {\bibinfo {author} {\bibfnamefont {G.}~\bibnamefont
  {Cao}}, \bibinfo {author} {\bibfnamefont {L.-Y.}\ \bibnamefont {He}},
  \bibinfo {author} {\bibfnamefont {J.}~\bibnamefont {Cao}}, \emph {et~al.},\
  }\href@noop {} {\emph {\bibinfo {title} {Multifractal detrended analysis
  method and its application in financial markets}}}\ (\bibinfo  {publisher}
  {Springer},\ \bibinfo {year} {2018})\BibitemShut {NoStop}%
\bibitem [{\citenamefont {O\ifmmode \acute{s}\else
  \'{s}\fi{}wi\ifmmode~\mbox{\c{e}}\else \c{e}\fi{}cimka}\ \emph
  {et~al.}(2006)\citenamefont {O\ifmmode \acute{s}\else
  \'{s}\fi{}wi\ifmmode~\mbox{\c{e}}\else \c{e}\fi{}cimka}, \citenamefont
  {Kwapie\ifmmode~\acute{n}\else \'{n}\fi{}},\ and\ \citenamefont {Dro\ifmmode
  \dot{z}\else \.{z}\fi{}d\ifmmode~\dot{z}\else
  \.{z}\fi{}}}]{PhysRevE.74.016103}%
  \BibitemOpen
  \bibfield  {author} {\bibinfo {author} {\bibfnamefont {P.}~\bibnamefont
  {O\ifmmode \acute{s}\else \'{s}\fi{}wi\ifmmode~\mbox{\c{e}}\else
  \c{e}\fi{}cimka}}, \bibinfo {author} {\bibfnamefont {J.}~\bibnamefont
  {Kwapie\ifmmode~\acute{n}\else \'{n}\fi{}}},\ and\ \bibinfo {author}
  {\bibfnamefont {S.}~\bibnamefont {Dro\ifmmode \dot{z}\else
  \.{z}\fi{}d\ifmmode~\dot{z}\else \.{z}\fi{}}},\ }\bibfield  {title} {\bibinfo
  {title} {Wavelet versus detrended fluctuation analysis of multifractal
  structures},\ }\href {https://doi.org/10.1103/PhysRevE.74.016103} {\bibfield
  {journal} {\bibinfo  {journal} {Phys. Rev. E}\ }\textbf {\bibinfo {volume}
  {74}},\ \bibinfo {pages} {016103} (\bibinfo {year} {2006})}\BibitemShut
  {NoStop}%
\bibitem [{\citenamefont {Ihlen}(2012)}]{ihlen2012introductionMFDFA}%
  \BibitemOpen
  \bibfield  {author} {\bibinfo {author} {\bibfnamefont {E.~A.}\ \bibnamefont
  {Ihlen}},\ }\bibfield  {title} {\bibinfo {title} {Introduction to
  multifractal detrended fluctuation analysis in matlab},\ }\href
  {https://doi.org/10.3389/fphys.2012.00141} {\bibfield  {journal} {\bibinfo
  {journal} {Frontiers in physiology}\ }\textbf {\bibinfo {volume} {3}},\
  \bibinfo {pages} {141} (\bibinfo {year} {2012})}\BibitemShut {NoStop}%
\bibitem [{\citenamefont {Hurst}(1951)}]{hurst1951long}%
  \BibitemOpen
  \bibfield  {author} {\bibinfo {author} {\bibfnamefont {H.~E.}\ \bibnamefont
  {Hurst}},\ }\bibfield  {title} {\bibinfo {title} {Long-term storage capacity
  of reservoirs},\ }\href@noop {} {\bibfield  {journal} {\bibinfo  {journal}
  {Trans. Am. Soc. Civil Eng.}\ }\textbf {\bibinfo {volume} {116}},\ \bibinfo
  {pages} {770} (\bibinfo {year} {1951})}\BibitemShut {NoStop}%
\bibitem [{\citenamefont {Mishra}\ and\ \citenamefont
  {Verma}(2010)}]{Mishra2010}%
  \BibitemOpen
  \bibfield  {author} {\bibinfo {author} {\bibfnamefont {P.~K.}\ \bibnamefont
  {Mishra}}\ and\ \bibinfo {author} {\bibfnamefont {M.~K.}\ \bibnamefont
  {Verma}},\ }\bibfield  {title} {\bibinfo {title} {Energy spectra and fluxes
  for rayleigh-b\'enard convection},\ }\href
  {https://doi.org/10.1103/PhysRevE.81.056316} {\bibfield  {journal} {\bibinfo
  {journal} {Phys. Rev. E}\ }\textbf {\bibinfo {volume} {81}},\ \bibinfo
  {pages} {056316} (\bibinfo {year} {2010})}\BibitemShut {NoStop}%
\end{thebibliography}%


\begin{thebibliography}{99}%











\bibitem{neelin2022precipitation}  
J. D. Neelin, C. Martinez-Villalobos, S. N. Stechmann, F. Ahmed, G. Chen, J. M.Norris, Y. H. Kuo, G. Lenderink,  ``Precipitation Extremes and Water Vapor'', Curr. Clim. Chang. Reports, \textbf{1(8)}, 17-33 (2022).


\bibitem{ole2010universality} 
 P. Ole, A. Deluca, Á. Corral, J. D. Neelin, ``Universality of rain event size distributions'', J. Stat. Mech. Th. and Expt. (2010).

 \bibitem{hasan2010simple}   
 M.M. Hasan, P.K. Dunn, ``A simple Poisson–gamma model for modelling rainfall occurrence and amount simultaneously'', Agri. and Forest Meteorol., \textbf{150(10)}, 1319-1330 (2010).


\bibitem{mandal2019reservoir} 
S. Mandal, R. Arunkumar, P.A. Breach, S.P. Simonovic, ``Reservoir operations under changing climate conditions: hydropower-production perspective'', J. Water Res. Plan. and Manage., \textbf{145(5)}, 04019016 (2019).


\bibitem{kurihara1970statistical} 
Y. Kurihara, ``A statistical-dynamical model of the general circulation of the atmosphere'', J. Atmos. Sci., \textbf{27(6)}, 847-70 (1970).

\bibitem{raju2020review} 
K.S. Raju, D.N. Kumar, ``Review of approaches for selection and ensembling of GCMs'', J. Water and Clim. Chang., \textbf{11(3)}, 577-99 (2020).


\bibitem{vallis1982statistical} 
G.K. Vallis, ``A statistical-dynamical climate model with a simple hydrology cycle'', Tellus, \textbf{34(3)}, 211-27 (1982).

\bibitem{hasselmann1976stochastic}
K. Hasselmann, ``Stochastic climate models part I. (Theory)'', tellus, \textbf{28(6)}, 473-85 (1976).

\bibitem{frankignoul1977stochastic}
C. Frankignoul, K. Hasselmann, ``Stochastic climate models, Part II Application to sea-surface temperature anomalies and thermocline variability'', Tellus,  \textbf{29(4)}, 289-305 (1977).
 
\bibitem{lemke1977stochastic} 
 P. Lemke, ``Stochastic climate models, part 3. Application to zonally averaged energy models'', Tellus, \textbf{29(5)}, 385-92 (1977).

\bibitem{peters2006critical} 
 O. Peters, J. D. Neelin, `` Critical phenomena in atmospheric precipitation'', Nature physics, \textbf{2(6)}, 393-6 (2006).

\bibitem{bak1987self}   
    P. Bak, C. Tang, K. Wiesenfeld, ``Self-organized criticality: An explanation of the 1/f noise'', Phys. Rev. Lett., \textbf{59(4)}, 381 (1987).

\bibitem{datasets}
https://disc.gsfc.nasa.gov/datasets/GPM-3IMERGHH-06/summary.

\bibitem{kurniawan2019distribution}
 V. Kurniawan, ``Distribution fitting on rainfall data in Jakarta'', IOP Con. Series: Mat. Sci. and Eng., IOP Publishing, \textbf{650(1)}, 012060 (2019).

\bibitem{deka2008use}
S. Deka, S. C. Kakaty, M. Borah, ``Use of probability distributions for the analysis of daily rainfall data of North East India'', MAUSAM, \textbf{59(4)}, 518-27 (2008).

\bibitem{cho2004comparison}
H. K. Cho, K. P. Bowman, G. R. North, ``A comparison of gamma and lognormal distributions for characterizing satellite rain rates from the tropical rainfall measuring mission'', J. App. Meteorol. and Clim., \textbf{43(11)}, 1586-97 (2004).

\bibitem{sreedhar2019fitting}
  B. R. Sreedhar,`` Fitting of probability distribution for analyzing the rainfall data in the state of Andhra Pradesh, India'', Int. J. Appl. Eng. Res. ,\textbf{14(3)}, 835--839 (2019). 
 
\bibitem{kedem1987lognormality}
  Kedem, Benjamin, L.S.Chiu, ``On the lognormality of rain rate'', PNAS, Natl. Acad. Sci., \textbf{84(4)}, 901-905 (1987).
  
\bibitem{foster2006precipitable}
  J. Foster, M. Bevis, W. Raymond, ``Precipitable water and the lognormal distribution'', J. Geophys Res: Atmos., Wiley Online Library, \textbf{111(D15)}, (2006).

\bibitem{ricciardulli2002local} 
 L. Ricciardulli, P. D. Sardeshmukh, ``Local time-and space scales of organized tropical deep convection'', J. clim.,\textbf{15(19)}, 2775-90 (2002).

\bibitem{kolmogorov1962refinement}
A. N. Kolmogorov, ``A refinement of previous hypotheses concerning the local structure of turbulence in a viscous incompressible fluid at high Reynolds number'', J. Fluid Mech., \textbf{13(1)}, 82-5 (1962).

\bibitem{torrence1998practical}
 C. Torrence,  G. P. Compo, ``A practical guide to wavelet analysis'', Bul. Ameri. Meteorol. society. \textbf{79(1)}, 61-78 (1998).

\bibitem{huang1998empirical}
 N. E. Huang,  Z. Shen, S.R. Long , M. C. Wu, H. H. Shih, Q. Zheng, N. C. Yen , C. C. Tung , H. H. Liu, ``The empirical mode decomposition and the Hilbert spectrum for nonlinear and non-stationary time series analysis'', Proceed.  Royal Soc. of London. Series A: math., phys. and eng. sci., \textbf{454(1971)}, 903-95 (1998 ).

\end{thebibliography}

\end{document}